\documentclass[aps,onecolumn,preprint,superscriptaddress,nofootinbib,floats,tightenlines]{revtex4}
\usepackage{amsmath,amssymb,color,mathrsfs, graphicx,verbatim,epsfig, bbm, wasysym, axodraw}
\usepackage[hyperfootnotes=false]{hyperref}
\usepackage{slashed}
\allowdisplaybreaks

\def\lsim{\mathrel{\rlap{\lower4pt\hbox{\hskip1pt$\sim$}}
    \raise1pt\hbox{$<$}}}
\def\gsim{\mathrel{\rlap{\lower4pt\hbox{\hskip1pt$\sim$}}
    \raise1pt\hbox{$>$}}} 
\newcommand{\vev}[1]{ \left\langle {#1} \right\rangle }

\newcommand{\be}{\begin{eqnarray}}
\newcommand{\ee}{\end{eqnarray}}

\def\addresses#1#2{\hbox to \hsize{\@tablebox{#1}\hfil\@tablebox{#2}}}
\def\@tablebox#1{\vtop{\hsize=5in \begin{flushleft} #1 \end{flushleft}}}

\def\beq{\begin{equation}}
\def\eeq{\end{equation}}
\def\bit{\begin{itemize}}
\def\eit{\end{itemize}}
\def\beqa{\begin{eqnarray}}
\def\eeqa{\end{eqnarray}}

\def\MadGraph{{\tt MadGraph}}
\def\MadGraph5{{\tt MadGraph5}}

\begin{document}

\baselineskip 0.6cm

\begin{titlepage}

\thispagestyle{empty}

\begin{flushright}
PITT PACC 1707
\end{flushright}

\begin{center}

\vskip 1cm

{\Large \bf Top-Tagging at the Energy Frontier}

\vskip 1.0cm
{\large Zhenyu Han$^{1}$, Minho Son$^{2}$, and Brock Tweedie$^{3}$}
\vskip 0.4cm
$^{1}$ {\it Institute for Theoretical Science, University of Oregon, Eugene, OR 97403, USA} \\
$^{2}$ {\it Department of Physics, Korea Advanced Institute of Science and Technology,\\ 291 Daehak-ro, Yuseong-gu, Daejeon 34141, Republic of Korea} \\
$^{3}$ {\it PITT PACC, Department of Physics and Astronomy, University of Pittsburgh, Pittsburgh, PA 15260}
\vskip 2.0cm

\end{center}

\noindent  At proposed future hadron colliders and in the coming years at the LHC, top quarks will be produced at genuinely multi-TeV energies. Top-tagging at such high energies forces us to confront several new issues in terms of detector capabilities and jet physics. Here, we explore these issues in the context of some simple JHU/CMS-type declustering algorithms and the $N$-subjettiness jet-shape variable $\tau_{32}$. We first highlight the complementarity between the two tagging approaches at particle-level with respect to discriminating top-jets against gluons and quarks, using multivariate optimization scans. We then introduce a basic fast detector simulation, including electromagnetic calorimeter showering patterns determined from {\tt GEANT}. We consider a number of tricks for processing the fast detector output back to an approximate particle-level picture. Re-optimizing the tagger parameters, we demonstrate that the inevitable losses in discrimination power at very high energies can typically be ameliorated. For example, percent-scale mistag rates might be maintained even in extreme cases where an entire top decay would sit inside of one hadronic calorimeter cell and tracking information is completely absent. We then study three novel physics effects that will come up in the multi-TeV energy regime: gluon radiation off of boosted top quarks, mistags originating from $g\to t\bar t$, and mistags originating from $q\to (W/Z) q$ collinear electroweak splittings with subsequent hadronic decays. The first effect, while nominally a nuisance, can actually be harnessed to slightly improve discrimination against gluons. The second effect can lead to effective $O(1)$ enhancements of gluon mistag rates for tight working points. And the third effect, while conceptually interesting, we show to be of highly subleading importance at all energies.

\end{titlepage}

\setcounter{page}{1}

\section{Introduction}
\label{sec:intro}

At energy frontier machines such as the upgraded LHC or a future 100 TeV proton collider, the top quark can be produced with highly relativistic velocities. Similar to relativistic bottom and charm quarks familiar from previous colliders, these relativistic top quarks will appear as jets, and discriminating them against copious light quark-jets and gluon-jets requires dedicated tagging algorithms. In the past several years, many different approaches to top-jet tagging have been developed, utilizing various aspects of jet substructure and specialized treatments of non-isolated leptons. For the dominant hadronic decays of the top quark, which we will focus on here, the general strategy is to exploit the high mass scales and characteristic three-body kinematic features, as well as more detailed aspects of the radiation pattern. Many of these approaches have now been tested against one another and in combination with one another, both in simulation and in collider data~\cite{Abdesselam:2010pt,Altheimer:2012mn,Altheimer:2013yza,Adams:2015hiv,CMStagger4,ATL-PHYS-PUB-2015-053,Aad:2016pux}.

However, the vast majority of such studies have focused on the $\approx$~1~TeV energy scales available to early LHC. As we look ahead to the future capabilities of hadron machines, we must contemplate much higher energies. The HL-LHC, for example, is expected to probe $t\bar t$ resonances up to 6~TeV~\cite{ATL-PHYS-PUB-2013-003}, which would already benefit from top-jet identification up to $p_T \simeq 3$~TeV. A 100~TeV proton collider could reach mass scales of 10's of~TeV. To give a sense of perspective, a top quark with $p_T \simeq 3$~TeV would decay into a patch of $\eta$-$\phi$ space with a characteristic radius $R \lsim 4m_t/p_T \sim 0.2$. This is barely large enough to be resolved within separate hadronic calorimeter cells at either ATLAS or CMS, and the relevant substructures live on even smaller angular scales. Future detectors are expected to have at least $O(1)$ finer angular resolution, but it is currently unclear whether the scaling in angular resolution will be able to match the dramatic shrinking in decay angles that will occur for top quarks with $\approx$~7 times higher energy. In principle, we would need to consider ``top-jets'' with $R \sim 0.03$. We are therefore faced with an immediate question of whether realistic detectors, both present and future, are capable of resolving boosted top quarks within their highest energy searches.

The question of detector performance is potentially compounded by several novel physics issues that appear at very high energies. First and foremost, the top quark will radiate just like an up or charm quark, and will be surrounded in a haze of its own QCD final-state radiation (FSR). Besides making a top-jet look much more like a light quark-jet, this top-FSR can sometimes confuse taggers by generating additional substructure. On the other hand, the distinctive ``quark-like'' radiation pattern potentially offers some extra discrimination power against gluon-jets. Second, at very high energy, gluons can split into a pair of top quarks, analogous to $g \to b \bar b$. While such $g \to t \bar t$ splittings in some sense yield ``genuine'' top-jets, analyses that search for signals of prompt top quark production would consider them as an additional background. Third, with $p_T \gg m_W$, light quark jets gain the opportunity undergo {\it weakstrahlung}, radiating $W$ and $Z$ bosons much as they do photons and gluons. This effect was studied for leptonic top-tagging~\cite{Rehermann:2010vq}, but, to the best of our knowledge, has not been addressed in the context of hadronic top-tagging.

Our goal here will be to study the above detector and physics effects for genuinely multi-TeV top quarks, in the hope of providing a more comprehensive picture of top-tagging at such high energy. We perform these studies within the context of JHU/CMS-type taggers~\cite{Kaplan:2008ie,CMStagger1,CMStagger2} and the powerful jet-shape variable $N$-subjettiness~\cite{Thaler:2010tr}. These two approaches have been shown to have complementary discriminating power in simulation studies~\cite{Adams:2015hiv,CMStagger3}. Loosely speaking, JHU/CMS taggers can capture the ``hard'' substructure of a jet, while $N$-subjettiness is capable of also probing its ``soft'' substructure. We consider optimizations of these two approaches independently of one another and in a simple combined tagger that directly incorporates both. We also discuss the possible merits of track-counting outside of the top decay cone, as a possible way to further improve discrimination against gluon-jets in analogy to light-quark/gluon discrimination~\cite{Gallicchio:2012ez,Gallicchio:2011xq}. Many other approaches to top-tagging also exist (reviewed in~\cite{Abdesselam:2010pt,Altheimer:2012mn,Altheimer:2013yza,Adams:2015hiv}), with various ways of exploiting hard and soft substructure, or combinations thereof, but we take the handful of well-studied approaches considered here as representative. There is also a growing interest in adapting the approaches of deep learning to the problem of top-tagging~\cite{deFavereau:2013fsa,Almeida:2015jua,Kasieczka:2017nvn,Pearkes:2017hku}. Employing these techniques at future colliders could be quite interesting (and possibly inevitable), but we reserve such advanced studies for the future.

Other papers~\cite{Calkins:2013ega,Schaetzel:2013vka,Larkoski:2015yqa} have also performed related studies of multi-TeV top-jets. In~\cite{Calkins:2013ega}, the degrading effects of both top-FSR and detector granularity were highlighted, as well as simple solutions: scale the active top-tagging jet radius as $1/p_T$ (an approach already coarsely applied in the original JHU tagger~\cite{Kaplan:2008ie}) and exploit the fine-grained electromagnetic calorimeter as a tracer of energy flow (an idea earlier advocated in~\cite{Katz:2010mr,Son:2012mb}). Here, both effects will be taken to further extremes, and the latter addressed in more realistic detail. We dub the above calorimeter-based reconstruction strategy {\it EM-flow}. Ref.~\cite{Schaetzel:2013vka} suggested an alternative approach to handling the detector granularity: use tracks as tracers of the energy flow, an approach we call {\it track-flow}. We will include as well a variation of this approach under the idealization of perfect tracking. We also consider combining both approaches to obtain a simple mock-up of full {\it particle-flow}, which exhibits improved resilience to charge-to-neutral fluctuations. (See~\cite{Bressler:2015uma} for a detailed discussion on the theoretical limitations of such approaches.) More recently,~\cite{Larkoski:2015yqa} applied both the scaled jet radius and track-flow ideas to study top-jets and individually quark/gluon-jets up to beyond 10~TeV, using the substructure approaches of $N$-subjettiness~\cite{Thaler:2010tr} and optimized energy correlation functions~\cite{Larkoski:2014zma}. Here, we will revisit some of the same issues, considering complementary substructure and detector reconstruction procedures, more aggressive optimizations, and inclusion of the novel high-$p_T$ physics effects. Some direct comparisons to the track-flow $N$-subjettiness results of~\cite{Larkoski:2015yqa} are also included.

Our main findings regarding detector/algorithm performance are as follows:
\begin{itemize}
\item  Particle-level top-tagging performance becomes approximately scale-invariant at multi-TeV energies. In this regime, the JHU/CMS tagger offers better discrimination against quark-jets than does $N$-subjettiness, whereas the reverse is true for discrimination against gluons. The relative differences in mistag rates are typically $O(10\%)$. A simple combined tagger can implement the best performances from both methods, and appears to allow for nearly simultaneous optimization for discrimination against quarks and gluons.
\item  The scale-invariant behavior is strongly broken by processing the jets through a detector. We explore this using a set of toy detector models with semi-realistic energy deposition patterns. While naive binning into coarse calorimeter cells is particularly detrimental to discrimination power, we show that the more refined reconstruction strategies introduced above offer the potential for much more stable behavior up to $O$(10~TeV) energy. For example, simply folding in higher-granularity information from the ECAL via EM-flow can by itself keep mistag rates at the percent scale.
\item  Tradeoffs between detector reconstruction and substructure algorithm at very high energy can also be nontrivial. $N$-subjettiness degrades more severely than JHU/CMS unless very high-resolution tracking is provided. The combined tagger adjusts itself to take advantage of whichever substructure variables are more strongly performing in each reconstruction scenario. In particular, particle-flow like reconstruction with imperfect tracking, processed through the combined tagger, leads to mistag estimates that are $O(1)$ lower than those predicted in~\cite{Larkoski:2015yqa} using track-flow and $N$-subjettiness.
\end{itemize}

And our main findings regarding physics issues are:
\begin{itemize}
\item  QCD FSR off of energetic top-jets is different than that off of prompt gluon-jets. Simply adding fat-jet track-counting as an additional substructure variable improves top/gluon discrimination by about 20\%.
\item  Collinear $g\to t\bar t$ splittings are a potentially important contribution, and can effectively enhance the mistag rates for gluons by $O$(0.1--1). This can be partially ameliorated using additional cuts such as reconstructed top quark energy fraction. The rate of this background increases logarithmically with energy. (This process should also be seriously studied as a background to leptonic boosted tops.)
\item  Collinear $q\to (W/Z)q$ splittings can effectively enhance the mistag rates for quarks, but only by at most $O$(10\%) for very tight working points. Its (small) importance remains static with increasing energy.
\end{itemize}

The next section reviews the JHU/CMS and $N$-subjettiness techniques which we have selected for study. Section~\ref{sec:particle} establishes their naive baseline performance at multi-TeV energies at particle-level. Section~\ref{sec:detector} then studies the impact of different detector granularity assumptions and reconstruction strategies, based in part on toy {\tt GEANT} simulations of the calorimeters. Section~\ref{sec:physics} proceeds to investigate the possible impact of top-FSR, $g\to t\bar t$ splittings, and weakstrahlung. We present our conclusions and outlook in~\ref{sec:conclusions}. An appendix discusses the details of our detector simulations and shows some plots illustrating the estimated detector effects on substructure distributions.

\section{Review of Substructure Methods}
\label{sec:taggers}

We utilize a JHU/CMS-type declustering top-tagger and the jet-shape variable $N$-subjettiness, described in the following subsections. As we will ultimately find, a simple combination of these two approaches yields a more robust ``combined tagger'' (serving as a basic example of the advantages of multivariate tagging approaches). The full set of clustering/declustering parameters and cut variables are summarized in Table~\ref{tab:ListVariables}, with further details in the descriptions below.

\begin{table}[t!]
\begin{center}
\scalebox{0.995}{
\begin{tabular}{ c|ll }
\hline 
\multicolumn{3}{c}{Combined top tagger} \\[0.1cm]  \hline
%
(De)clustering parameters \quad &\quad $\beta_R$, $\beta_r$, $\delta_p$  where & \\ [0.35cm] 
%
%
&\quad $R(p_T) = \beta_R \times \frac{m_t}{p_T}$ &: jet radius \\[0.35cm]
&\quad $\delta_r(p_T) = \beta_r \times {m_t \over p_T}$ &: cut on subjet collinearity\\[0.35cm]
&\quad $\delta_p(p_T) = \delta_p$=const. &: cut on subjet softness \\[0.35cm]
%
%
Cut variables &\quad $N_{\rm subjets}$ &: number of subjets  \\ [0.3cm] 
&\quad  $m_{\rm subjets}$ &: invariant mass of all subjets \\ [0.3cm] 
&\quad $m_{\rm min}$ &: minimum pairwise mass   \\ [0.3cm] 
&\quad $\tau_{32}\equiv \tau_3/\tau_2$ &: ratio of 3-and 2-subjettiness   \\ [0.3cm] 
\hline
\end{tabular}
} 
\caption{\label{tab:ListVariables}
List of substructure variables in our combined top-tagger. This combines a declustering JHU/CMS-type top-tagger and the jet-shape variable $N$-subjettiness. See text for detailed descriptions.}
\end{center}
\end{table}

\subsection{JHU/CMS (declustering)}

JHU/CMS-type top-taggers~\cite{Kaplan:2008ie,CMStagger1,CMStagger2,Tweedie:2014yda} are immediate descendants of the jet substructure approach introduced in~\cite{Butterworth:2008iy}. Particles or detector elements are first clustered via the Cambridge/Aachen (C/A) sequential recombination algorithm~\cite{Dokshitzer:1997in,Wobisch:1998wt}, which at hadron colliders uses $\Delta R \equiv \sqrt{\Delta\eta^2 + \Delta\phi^2}$ as the distance measure and is characterized by a single jet radius $R$. A candidate jet is then systematically {\it declustered}, serving two purposes:  contaminating ``soft'' radiation is groomed away, and ``hard'' subjets are identified. The subjets serve as our proxies for partonic quarks or gluons at some resolution scale set by declustering parameters. In the case of top quarks, these ideally map to the three decay quarks. Subsequently, multibody kinematic cuts can be applied (subjet counting, subjet-pair masses, reconstructed decay angles, etc). It is rather uncommon for a QCD jet, however processed, to mimic all of the kinematic features characteristic of a top decay, and therein lies the discrimination power.

The operation of the taggers proceed in several stages. The basic operation is a recursive attempt to break a given jet (or subjet) into two hard subjets: 
\begin{enumerate}
\item  Reverse the clustering one stage, resolving branches $j_a$ and $j_b$ (both of which are 4-vectors obtained from all prior $2 \to 1$ clusterings). If there was only one particle to begin with, the subjet search has trivially failed.
\item  Check if the branches are {\it collinear}: $r(j_a,j_b) < \delta_r$, where $r$ is some angular distance measure and $\delta_r$ is a predefined declustering parameter. If collinear, the two branches are considered unresolvable, and again the subjet search has failed.
\item  Check if the branches are {\it soft}: $p_T(j_{a,b})/p_T(J) < \delta_p$, where  $\delta_p$ is another predefined declustering parameter, and $J$ is the entire original jet before any declustering steps. If both branches are soft, then the jet has been completely disassembled into soft radiation, and yet again the subjet search has failed. If one branch is soft and one is hard, throw away the soft branch and continue declustering the hard branch (go back to step~1). If both are above this threshold, then the subjet search has succeeded: both ``hard'' branches are promoted to subjets, and the declustering is stopped. 
\end{enumerate}
If run only once, this procedure is already well-adapted to finding two-body decays such as Higgs, $W$, and $Z$ bosons (\cite{Butterworth:2008iy} originally applied a variation of it to $h\to b\bar b$). To find a three-body top quark decay, it needs to be run one more time. Assuming that the initial subjet search was a success, the two subjets themselves are then declustered via the above steps (still using the original jet $J$ to set the reference $p_T$ scale in step~3). A subjet that fails declustering is simply reconstituted. Depending on the outcomes of these two secondary declusterings, we may have either two, three, or four final subjets. Jets that successfully break into at at least three subjets are considered to be good top candidates. Already at this stage, simple subjet counting serves as a good discriminator against QCD jets.

There is still some freedom in defining the collinear distance measure $r(j_a,j_b)$, as well as the parameters $R$, $\delta_r$, and $\delta_p$. In~\cite{Kaplan:2008ie}, the declustering was optimized on an assumed perfect calorimeter grid, $r$ was defined as the Manhattan distance $|\Delta\eta|+|\Delta\phi|$, and $\delta_r$ was chosen to be a fixed number comparable to the calorimeter cell size. In~\cite{CMStagger2}, the usual Pythagorean distance $\Delta R$ was used, and $\delta_r$ was allowed to shrink linearly with the $p_T$ scale of the jet. The choice of distance measure is to some extent a minor detail, but the evolution of the $\delta_r$ threshold with $p_T$ will be very important. Here we take an approach more similar to~\cite{CMStagger2}, using the Pythagorean distance measure $r(j_a,j_b) \equiv \Delta R(j_a,j_b)$, but defining $\delta_r$ to scale inversely with the jet $p_T$ or some proxy thereof. We apply a similar philosophy to the jet radius. Together,
\begin{eqnarray}
R(p_T) & \,\equiv\, & \beta_R \times \frac{m_t}{p_T} \nonumber \\
\delta_r(p_T) & \,\equiv\, & \beta_r \times \frac{m_t}{p_T} \nonumber \\
\delta_p(p_T) & \,\equiv\, & \delta_p \,=\, {\rm const} \, .  \label{eq:betas}
\end{eqnarray}
From here forward, this defines our set of {\it (de)clustering parameters}: $\beta_R$, $\beta_r$, and $\delta_p$.\footnote{We also point out that there is further freedom in recombining or further declustering the subjets found from this nominal JHU procedure, in order to improve the association between the subjets and the quarks in the top decay. This adds steps to the algorithm, but can have further advantages for applications such as polarization measurement~\cite{Tweedie:2014yda}. We have found that the modified approach of~\cite{Tweedie:2014yda} maintains nearly equivalent discrimination power against QCD jets as that obtainable by the default approach studied here, while offering the additional benefit of enhancing discrimination between left-handed and right-handed chiral tops. However, as polarization is outside the scope of the present article, we reserve discussion of these issues for future work.}

With subjets in-hand, whatever the exact procedure to obtain them, the next question is what multibody kinematic cuts to apply. The original JHU tagger first demands that the 3/4-subjet system mass, $m_{\rm subjets}$, lies within a window about $m_t$. All subjet-pairs are then formed, and the one closest to $m_W$ is identified as the $W$-candidate.\footnote{Methods that can utilize dedicated subjet $b$-tagging, even a very loose version, would of course do better by both breaking the combinatoric ambiguity and adding additional flavor discrimination against backgrounds (see, e.g.,~\cite{CMS-PAS-BTV-15-001,CMS-PAS-BTV-15-002,ATLAS-CONF-2016-001}). However, given the uncertain situation of $b$-tagging at very high-$p_T$, especially at future colliders, we as usual defer on this issue and assume that the $b$-subjets cannot be independently identified.} This system is then also subjected to a mass window cut. Finally, a one-sided cut is applied on the $W$-candidate's helicity angle, defined as the decay angle within the $W$ rest frame relative to parent top's momentum vector. This set of JHU cuts is specified by five parameters: upper and lower top-candidate mass, upper and lower $W$-candidate mass, and helicity angle cut. With the CMS tagger, the $W$ reconstruction step is bypassed, and instead subjet-pairs are formed amongst only the three hardest (excluding any fourth subjet), and the minimum pairwise mass $m_{\rm min}$ is determined. This variable also exhibits a $W$ mass peak, although all events tend to be drawn to smaller values by construction. Subsequently, a one-sided cut is placed on $m_{\rm min}$. This full set of cuts is specified by only three parameters: upper and lower top-candidate mass, and minimum subjet-pair mass cut.\footnote{Technically, another difference is that CMS uses the ungroomed original jet mass, instead of the mass of the collection of subjets after declustering. We continue to use the latter, which expect to be advantageous, though we have not systematically studied the impact of this choice.} Both approaches have been shown to yield comparable performance in optimized simulation studies~\cite{Abdesselam:2010pt}. We have independently verified this behavior at both particle-level and detector-level over a broad range of top $p_T$'s, against both quark-jets and gluon-jets. For the remainder of the main paper, we use the simpler three-parameter CMS cut scheme.

\subsection{N-subjettiness (jet-shape)}

While declustering-based approaches to top-tagging are quite powerful by themselves, they hardly utilize the full information contained in the substructure of the jet. One major difference between top-jets containing hard subjets and QCD-jets containing hard subjets is that, for the former, the subjets are usually formed from showered quarks, whereas for the latter, most of the subjets arise from showered gluons. These gluon-subjets are more ``diffuse.'' Another difference is the structure of the color connections and the phase space available for the shower.  A jet-shape variable that capitalizes on these differences is the $N$-subjettiness ratio $\tau_{32} \equiv \tau_3/\tau_2$~\cite{Thaler:2010tr}. Here, the variables $\tau_N$ are defined as
\beq
\tau_N \,\equiv\, \frac{\underset{\hat j_1,...,\hat j_N}{\min}\left[\sum_i p_T(i) \min \{\Delta R(i,\hat j_1),...,\Delta R(i,\hat j_N)\}\right]}{\sum_i p_T(i) R}
\eeq
In this formula, $i$ labels the jet constituents. The $N$ unit vectors $\hat j_1,...,\hat j_N$ represent candidate subjet axes. The numerator is a weighted sum over the constituent $p_T$'s, with the weight equal to the $\eta$-$\phi$ distance from the closest candidate axis (approximately the sum of splitting $k_T$'s relative to these axes). The axes are chosen so as to minimize this sum. The denominator is effectively an unweighted sum over the constituent $p_T$'s, (essentially the full jet $p_T$), multiplied by the jet radius $R$ for normalization. This term cancels out in the ratio $\tau_{32}$. We do not perform the full numerical minimization over candidate axes~\cite{Thaler:2011gf}, but approximate it using single-pass $k_T$ clustering with the ``winner-take-all'' recombination scheme~\cite{Larkoski:2014uqa}. As for JHU/CMS, we apply $N$-subjettiness only on constituents within a tag-cone that shrinks with $p_T$, as per Eq.~\ref{eq:betas}.

Combining $N$-subjettiness with JHU/CMS is known to form a tagger that is more powerful than either individually~\cite{Adams:2015hiv,CMStagger3}. When performing such a combination, we nominally define the $N$-subjettiness variables before applying the declustering stages of JHU/CMS, which shed some of the jet's soft radiation. However, we have also checked the performance of $\tau_{32}$ on the union of subjet constituents after declustering, and found it to be nearly identical. This suggests that $N$-subjettiness is adding information about the distribution of particles inside the JHU/CMS subjets, rather than in-between them. There is also significant overlap between an $N$-subjettiness cut and the possible kinematic cuts on the hard subjets, including discriminating variables not directly exploited in the JHU/CMS tagger, such as the relative $p_T$ of the softest or next-to-softest subjet. We have found that $N$-subjettiness is more powerful in combination with JHU/CMS than simply defining JHU/CMS with these additional hard kinematic variables. Conversely, we have found that, while a strong cut on $\tau_{32}$ in combination with a top-jet mass window can already define a powerful tagger, the additional grooming, discrete subjet-counting, and kinematic variables provided by JHU/CMS yields even greater discriminating power.

\section{Baseline Performance at Particle-Level}
\label{sec:particle}

We establish our baseline performance evaluations using particle-level Monte Carlo data. The simulations are all performed at a nominal 100~TeV $pp$ collider, though our lower-$p_T$ results should apply as well to the LHC.\footnote{The structure of the underlying event may be somewhat different between the 100~TeV and 14~TeV colliders, and the different PDFs might lead to somewhat different patterns of initial-state radiation. Given the very high energy scales at which we work, we expect any such differences to have little practical importance. Similarly, we neglect the contributions from pileup, which should have minor impact on the hard substructure of the event after even basic jet-cleaning strategies are applied. (Though some impact might be expected on substructure methods sensitive to aspects of the soft radiation pattern or very soft subjets.) See~\cite{Tweedie:2014yda} for a simple study that indicates the robustness of JHU against fairly pessimistic pileup and with fairly simplistic jet-cleaning.} ``Pure'' partonic samples of top, quark, and gluon are defined via the processes $q\bar q \to t\bar t$, $qg \to qZ$, and $q\bar q \to gZ$, with the $t\bar t$ sample decayed into the $\mu$+jets channel and the $Z$ decayed invisibly in the latter two. The hard partons are forced to be central ($|\eta| < 1$) and are generated within specific narrow slices of $p_T$. For all of what follows, ``tag rate'' will be defined using the full sample size at a given $p_T$ as the denominator. The samples are generated using {\tt PYTHIA8}~\cite{Sjostrand:2007gs}, utilizing its default $p_T$-ordered shower, hadronization, and underlying event models. Each sample consists of 100k events. Weak showering and $g\to t\bar t$ are not incorporated at this stage, and QCD FSR off of the top is fixed on, which is the standard configuration for most top-tagging studies to date. (The effects of changing these configurations are to be investigated in Section~\ref{sec:physics}.)

Jet reconstruction and declustering are performed within the {\tt FastJet}~\cite{Cacciari:2005hq} framework. Mini-isolated~\cite{Rehermann:2010vq} leptons are first removed from the event record (isolation radius (15~GeV)/$p_T(l)$, isolation threshold 90\%) to reduce the chance of picking up a semileptonic top decay. The remaining particles are then clustered with anti-$k_T$~\cite{Cacciari:2008gp} at a large radius of $1.0$, and the hardest ``fat-jet'' is identified. The $p_T$ of this fat-jet sets our scale for defining $R$ and $\delta_r$ (via the coefficients $\beta_R$ and $\beta_r$ defined in Eq.~\ref{eq:betas}). The fat-jet's constituents are then reclustered with the C/A algorithm at the radius $R$, and the hardest new small-radius jet thus formed is selected for top-tagging.

It is common in substructure studies to perform optimization scans over mixed samples of quark and gluon jets, e.g. within dijet production. Since a top-tag is a rather multi-purpose tool that might be applied in situations with different quark/gluon-jet background compositions, we prefer to treat them as independent objects, at least in the sense as they are defined in the parton shower. As such, we are already faced with a question of whether a single tagger configuration is even adequate to simultaneously optimize discrimination against both quarks and gluons. To start, we therefore run separate optimizations on each. We scan over (de)clustering parameter choices and rectilinear cut thresholds, and for a given bin in top tag rate, seek out the minimum mistag rate. This defines the usual ROC curves in the plane of tag/mistag rate.

\begin{figure*}[t!]
\begin{center}
\includegraphics[width=0.48\textwidth]{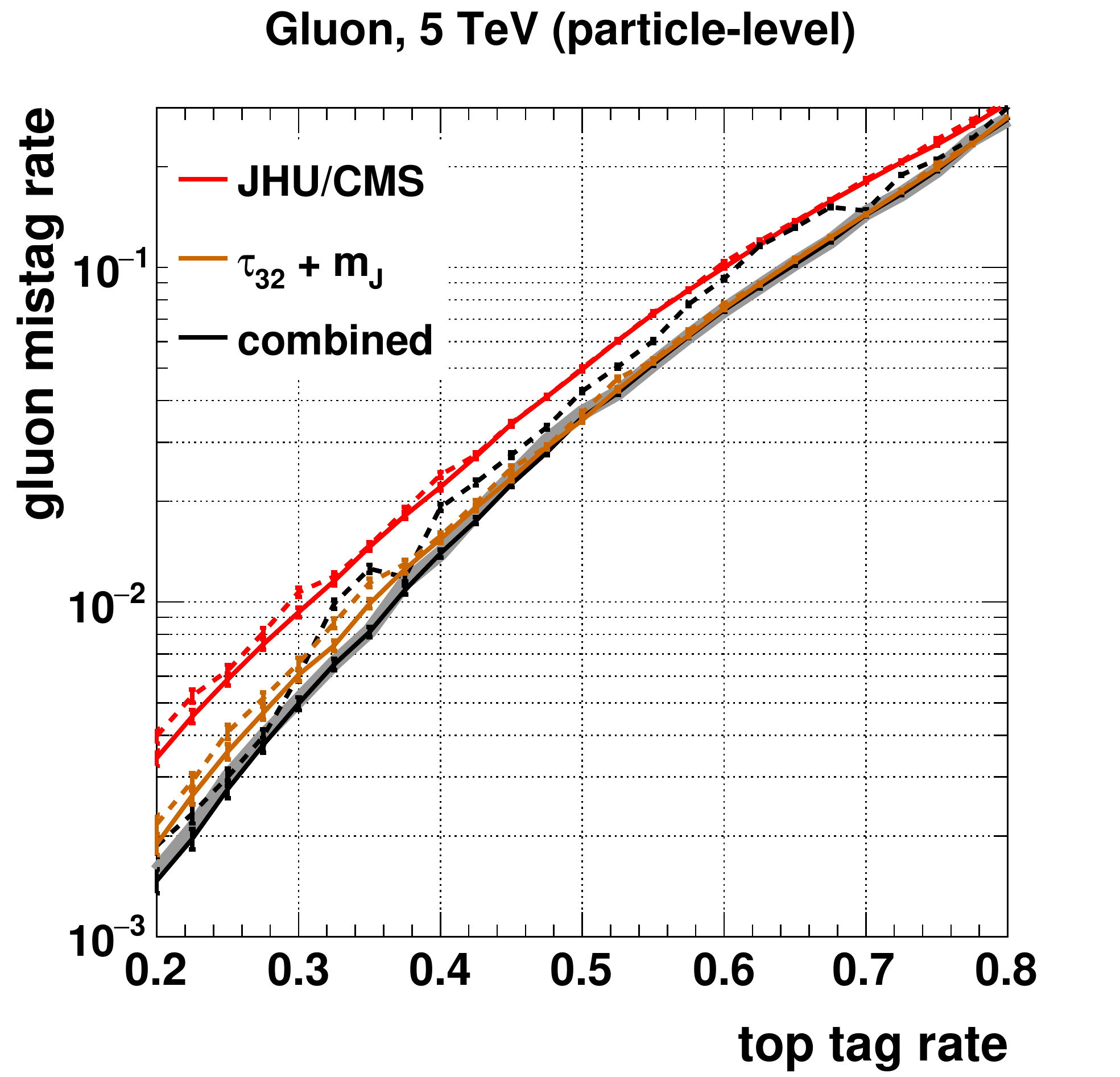} \hspace{0.2cm}
\includegraphics[width=0.48\textwidth]{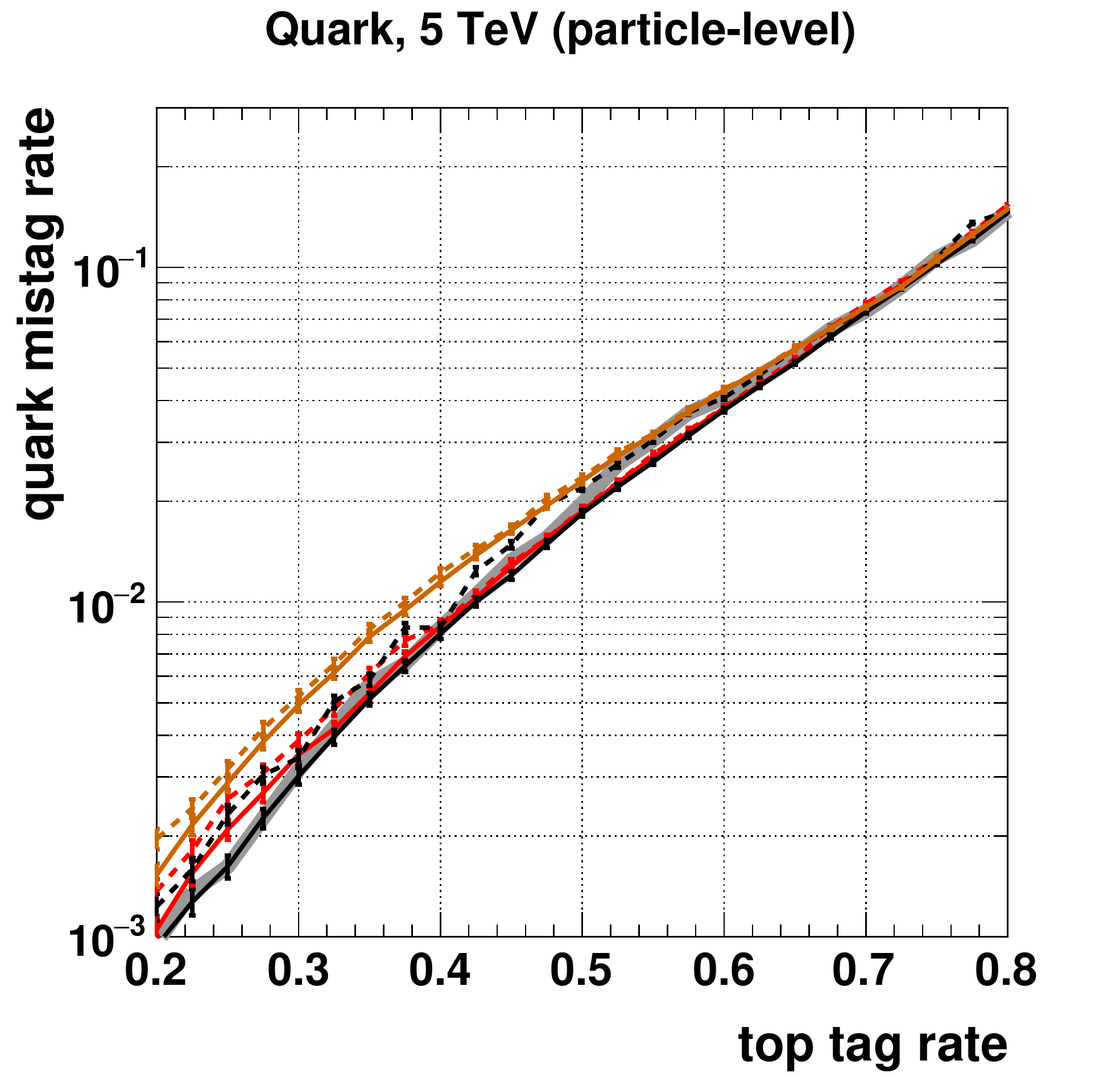}
\caption{Optimized tag/mistag rates for 5~TeV particle-level simulations using different substructure methods. Discrimination against gluons ({\bf left}) and quarks ({\bf right}) are displayed separately. Dashed lines show, e.g., the gluon mistag rates using quark-optimized parameters, and vice versa. The thick gray background line on each plot corresponds to the combined tagger optimized on a 50/50 admixture of gluon- and quark-jets. Approximate Monte Carlo error bars are computed using simple $1/\sqrt{N}$ counting statistics on the mistag rates, and are highly correlated.}
\label{fig:5TeV}
\end{center}
\end{figure*}

Fig.~\ref{fig:5TeV} shows the ROC curves for our 5~TeV samples, including as well the gluon mistag rates obtained with the parameters that minimize the quark mistags, and vice versa. We separately optimize the JHU/CMS declustering tagger, a jet-shape tagger based on $\tau_{32}$ supplemented with an ungroomed (but small-radius) top-jet mass window, and a combined tagger that adds a $\tau_{32}$ cut to the JHU/CMS tagger. For most of the displayed efficiency range for both gluons and quarks, and for all taggers, the optimized jet-radius slope is $\beta_R \simeq 4$. For the JHU/CMS tagger and combined tagger, we also typically find stable declustering parameters, $\beta_r \simeq 0.7$ and $\delta_p \simeq 0.03$. The shapes of the ROC curves are instead dominated by the subjet kinematics and jet-shape cuts, with large variations in $m_{\rm min}$ and $\tau_{32}$ versus efficiency. The optimized subjet-sum mass or top-jet mass cuts also vary, but less dramatically. The optimized window is approximately $m_{\rm subjets} \in [140,200]$~GeV in the vicinity of 50\% top-tag efficiency.

One can immediately observe from Fig.~\ref{fig:5TeV} that the gluon mistag rates are larger than the quark mistag rates by about a factor of 2--3, which owes to their higher splitting rates into hard subjets via QCD showering. It is also clear that there is a larger range of tagger performances for the gluons, growing in size to about a factor of two towards more aggressive tagging configurations. Interestingly, the relative performance between the individual JHU/CMS and $N$-subjettiness taggers flips between gluons and quarks. The difference is automatically picked up on by the combined tagger, which acts approximately like a pure $N$-subjettiness tagger for gluons and like a pure JHU/CMS tagger for quarks. This tendency can be seen to some extent when the combined tagger optimized on quarks is applied to gluons, or vice versa. In particular, the gluon-optimized combined tagger behaves very similarly to the $N$-subjettiness tagger for top-tagging efficiencies above 45\%, whether applied to gluon-jets or quark-jets. For the quark-optimized combined tagger applied to gluon-jets, there is still a noticeable, if highly fluctuating improvement over JHU/CMS for most of the available efficiency range. This behavior results from the fact that the quark optimization still benefits slightly from folding in some $\tau_{32}$, though with rather shallow optimization minima in the space of cuts. By contrast, the individual taggers appear to trivially allow for approximately simultaneous optimization between gluons and quarks. As far as we are aware, this is the first demonstration that gluon-jets and quark-jets exhibit such different behaviors under declustering and jet-shape approaches, at least within the context of the two specific taggers that we picked. This result suggests that aggressive combined taggers could benefit from re-optimization for different applications with different gluon/quark admixtures.

As a simple example of approximately simultaneous optimization of the combined tagger, we re-run the optimization on a 50/50 admixture of gluon-jets and quark-jets, with the result displayed by the thick gray background line in Fig.~\ref{fig:5TeV}. Since the mistag rates are anyway dominated by gluons, these unsurprisingly stay close to their best discrimination, naively dominated by $N$-subjettiness cuts. However, for top-tag rates at and below 50\%, the quark mistags now also come out close to their best discrimination, which was naively dominated by JHU/CMS. Clearly, there is a near-ideal compromise in the expanded space of substructure parameters. This compromise technically becomes less favorable for quark discrimination at higher top-tag rates, though anyway in the region where the $N$-subjettiness and JHU/CMS performances are starting to merge.

\begin{figure*}[t!]
\begin{center}
\includegraphics[width=0.48\textwidth]{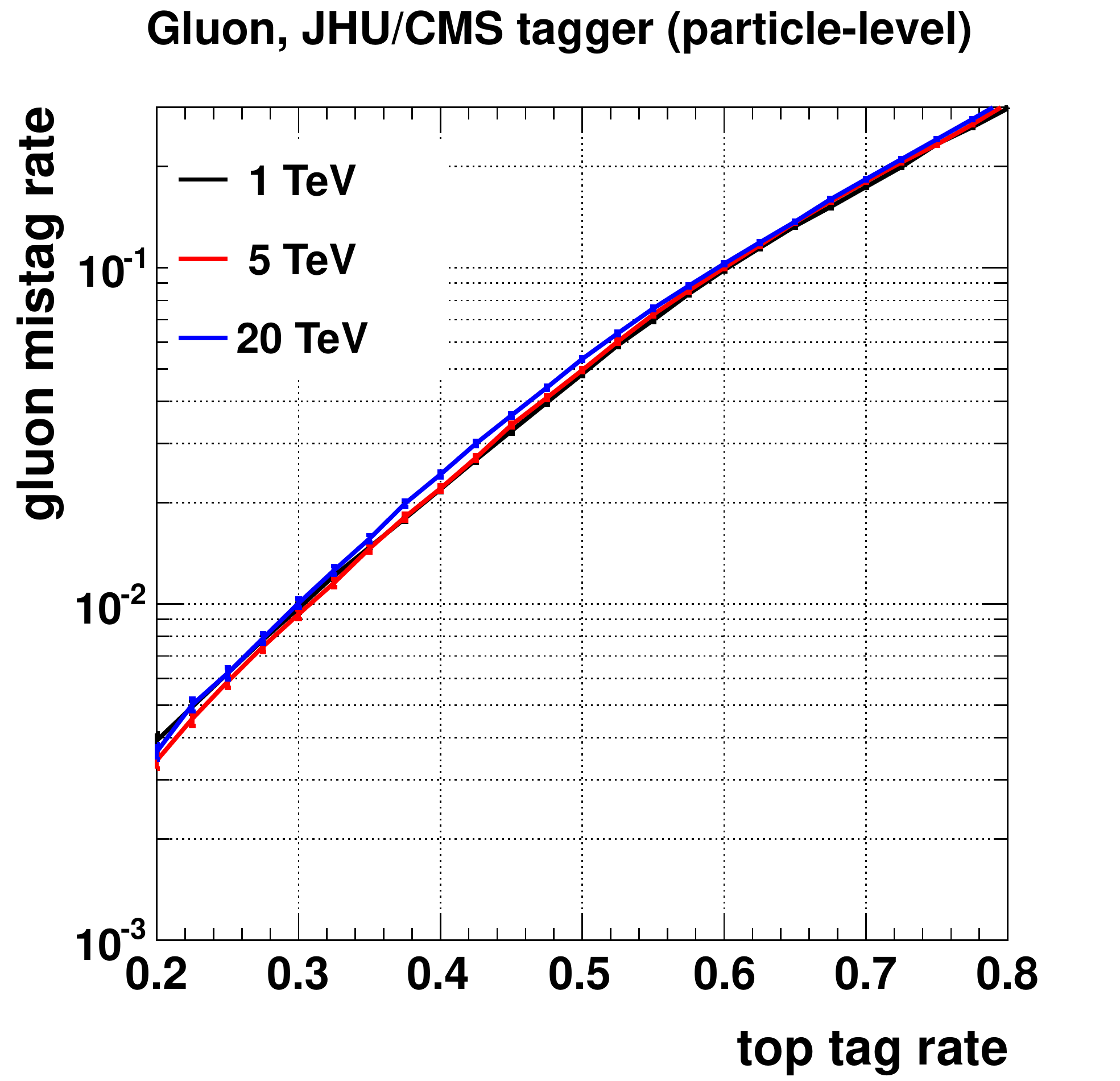} \hspace{0.2cm}
\includegraphics[width=0.48\textwidth]{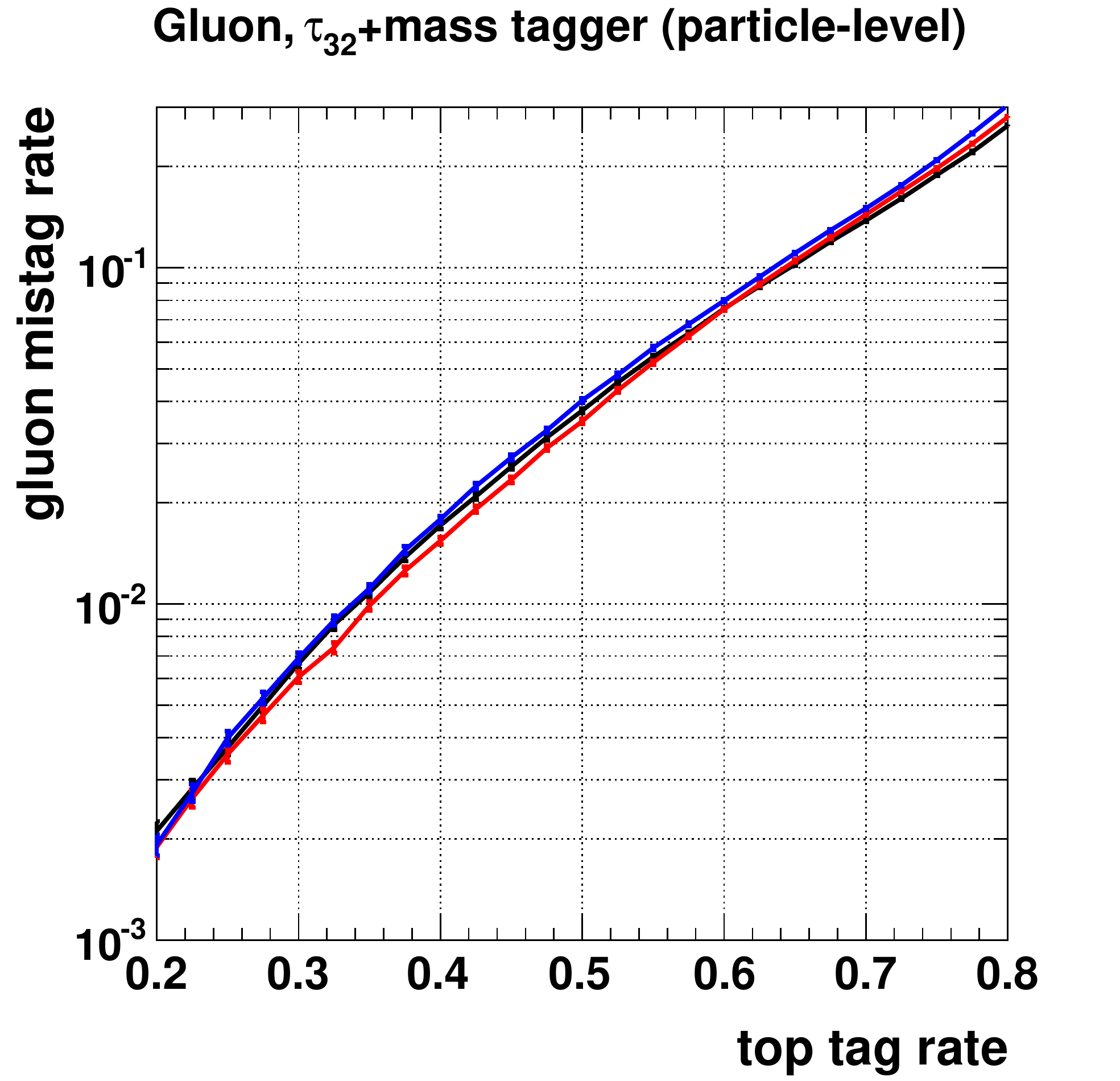} \\
\vspace{0.4cm}
\includegraphics[width=0.48\textwidth]{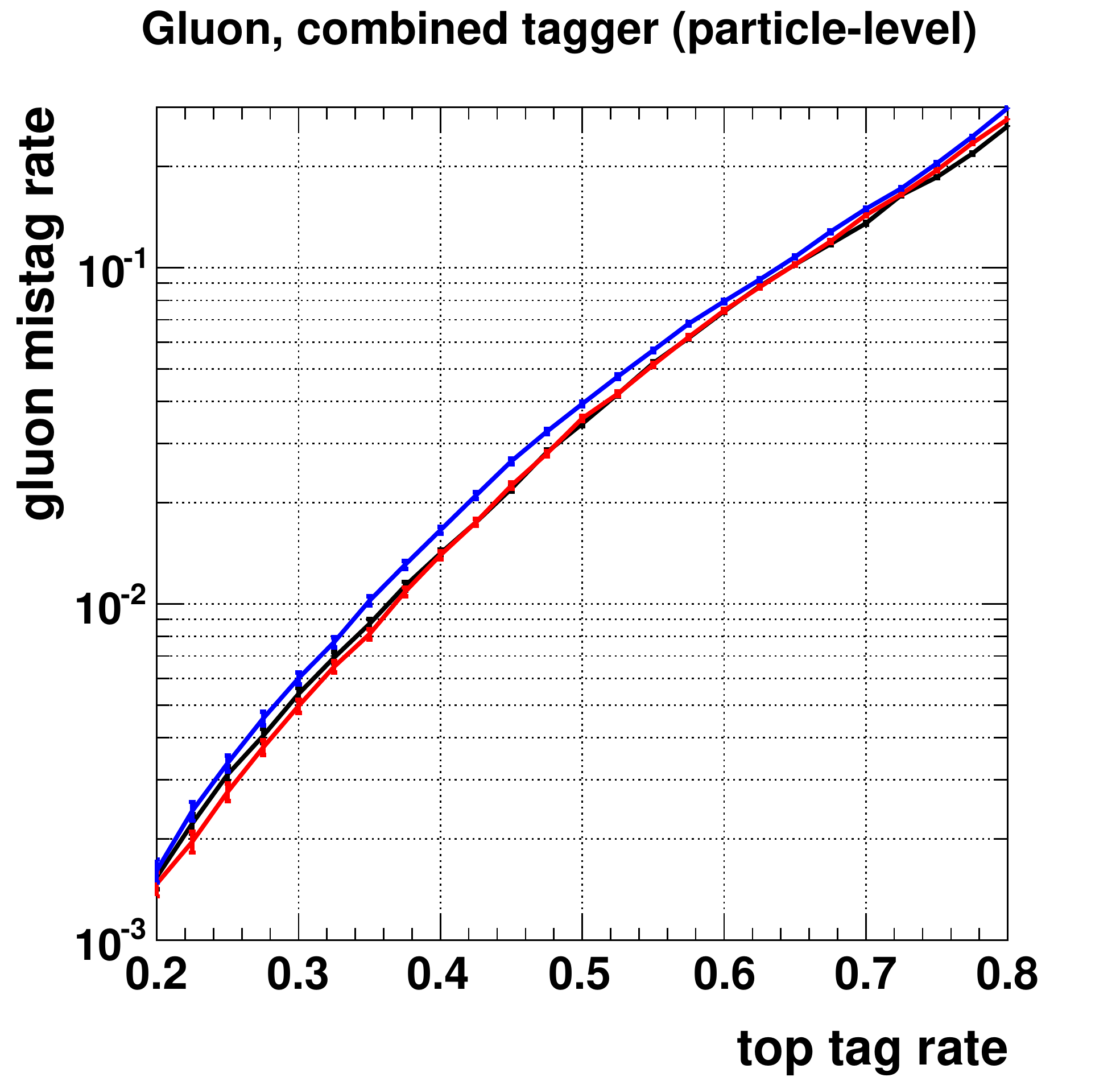}
\caption{Optimized tag/mistag rates against gluons for 1~TeV, 5~TeV, and 20~TeV particle-level simulations. Discrimination with only the JHU/CMS tagger ({\bf upper left}), with the $N$-subjettiness-based tagger, ({\bf upper right}), and with the combined tagger ({\bf bottom}) are displayed separately.}
\label{fig:1-5-20TeV}
\end{center}
\end{figure*}

While the above results use $p_T = 5$~TeV as a benchmark, we point out that the quantitative behavior at particle level is rather stable as a function of $p_T$ within the $O$(1--10)~TeV range of interest. We illustrate this for the three taggers, optimized and applied to gluon-jets, in Fig.~\ref{fig:1-5-20TeV}. (We obtain nearly identical behavior for quark-jets.) A small degradation of discrimination power can be observed at the highest $p_T$ that we study, 20~TeV. The effect appears to be due to a slight reduction in top-jet efficiency for a given set of cuts, in particular due to a leakage of events to more ``gluon-like'' regions in the space of top-tagger variables, with higher $\tau_{32}$ and/or lower $m_{\rm min}$. The gluon-jet efficiency, on the other hand, stays approximately constant as a function of $p_T$ for a given set of cuts. The optimization of the other parameters and cuts is also otherwise largely unchanged. In particular, both $\beta_R$ and $\beta_r$ stay fixed, indicating a simple $1/p_T$ scaling of the optimized jet radius and minimum subjet radius. 

The degrading of particle-level top-tagging efficiencies at higher $p_T$ is a first hint that the top-jets are starting to become more polluted with their own pre-decay FSR radiation. However, the effect is rather modest, and to larger extent we expect the $p_T$-evolution of these taggers to be dominated by the detector effects to be discussed in the next section. The physical consequences of top-FSR, as well as the possibility of further improving discrimination in ways that may evolve with $p_T$ by folding in more global information about the jet containing the top quark, will be discussed in detail in Section~\ref{sec:topFSR}.

\section{Detector Effects}
\label{sec:detector}

Detectors approach as close as possible to particle-level resolution within technological and budgetary constraints, but the inevitable mismatch between detector-level objects and particle-level objects can become a crucial limiting factor for jet substructure at very high energy. Here we make some preliminary investigations into the possible degrading effects from processing our jets through semi-realistic detector mock-ups, with a wide range of assumed performances. The aim here is threefold. Primarily, we would like to make some informed forecasts of what top-tagging quality might reasonably assumed at the upgraded LHC and at a future hadron collider, for the purposes of facilitating phenomenological studies of new physics searches. Secondly, we would like develop an understanding of how much discriminating power can be recovered by combining information from different detector subsystems and different tagging algorithms. Finally, with an eye toward future detector design, we would like to get an initial quantitative sense of to what extent improvements over current technology might be useful.

\subsection{Detector reconstruction strategies and models}
\label{sec:detector_strategies}

The basic inputs into detector-level jet substructure are hadronic calorimeter (HCAL) cells, electromagnetic calorimeter (ECAL) cells, and tracks. In many phenomenological studies, the HCAL is taken to define the ultimate cutoff in angular resolution, which at the LHC is $\Delta\eta\times\Delta\phi \simeq 0.1\times 0.1$. It has been pointed out several times before that this is far too conservative, and that boosted object reconstructions can benefit greatly from folding in the information available in either the ECAL~\cite{Katz:2010mr} or the tracker~\cite{Schaetzel:2013vka}. The former offers 4--5 times finer angular resolution at the LHC, and the latter in principle offers resolution down to angles of $O(10^{-3})$. CMS has applied variations on its particle-flow reconstructions, which combine information from all three systems, to the problem of boosted $W$-tagging~\cite{CMSboostedW} in full simulation. That study found only modest weakening of performance up to $p_T \simeq 3.5$~TeV, where the typical $\Delta R$ between quarks is $\sim 0.05$, using an updated treatment of particle-flow photons and advances in tracking algorithms. According to~\cite{CMSboostedW}, this performance is largely driven by the ECAL rather than the tracking at the highest energies, owing to degrading energy resolution and reconstruction efficiency on the tracks as they become stiffer and more collinear with each other. Presumably, this situation could still change with additional developments, and the analogous situation at future colliders remains to be determined.

For our own investigations, we will for the most part not attempt to invoke a detailed model of the performance of tracking, especially since it appears to be quite complex and possibly contingent on algorithm development beyond our scope. Instead, we will mainly operate on two extreme assumptions that bracket reality: tracking either works perfectly, or not at all. However, we will make some comparisons below to the parametrized tracking performance studied in~\cite{Larkoski:2015yqa}. We will also not employ any sophisticated particle-flow treatments in the manner of CMS, which require very detailed knowledge of the detector performance. Instead, we will focus on fairly minimalistic reconstruction strategies, which we hope will capture the main benefits of particle-flow type reconstructions while staying slightly conservative.

All of our reconstructions are based on generalizations of the trick introduced in~\cite{Katz:2010mr}. There, ECAL cells were locally rescaled to the energy of the full calorimeter, and the HCAL cells discarded. In~\cite{Son:2012mb}, this procedure was more carefully defined for realistic calorimeters, given the presence of energy-sharing between nearby calorimeter cells. The entire collection of ECAL and HCAL cells are first clustered into {\it mini-jets} with the anti-$k_T$ algorithm with $R$ comparable to the HCAL cell size. Here we take this $R$ to be $1.2$ times larger than an HCAL width. Within each mini-jet, a scaling coefficient $(E_{\rm ECAL}+E_{\rm HCAL})/E_{\rm ECAL}$ is defined, and applied to the ECAL cells. These rescaled ECAL cells then serve as the ``particle'' inputs to subsequent jet clustering and substructure. In~\cite{Schaetzel:2013vka}, a similar trick was suggested, using tracks instead of ECAL cells, effectively rescaling them by $(E_{\rm ECAL}+E_{\rm HCAL})/E_{\rm tracks}$. We refer to the former trick as {\it EM-flow}, and the latter as {\it track-flow}.

\begin{figure*}[t!]
\begin{center}
  \includegraphics[width=0.28\textwidth]{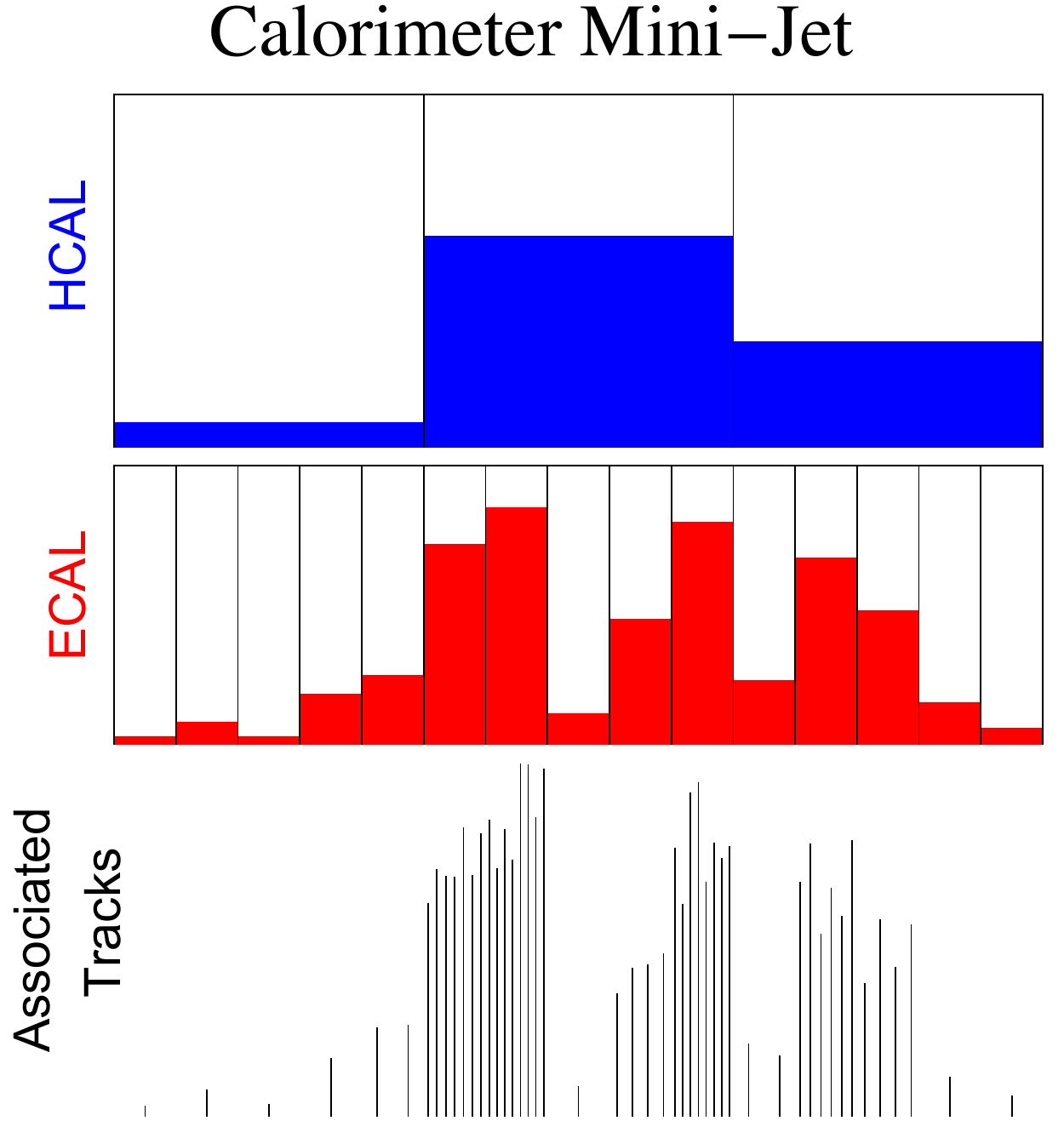} \\[30pt]
  \includegraphics[width=0.28\textwidth]{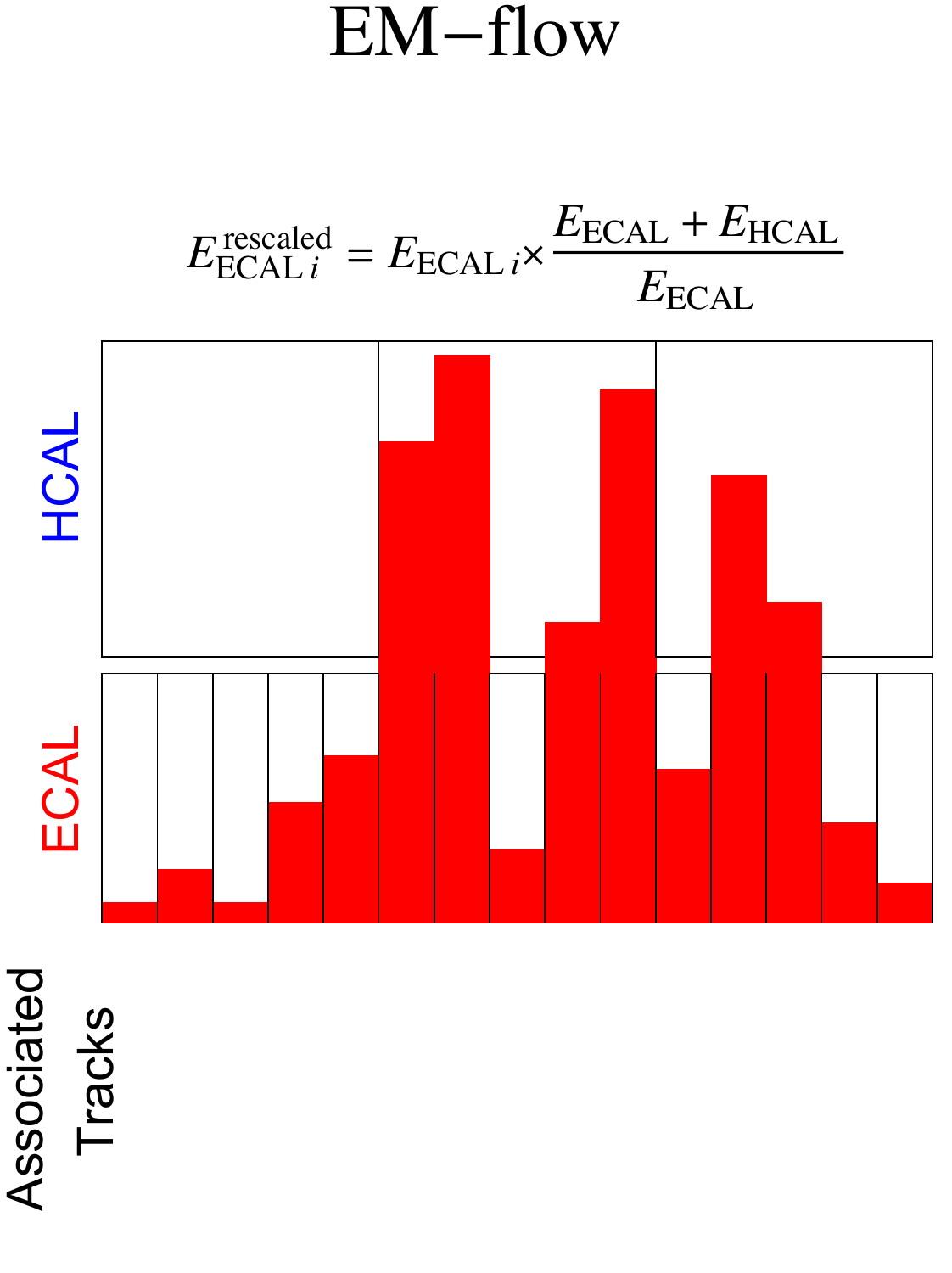} \hspace{0.3cm}
  \includegraphics[width=0.28\textwidth]{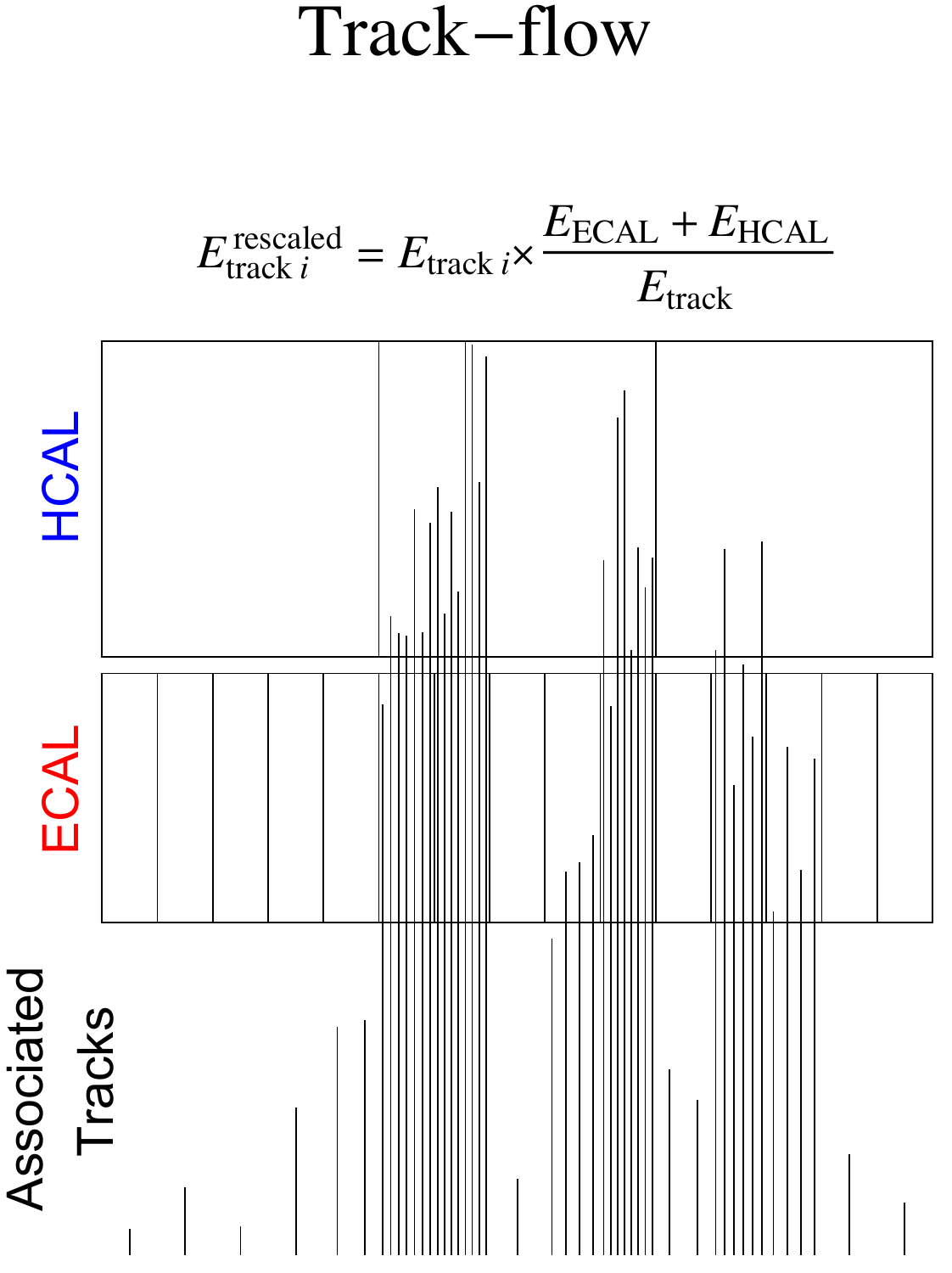} \hspace{0.3cm}
  \includegraphics[width=0.28\textwidth]{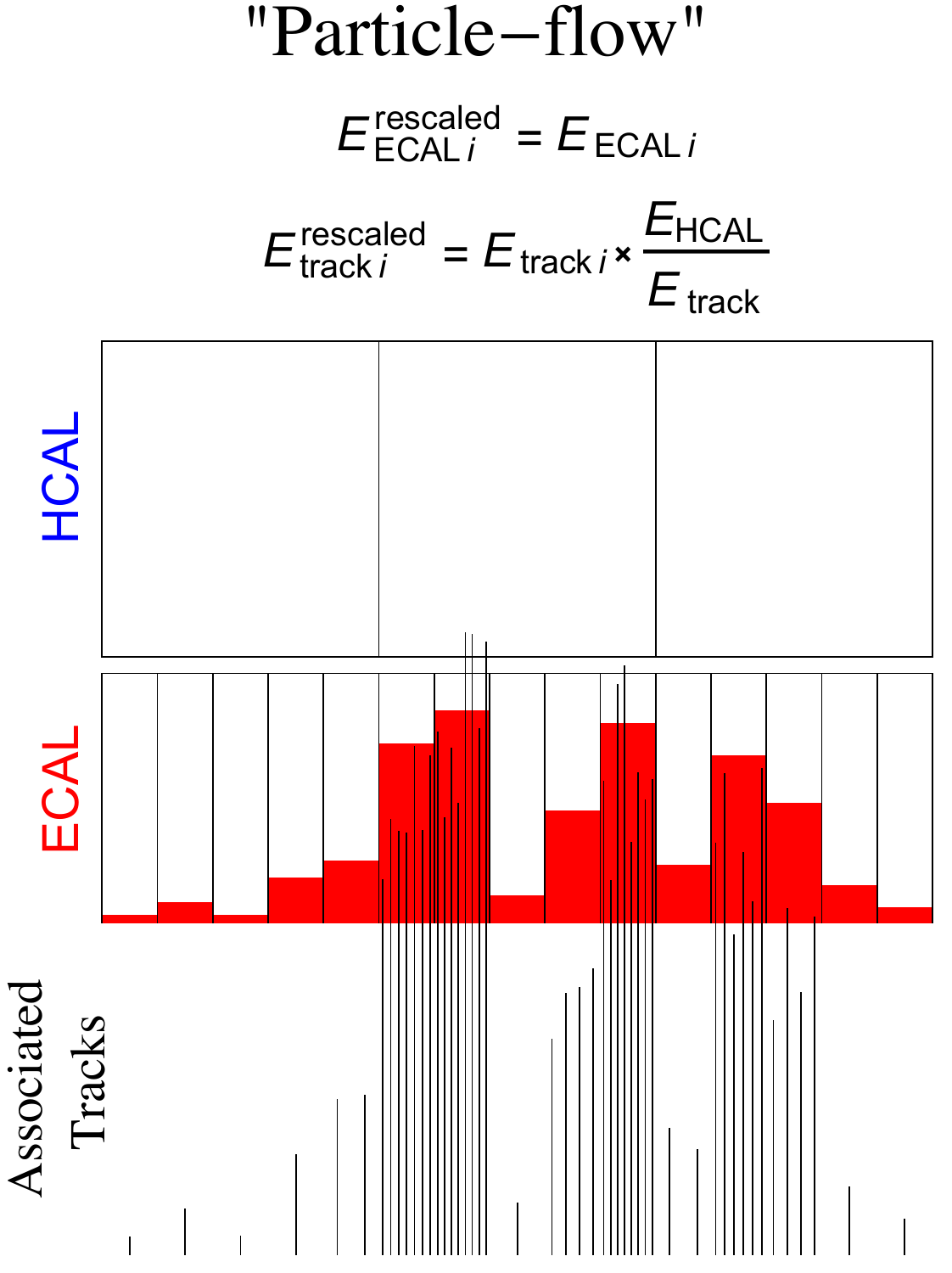}
\caption{Cartoons of our three detector reconstruction approaches. ECAL cells (red) and HCAL cells (blue) are first clustered into {\it mini-jets} of radius $1.2\times\Delta\phi_{\rm HCAL}$. Collections of ECAL cells and/or tracks are then rescaled up in energy, and the HCAL cells deleted.}
\label{fig:toy-detectors}
\end{center}
\end{figure*}

Both of these methods are strongly susceptible to local fluctuations in the charged-to-neutral content of the jet. Despite this, they have been shown to yield perhaps surprisingly good performance when applied to substructure-sensitive observables such as the jet mass, and are certainly better than using raw calorimeter cells as inputs. However, in the fortuitous case of both high-quality tracking and a high-granularity ECAL, combining the two should be even better. Physically, then, the only lost information is the detailed angular distribution of the long-lived neutral hadrons in the jet (mostly neutrons and $K_L$), which leads to a small irreducible loss of performance~\cite{Bressler:2015uma}. Since the HCAL is actually mostly double-counting the track energy, in combination with a subdominant component of long-lived neutral hadron energy, we effectively replace the HCAL with the tracks by rescaling them by $E_{\rm HCAL}/E_{\rm tracks}$, within mini-jets as defined above. The ECAL cells are left as-is. This defines our highly simplified ``{\it particle-flow}'' procedure. Of course, realistic particle-flow is often used to instead leverage the high precision of tracker energy measurements relative to the nominally poorer energy measurements in the calorimeters. However, the situation may actually become reversed at very high energies. In any case, we will indeed demonstrate that our simplified procedure can yield significant tagger performance gains.

All three procedures (EM-flow, track-flow, ``particle-flow'') are illustrated in Fig.~\ref{fig:toy-detectors}.

While our tracking inputs into these procedures (when tracks are available) are just particle-level charged hadrons, our modeling of the calorimeter is more rigorous.\footnote{Throughout, we neglect the effect of the detector's magnetic field on the charged particle trajectories, which we expect to be quite small at such high energies. Moreover, for any softer particles that do become well-separated at the scale of the calorimeter cell size, precision tracking is expected to work without significant degradation.} The ECAL is modeled using {\tt GEANT}~\cite{Agostinelli:2002hh}, and incorporates detailed angular deposition patterns, energy smearing, and deposits from charged and neutral hadrons due to nuclear interactions. $O(20\%)$ of the jet energy becomes absorbed in the ECAL due to this last effect, in fact comparable to the fraction of energy captured from the canonically electromagnetic $\pi^0\to\gamma\gamma$. On average, the ECAL carries around half of the total jet energy. The HCAL is modeled using a simpler parametrization, which should capture the most relevant spatial and energy smearing effects there. The full description of the model, as well as a validation against CMS's high-$p_T$ $W$-jet studies, can be found in Appendix~\ref{sec:detectorModel}.

\begin{table}[t]
\centering
\begin{tabular}{c|cccc}
  Model \    & Tracking          & \ ECAL cell size\  & \ HCAL cell size     \\
\hline
   FCC1 \    & \ perfect/absent \   & $0.01\times0.01$ & $0.05\times0.05$  \\
   FCC2 \    & \ perfect/absent \   & $0.005\times0.005$ & $0.05\times0.05$  \\
\end{tabular}
\caption{Summary of our benchmark FCC detector models. See text for more details.} 
\label{tab:FCCdetectors}
\end{table}

Our baseline detector configuration for a Future Circular Collider (FCC) has a CMS-like calorimeter with an ECAL composed of $2.2 \times 2.2 \times 23$~cm lead tungstate crystals with no longitudinal segmentation.\footnote{The exact depth of the crystal will not be crucial. While a realistic FCC detector would use longer crystals than CMS, the necessary containment depth only scales logarithmically with particle energies.} The crystals are assumed to be arranged around a barrel with inner radius roughly two times larger than CMS. This leads to calorimetry with roughly twice as good angular granularity as CMS. Slightly rounding-up the cell sizes, we choose $\eta$-$\phi$ widths of $0.01$ for the ECAL. (Strictly speaking, this corresponds to an inner ECAL radius that is 1.7 times larger than CMS, or about 2.2~m.) For the HCAL, we assume that the geometry and materials also allow for a similar improvement in angular resolution, and again analogous to CMS make each HCAL cell encapsulate a $5\times 5$ grid of ECAL cells. This leads to an HCAL cell $\eta$-$\phi$ width of $0.05$. We refer to this ECAL/HCAL setup as our ``FCC1'' detector.

We also consider the possibility of using more refined calorimetry, as CMS technology will inevitably be superseded in the coming decades. In principle, the ideal would be tracking calorimeters with a high degree of both angular and longitudinal segmentation, in which the development of the cascade of each particle can be followed in full detail~\cite{Chekanov:2016ppq}. This might return us close to a particle-level picture. But even a somewhat more conventional calorimeter with longitudinal segmentation, and finer transverse granularity at inner radii, would be useful for effectively improving the angular resolution. However, rather than employ a detailed model of such calorimeters or advanced methods to interpret the cascade shapes, we simply take the average between ``perfect'' angular resolution and the conservative FCC1 setup above, namely a longitudinally-integrating ECAL with $\eta$-$\phi$ cells of size $0.005$. Effectively, this would correspond to building the same type of ECAL two times farther away from the beampipe.\footnote{We have also run tests with an artificial ``pure tungsten'' calorimeter, with physical cell dimensions of $1.1 \times 1.1 \times 10$~cm. This exploits the smaller Moli\`ere radius and radiation length of pure tungsten relative to lead tungstate, the former being 9.3~mm versus 19.6~mm. Results come out practically identical to a CMS-like ECAL with enlarged inner radius. We do point out that the more realistic calorimetry of~\cite{Chekanov:2016ppq} is a silicon-tungsten sandwich, with individual cell transverse sizes explored down to $0.3 \times 0.3$~cm. The sPHENIX collaboration has also proposed a tungsten sampling calorimeter with accordion geometry and an effective Moli\`ere radius of 15.4~mm~\cite{sPHENIX1}, about halfway between pure tungsten and lead tungstate.} As for the HCAL, we very conservatively maintain the same configuration as before, namely $0.05$ cells. The exact HCAL resolution will be a subdominant factor in what follows, though of course more refined hadronic calorimetry would only help. (In the simple case where {\it both} the ECAL and HCAL see further factor-of-two improvements in angular resolution, our FCC1 results will approximately apply with an overall rescaling of the energy.) We call this configuration, with improved ECAL, our ``FCC2'' detector. 

The parameters of the two benchmark detectors are summarized in Table~\ref{tab:FCCdetectors}.

\subsection{Tagger performances within the detectors}

\begin{figure*}[t!]
\begin{center}
\includegraphics[width=0.48\textwidth]{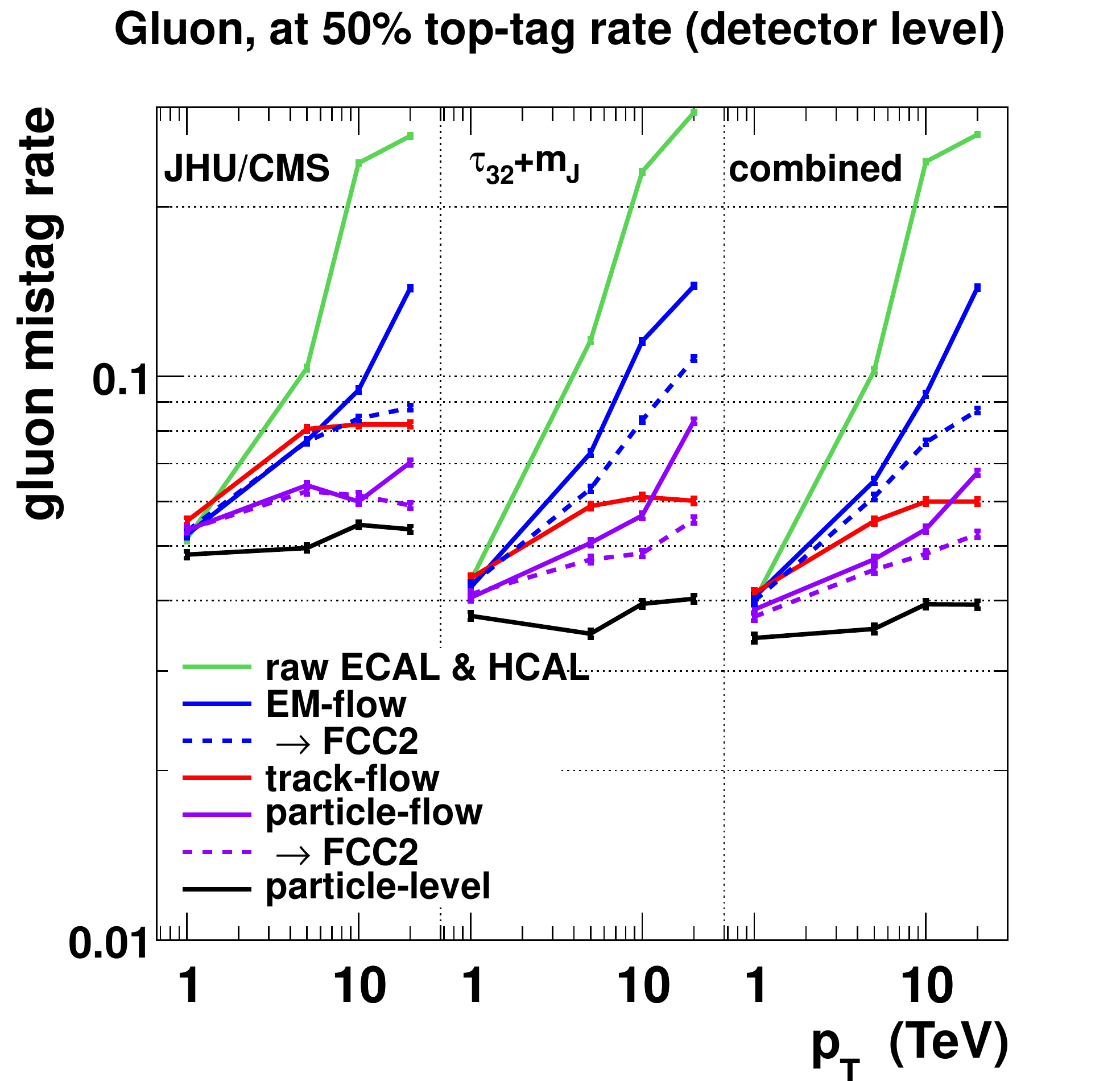} \hspace{0.2cm}
\includegraphics[width=0.48\textwidth]{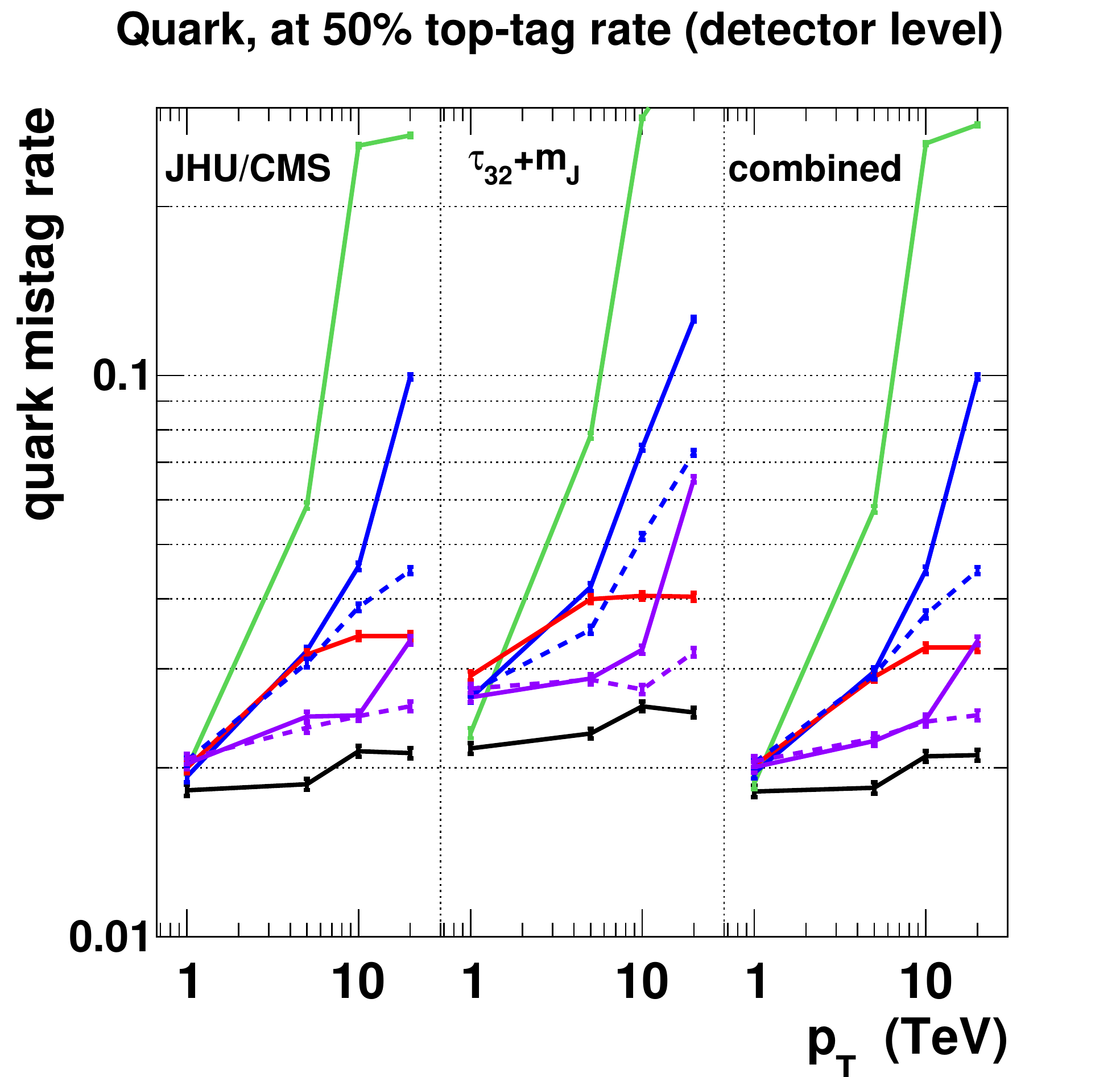}
\caption{Optimized detector-level tag/mistag rates against gluons ({\bf left}) or quarks ({\bf right}) as a function of fat-jet $p_T$ and top-tagger. Different curves represent different detector reconstruction and detector configuration assumptions. (The FCC2 results can also be approximately applied to the LHC by halving the $p_T$.)}
\label{fig:FCC_50percent}
\end{center}
\end{figure*}

With our detector simulation and reconstruction methodology established, we revisit top/gluon and top/quark discrimination. To facilitate comparisons, we start by focusing on the mistag rate in a fixed slice of 50\% top-tag rate. We continue to independently optimize discrimination against gluons and quarks.\footnote{For our detector-level optimization scans, we fix $\beta_R = 4$, as this was universally optimal in our particle-level scans, and saves some time on the computationally more expensive detector simulation. The exact same coefficient was also used in~\cite{Calkins:2013ega} and~\cite{Larkoski:2015yqa}. The optimized values of the other (de)clustering parameters are also approximately unchanged relative to particle-level. Typically, most of the degradation of performance under detector conditions arises from worsening resolution on $m_{\rm min}$ and/or $\tau_{32}$.} As we saw above (and as continues to hold within the detector), optimization against gluons anyway yields $O(1)$ smaller mistag rates for quarks, such that in a roughly evenly-mixed sample of gluon- and quark-jets, the gluon optimization is more important. The quark-optimized results, on the other hand, become relevant in cases with highly quark-dominated backgrounds, which especially includes background events with the highest-$p_T$ jets, due to slower falloff of valence quark parton distribution functions.

Fig.~\ref{fig:FCC_50percent} displays the predicted mistag rates for each individual top-tagger as a function of fat-jet $p_T$, spanning from 1~TeV up to 20~TeV. (For some reference kinematic plots at 10~TeV, see Appendix~\ref{sec:detectorModel}.) We can immediately contrast the approximate stability of particle-level tagging against the severe instability of raw calorimetry with individual HCAL and ECAL cells. This is not unexpected, as even HCAL cells of angular size $\sim$~0.05 have no hope of resolving a top decay at $O$(10~TeV) energies. These two extremes set the broadest boundaries in which we can expect to find realistic performance with our chosen top-taggers. In between, we display the results of the EM-flow, track-flow, and ``particle-flow'' strategies. The first is mainly relevant in cases with very poor tracking, and the other two assume perfect tracking. The default results are shown assuming the FCC1 detector configuration, and the improvements to EM-flow and particle-flow available from the FCC2 detector are also indicated. In either case, the performance is typically bracketed by EM-flow and particle-flow.

The advantage of pursuing a more refined FCC2-style ECAL is clear, especially if the tracking is de-emphasized and calorimetry becomes the main option. At 20~TeV, it can recover roughly a factor of two in lost discrimination power for pure EM-flow. Even if near-perfect tracking is developed, such that track-flow remains stable with growing energy, an ECAL with an additional $O(1)$ angular refinement can be combined with the tracking to form a particle-flow reconstruction that is consistently more powerful than either EM-flow or track-flow individually.

All together, there remains an $O(1)$ range of possible performances under the different detector reconstruction and detector configuration assumptions. Still, we take this to be a good sign. The jets studied here are an order of magnitude more energetic than what is available at the LHC, but we have seen that the detectors do not need to be an order of magnitude better to prevent catastrophic failure of top-tagging. Note as well that our 5~TeV FCC1 results should serve as a good proxy for 2.5~TeV jets at the LHC. Here the range of predicted performances is even smaller, and we will be surprised if top-tagging at this energy proves to be qualitatively more difficult than at the well-studied 1~TeV vicinity.

We can also see in Fig.~\ref{fig:FCC_50percent} the relative performances of the different tagging algorithms under detector conditions. On the whole, the $\tau_{32}$+mass tagger continues to perform better than the JHU/CMS tagger for gluons, and vice versa for quarks. However, $N$-subjettiness exhibits more severe performance losses in the absence of perfect tracking. In particular, gluon discrimination becomes comparable to JHU/CMS already at 5~TeV. However, these issues are ameliorated by running the combined tagger, and even more so with more refined calorimetry.

\begin{figure*}[t]
\begin{center}
\includegraphics[width=0.48\textwidth]{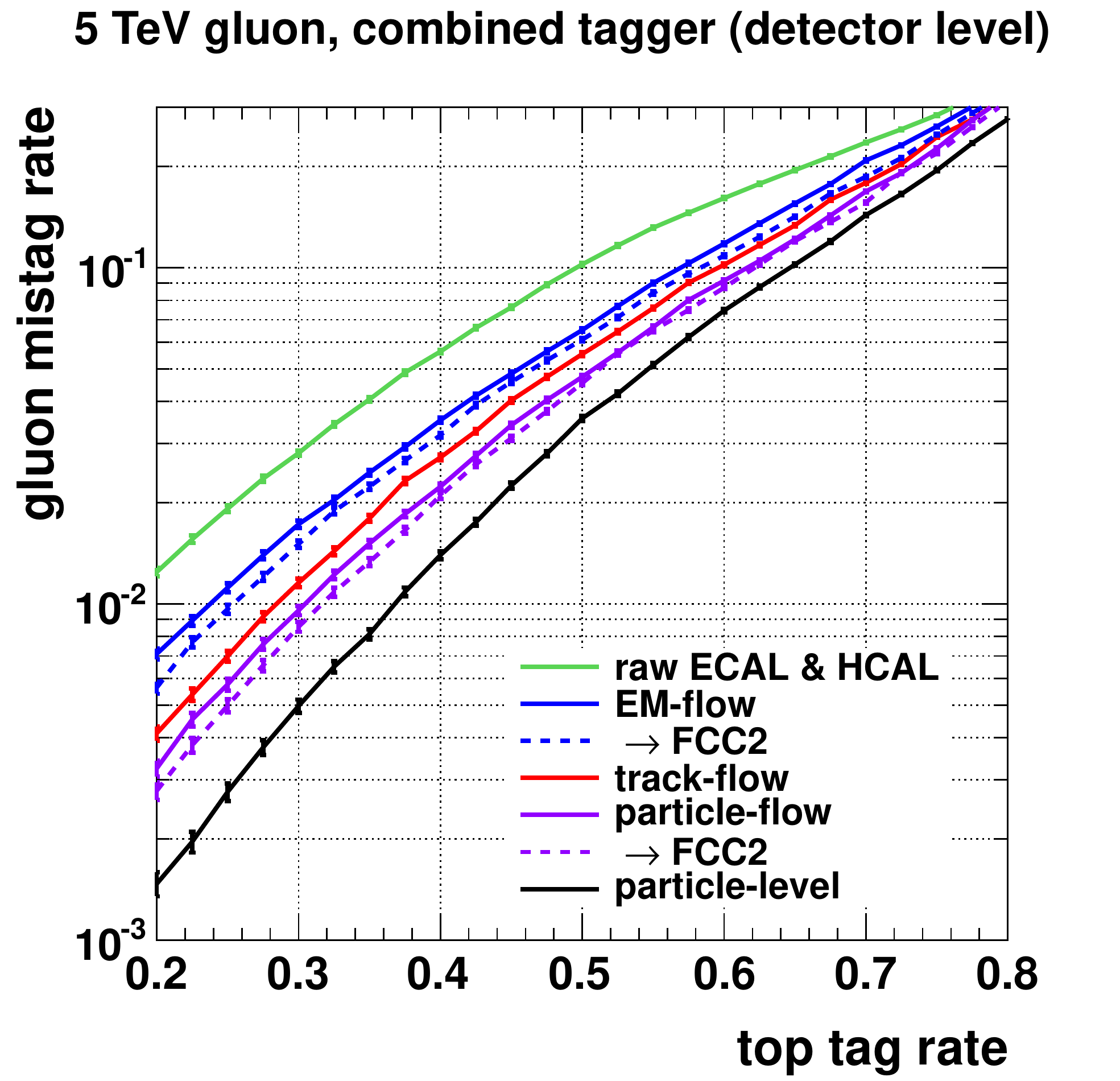} \hspace{0.2cm}
\includegraphics[width=0.48\textwidth]{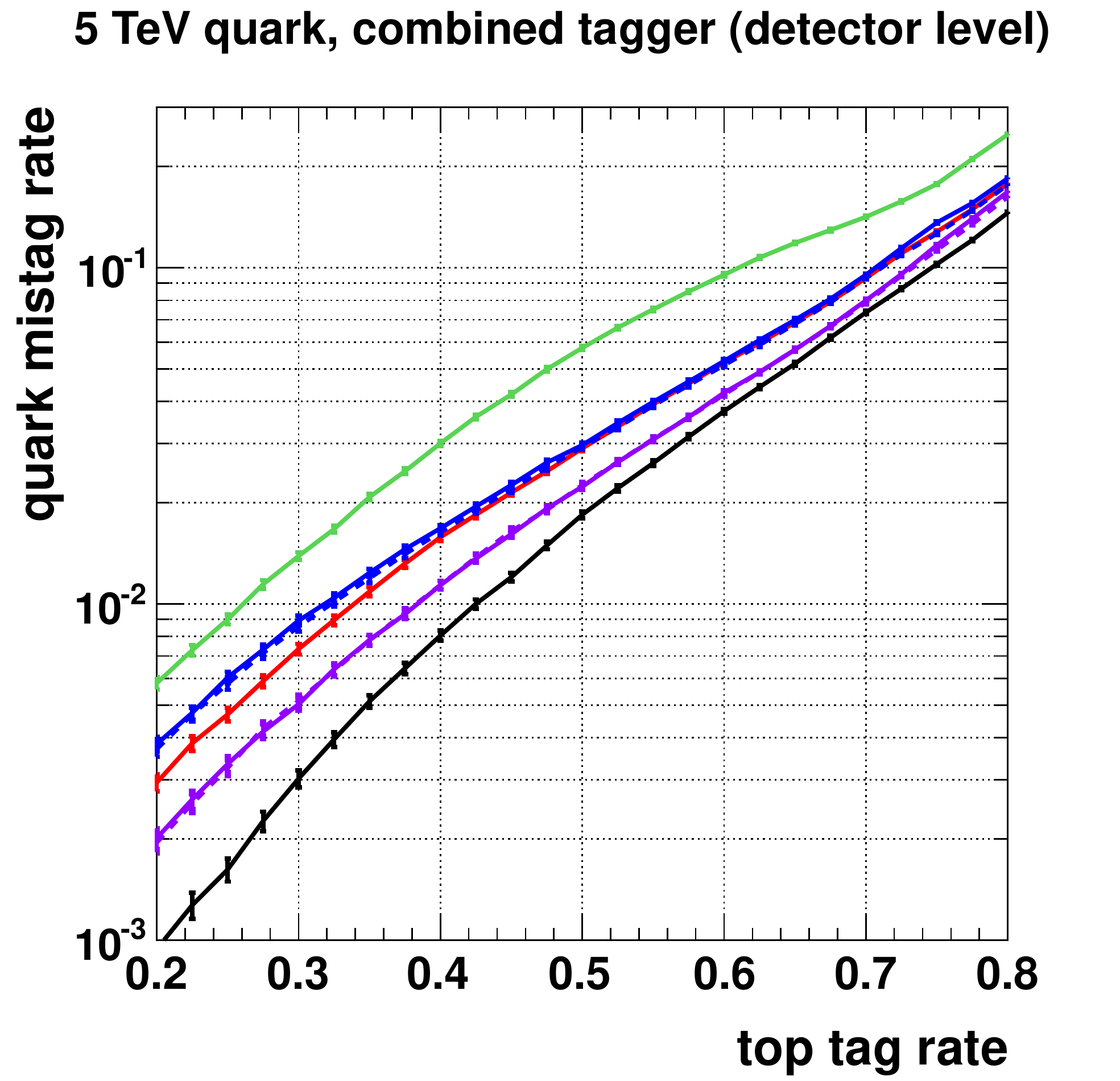} \\
\vspace{0.4cm}
\includegraphics[width=0.48\textwidth]{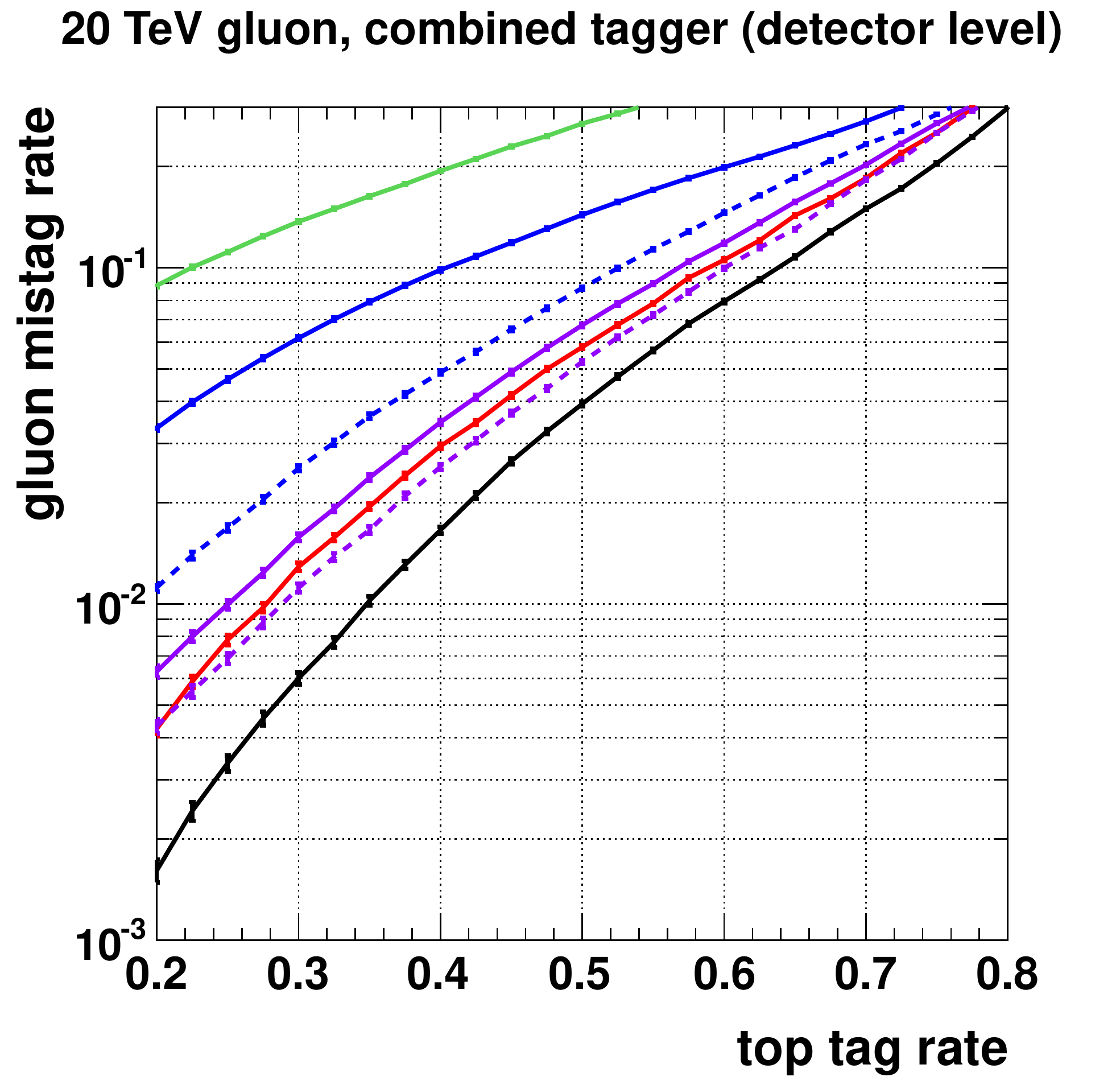} \hspace{0.2cm}
\includegraphics[width=0.48\textwidth]{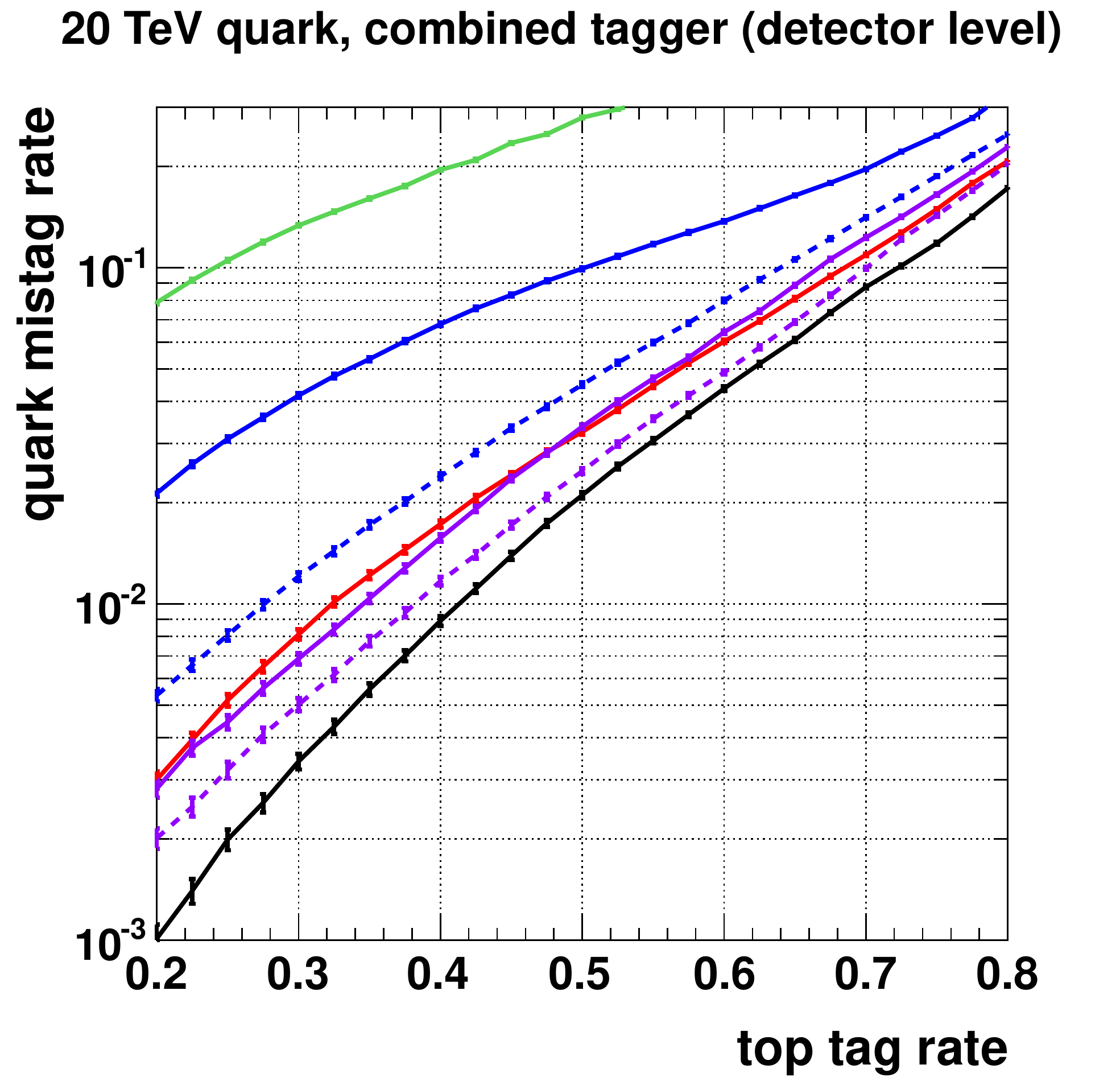}
\caption{Optimized tag/mistag rates against gluons ({\bf left}) and quarks ({\bf right}) for 5~TeV ({\bf top}) and 20~TeV ({\bf bottom}) detector-level simulations, using the combined top-tagger. (The FCC2 results can also be approximately applied to the LHC by halving the $p_T$.)}
\label{fig:FCC1_ROCs}
\end{center}
\end{figure*}

To provide a broader perspective on the possible performance at different top-tag working points, we also provide a few representative ROC curves for the combined tagger in Fig.~\ref{fig:FCC1_ROCs}. The trends seen in Fig.~\ref{fig:FCC_50percent} at fixed 50\% top-tag efficiency essentially extrapolate unchanged.

\begin{figure*}[th!]
\begin{center}
\includegraphics[width=0.48\textwidth]{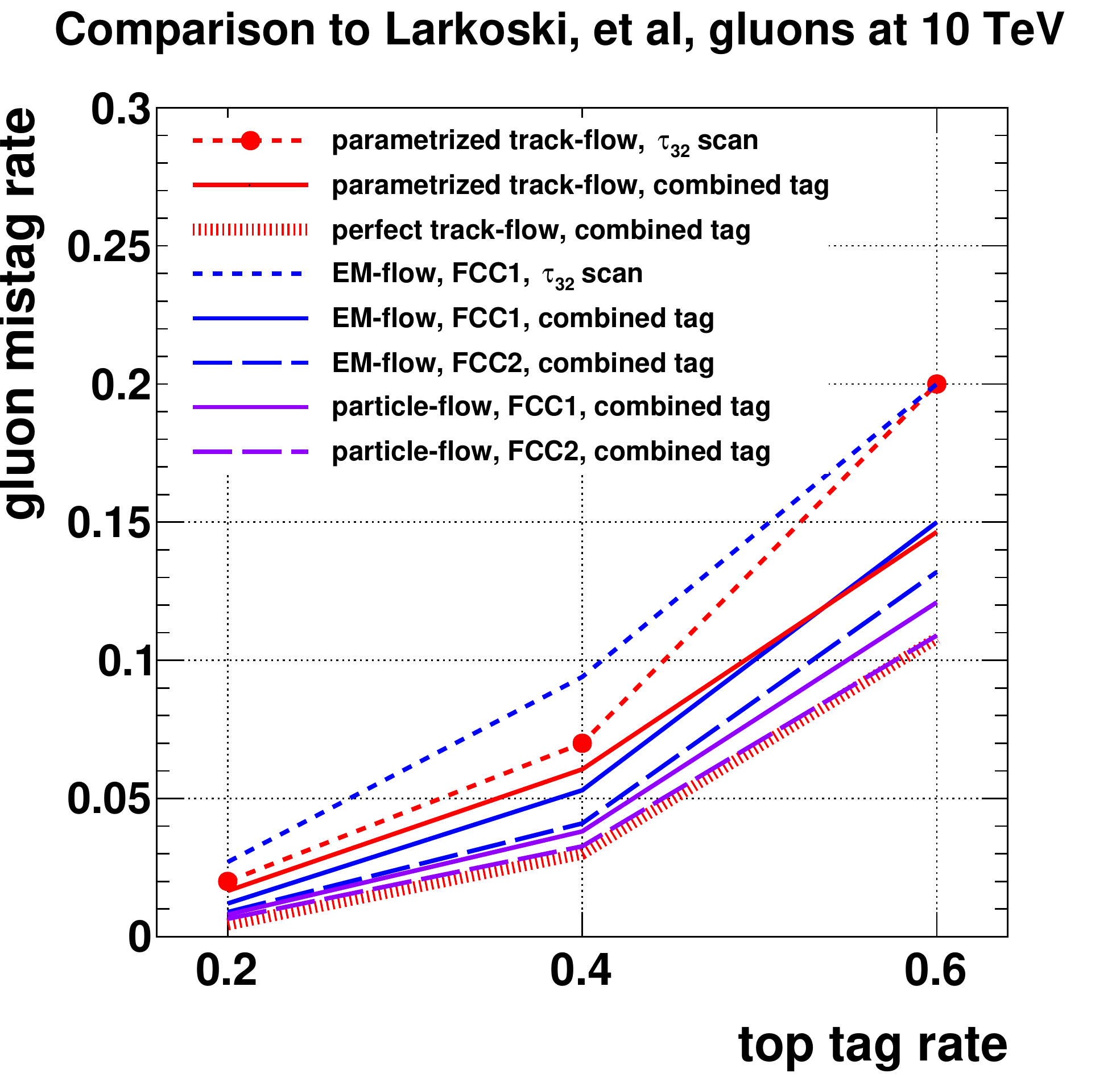} \hspace{0.2cm}
\includegraphics[width=0.48\textwidth]{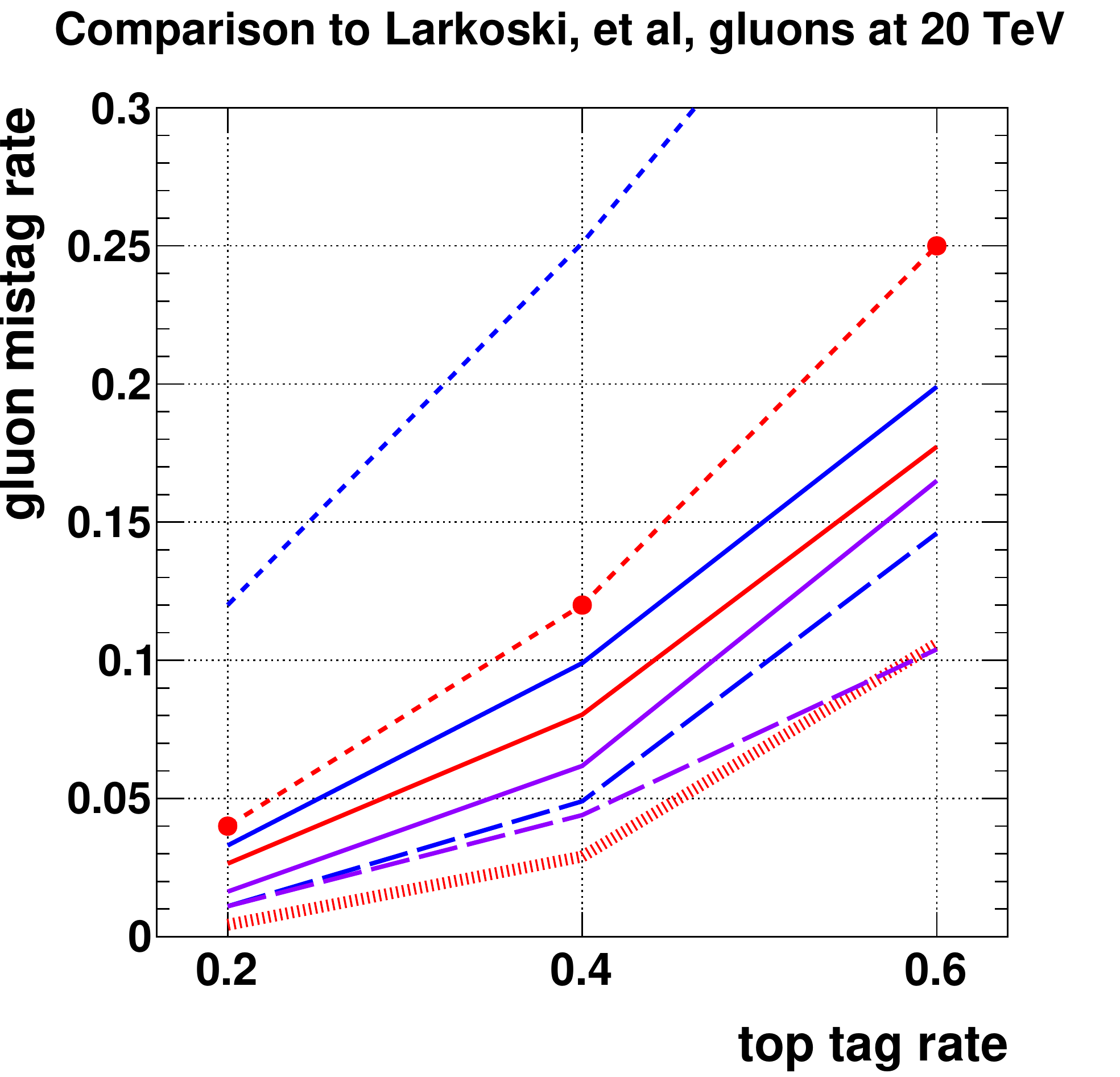}
\caption{Performance comparisons against the estimates of Larkoski, et al~\cite{Larkoski:2015yqa}, at 10~TeV ({\bf left}) and 20~TeV ({\bf right}). The ``parametrized track-flow'' uses their track reconstruction parametrization, and their original analysis is indicated by the circular markers. We additionally run this tracker parametrization within several of our own analyses, obtaining results that approximately model imperfections in the tracker and both calorimeters. Note that the perfect track-flow of our original analysis (thick and striped red line near the bottom) is dominated by $N$-subjettiness, and exhibits very similar performance under both our combined tagger and the simpler $\tau_{32}$-scan.}
\label{fig:LarkoskiComparison}
\end{center}
\end{figure*}

In~\cite{Larkoski:2015yqa}, a similar study has been made using a (conservative) parametrized model of tracking performance and a track-flow style of reconstruction, and focusing on $N$-subjettiness as a discriminator against gluon-jets. This study had a much less detailed model of the calorimeter and did not explore the possible benefits of incorporating the highly resolved ECAL cells. Since our own main study neglects details of the tracking, we can perform some informative comparisons. We have also implemented this parametrized tracking model, validated against the results of~\cite{Larkoski:2015yqa}, and used it to investigate the possible benefit of adding ECAL information and/or declustering-style substructure observables.\footnote{We thank Michele Selvaggi for assistance in reproducing their model. Note also that while the studies of~\cite{Larkoski:2015yqa} are based on the {\tt PYTHIA6} $p_T$-ordered shower, whereas our's are based on the {\tt PYTHIA8} $p_T$-ordered shower, we have closely reproduced the reported mistag rates consistently using both showers. A more careful study of performance ambiguities due to different showering models would nonetheless be warranted in the future, but are largely orthogonal to the energy-scaling and detector issues investigated here.} We display the results of these comparisons in Fig.~\ref{fig:LarkoskiComparison}, for gluon-jets at 10~TeV and 20~TeV. The substructure approach of~\cite{Larkoski:2015yqa} uses a fixed jet-mass window $m_J = [120,250]$~GeV and scans over $\tau_{32}$ to determine tag/mistag rates. We have also applied this approach to make some of our comparisons more direct, but include as well our optimized combined tagger.

One can immediately see the impact of the tracking imperfections in Fig.~\ref{fig:LarkoskiComparison}. Compared to an over-idealized perfect track-flow, the parametrized imperfect track-flow leads to approximately two times higher mistag rates at 10~TeV, and 2--4 times higher mistag rates at 20~TeV. Note that for the perfect track-flow, $\tau_{32}$ is the single most powerful discriminator amongst the variables studied here, such that the combined tagger also practically acts as a simple $\tau_{32}$ scan with a loose top-jet mass window (not plotted, though see Fig.~\ref{fig:FCC_50percent}, left panel). However, once tracking imperfections are introduced, adding the JHU/CMS substructure cuts proves beneficial, especially at higher-efficiency working points. Though $\tau_{32}$ by itself can be almost maximally powerful for gluon discrimination, that behavior appears not to be robust to the loss of very high-quality tracking information. Hybridizing with additional substructure observables then becomes an important strategy for helping to retain discrimination power.

For EM-flow, defined using our calorimeter parametrizations discussed above and in Appendix~\ref{sec:detectorModel}, we can also see that the $\tau_{32}$ scan is non-optimal, and even less robust to energy scaling. Nonetheless, at 10~TeV, it yields performance very comparable to~\cite{Larkoski:2015yqa}. Again, the benefits of adding more substructure variables is obvious. With the more fully-optimized combined tagger, EM-flow exhibits better performance than the estimates of~\cite{Larkoski:2015yqa}, and more stable $p_T$-dependence. If the ECAL granularity of our FCC2 model can be achieved, the performance improves yet again, uniformly beating the parametrized track-flow, and by itself approaching close to perfect track-flow. We also re-run the optimized tagger using our ``particle-flow'' reconstruction, folding together the imperfect tracking and imperfect calorimetry. It remains robustly more powerful than using track-flow or EM-flow individually.

We have seen, then, that even in the complete absence of a working tracker and using existing calorimeter technology, top-tag performance can be maintained well above 10~TeV without catastrophic degradation of performance relative to lower energies. We expect that a truly sophisticated combination of calorimetry with tracking, whatever its ultimate quality, should do even better. We therefore remain optimistic that even modest improvements in detector technology and reconstruction algorithms will allow top-tagging to remain quite robust at the FCC.

\section{Physics Effects}
\label{sec:physics}

The above discussion of detector effects was confined to a standard physics setup that includes final-state radiation within jets that contain top quarks, but it did not explore the consequences of this radiation. The standard setup also does not include genuinely new showering effects that begin to open up at multi-TeV energies: $g\to t\bar t$ splittings and EW showering $q\to q(W/Z)$. In this section, we return to particle-level to address these orthogonal issues.

\subsection{QCD radiation off of top quarks}
\label{sec:topFSR}

Before it decays, a boosted top quark will copiously radiate gluons, just as would any other relativistic quark. This radiation is largely confined to the region $k_T \gsim m_t$, which is the familiar dead cone effect for massive quarks. Conveniently, the top's decay products are confined to a complementary region $k_T \lsim m_t$. So at first pass, the structure of radiation before and after the top's decay are well-separated, and can be treated independently. This feature is exploited by the use of a shrinking radius for the active top-tag area~\cite{Kaplan:2008ie,Calkins:2013ega,Larkoski:2015yqa}.

Of course, strictly speaking, the separation is not perfectly clean. As pointed out in~\cite{Calkins:2013ega}, even a shrinking top-tagging radius still picks up some semi-hard FSR, leading to $O(10\%)$ of tops being reconstructed with spurious substructure and with groomed top-jet masses well above $m_t$. To what extent this is a problem depends on the goals of a particular analysis. Substructure methods to ameliorate confusion between FSR and decay products have been explored in~\cite{Tweedie:2014yda}, demonstrating appreciable gains in top reconstruction quality and in particular discrimination between different boosted top chiralities. However, for our purposes here, we would primarily like to obtain an understanding of what role this extra radiation might play in the problem of discrimination against gluon and light-quark jets.

As a naive study, we can consider re-running our optimization scans of Section~\ref{sec:particle} with $t\to tg$ turned off in {\tt PYTHIA8}. Doing so with the full set of variables turns out to be numerically meaningless, but conceptually enlightening. Run in this manner, the optimization scan seeks to use as large of a top-jet radius slope $\beta_R$ as possible, exploiting the fact that the region $\Delta R \gsim 4 m_t/p_T$ is largely free of radiation for the ``color-singlet'' top quark, but full of radiation for the colored gluon and light quarks. In effect, the problem of top-tagging begins to share features with that of $\tau$-tagging. The result is an unphysical order-of-magnitude reduction in mistag rates at fixed top-tag rate, which becomes progressively more pronounced at higher energies. (Of course, such a situation {\it does} apply in the context of boosted electroweak boson tagging, and large tag-jet radii were advocated in~\cite{Katz:2010mr}.)

\begin{figure*}[t!]
\begin{center}
\includegraphics[width=0.48\textwidth]{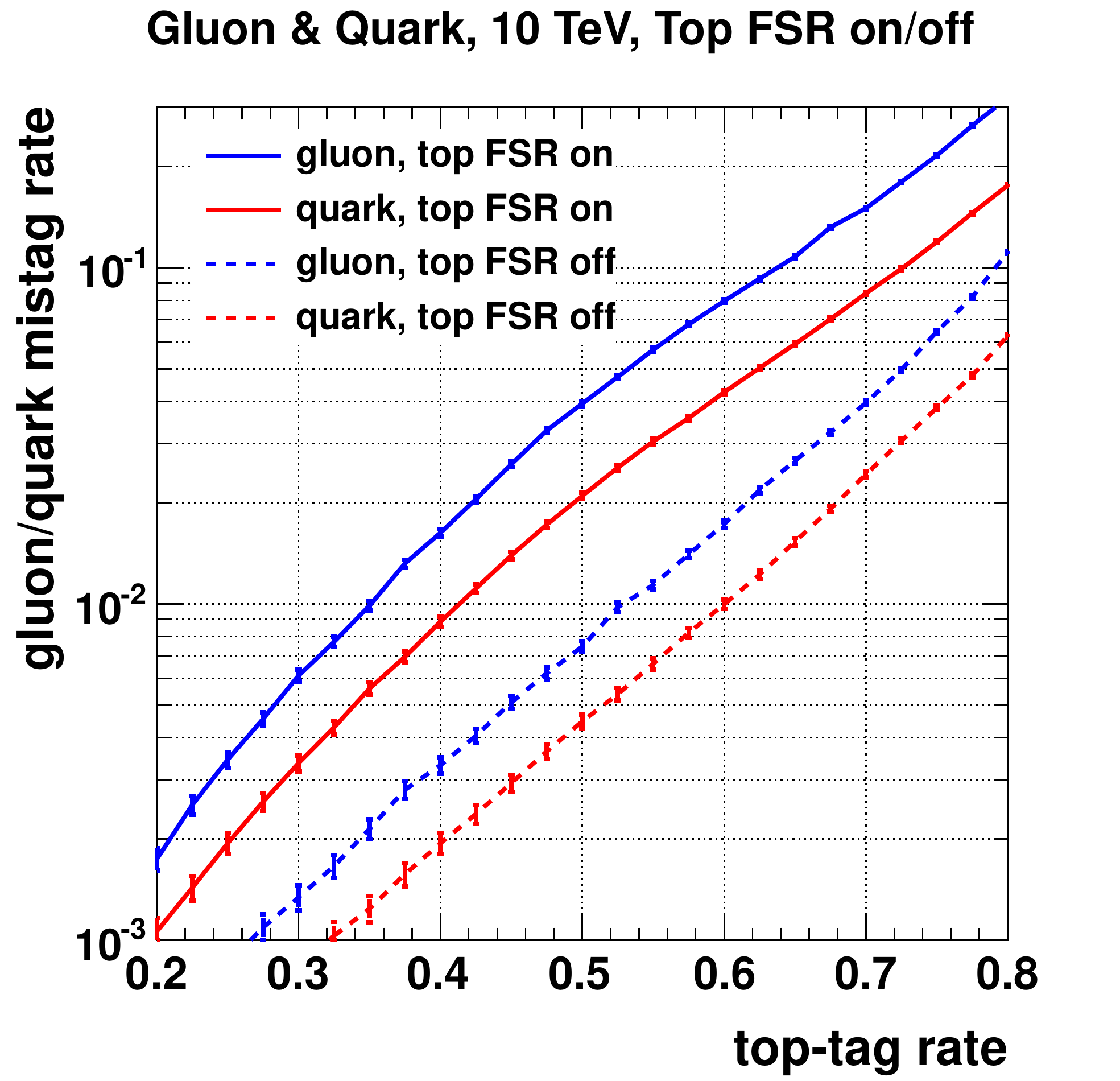} \hspace{0.2cm}
\includegraphics[width=0.48\textwidth]{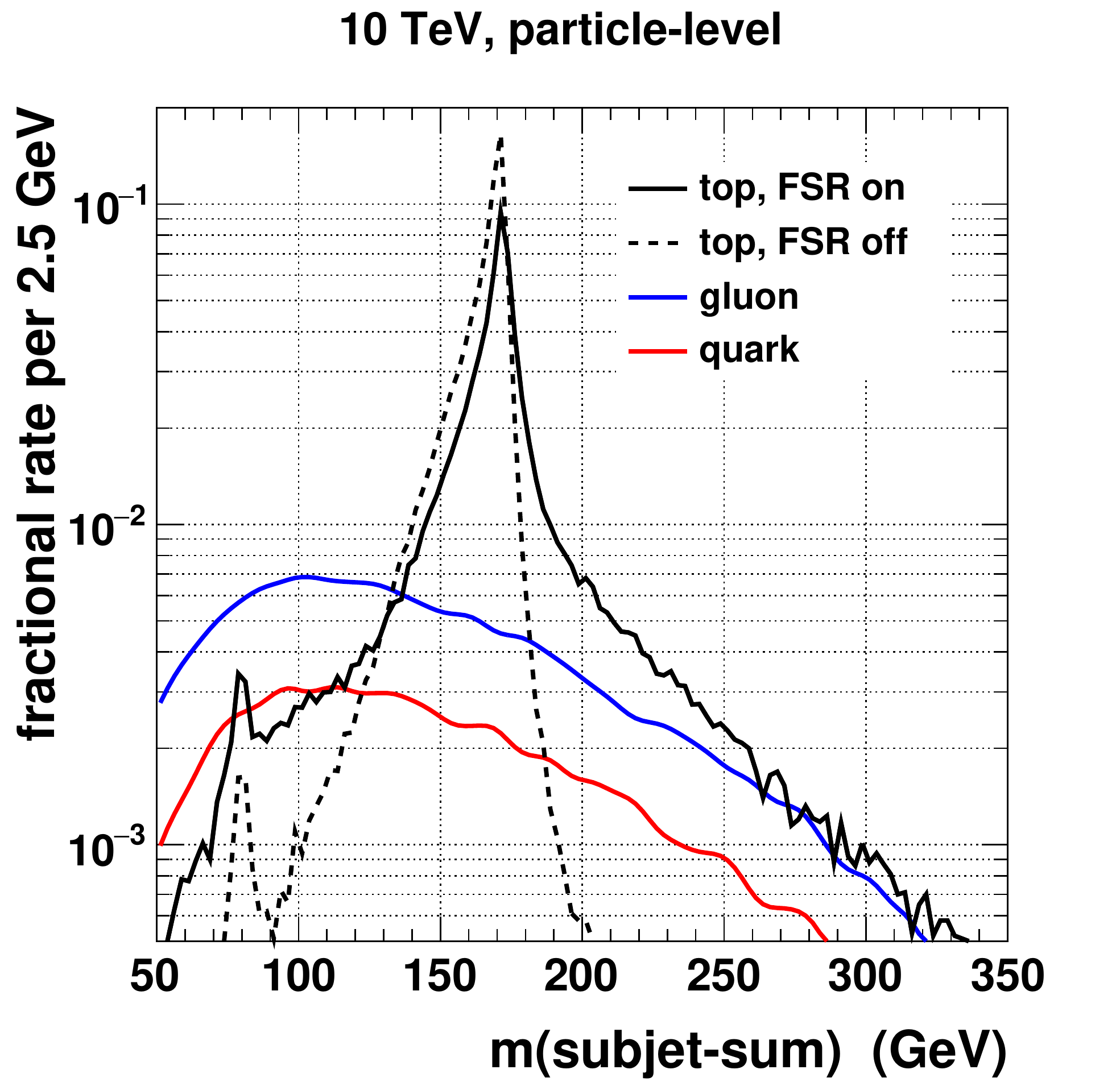}
\caption{Particle-level tag/mistag rates against gluons and quarks at 10~TeV using the combined top-tagger (fixing $\beta_R \equiv 4$), with $t\to tg$ FSR radiation turned on or off ({\bf left}). Distributions of subjet-sum mass after JHU/CMS declustering with $\delta_p = 0.03$ and $\delta_r = 0.7$, and $N_{\rm subjets} \ge 3$, but before other substructure cuts ({\bf right}).}
\label{fig:topFSR}
\end{center}
\end{figure*}

Still, to develop some numerical sense for how much the radiation is affecting the tagging within the known relevant region $\Delta R \lsim 4 m_t/p_T$, we can re-run the combined tagger scans with $\beta_R \equiv 4$. The result of this analysis at 10~TeV is shown in Fig.~\ref{fig:topFSR}, where we see that the improvement in top-tag rate at a fixed mistag rate would be $O(1)$, and that the decrease in mistag rate at a fixed top-tag rate is a dramatic factor of $\approx 5$. Most of this improvement arises from the simple fact that the top mass peak becomes much tighter, which is also shown in Fig.~\ref{fig:topFSR}. The discrimination is also improved somewhat due to a tighter $m_{\rm min}$ distribution and generally smaller values of $\tau_{32}$. Of course, these features would to some extent become washed-out by the detector effects. However, it is clear that FSR off of top quarks can be a very important limiting factor in top-tagging. We note that little critical attention has been paid to how this radiation is modeled or might be ameliorated/exploited in tagging, and that these points deserve further attention (though see~\cite{Joshi:2012pu,Maltoni:2016ays} as well as the references above).

\begin{figure*}[t!]
\begin{center}
\includegraphics[width=0.48\textwidth]{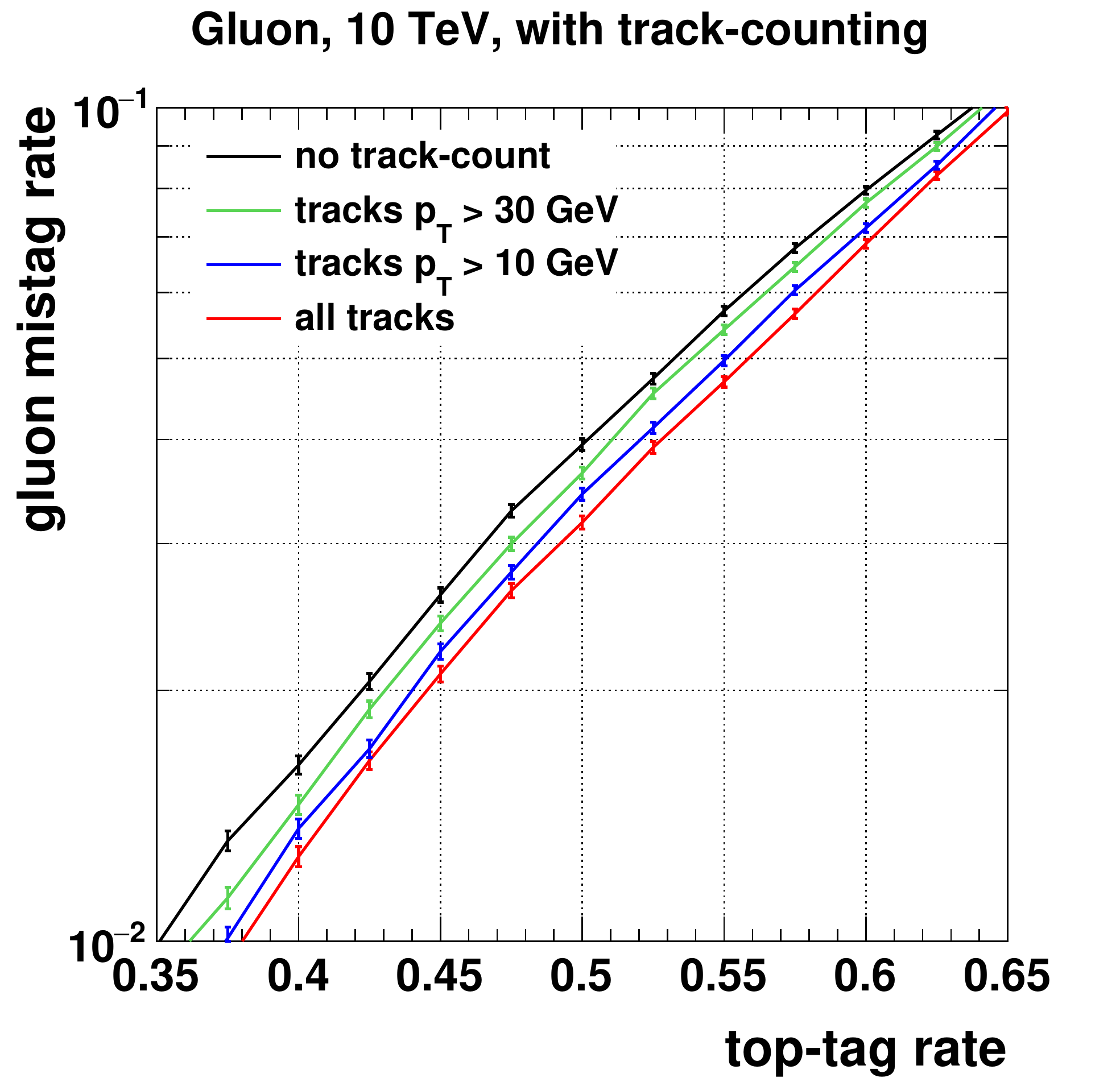} \hspace{0.2cm}
\includegraphics[width=0.48\textwidth]{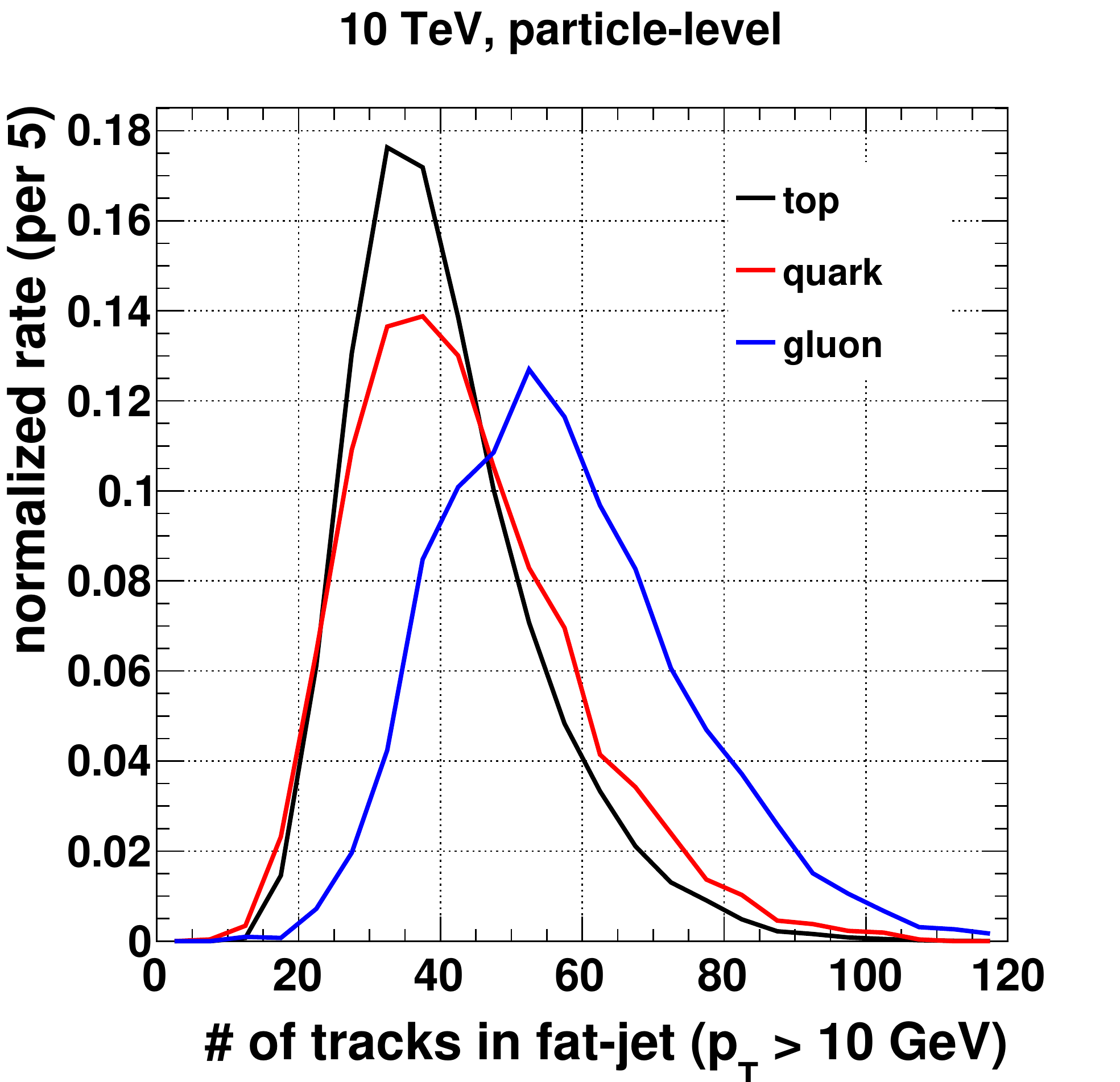}
\caption{Particle-level tag/mistag rates against gluons at 10~TeV using the combined top-tagger, including fat-jet track counts in the cut scan~({\bf left}). Normalized distributions of track counts inside the fat-jet at 10~GeV $p_T$ threshold, with (de)clustering parameters as in Fig.~\ref{fig:topFSR}, and substructure cuts $N_{\rm subjets} \ge 3$, $m_{\rm subjets} \in [140,210]$~GeV, $m_{\rm min} > 10$~GeV, and $\tau_{32} < 0.65$~({\bf right}).}
\label{fig:trackCounting}
\end{center}
\end{figure*}

One definite opportunity that immediately presents itself is the possibility of treating top quarks as ``light quarks'' in the context of quark/gluon discrimination. Because the top is color-triplet and the gluon is color-octet, the wide-angle radiation of the latter will tend to be more pronounced. A simple and common measure of this effect is the number of tracks contained in the jet. For this purpose, we would want to capture as much radiation as possible, and therefore count the tracks within the initial $R=1.0$ anti-$k_T$ fat-jet, before reclustering and substructure. As seen in Fig.~\ref{fig:trackCounting}, when we add this variable to our multivariate rectilinear cut scan for 10~TeV jets, we find that the re-optimized mistag rates can be modestly reduced by a relative factor of about 20\% when all tracks are included. The improvement is essentially orthogonal to the other cuts and (de)clustering parameters. More realistically, especially given the presence of pileup and high magnetic fields in the inner detector, only tracks above some $p_T$ threshold might be useful. We therefore show as well the results assuming baseline track $p_T$ thresholds of 10~GeV or 30~GeV. The improvement becomes less pronounced, though the 10~GeV threshold still maintains most of the gains. Fig.~\ref{fig:trackCounting} also shows the raw track-count distributions, with a track $p_T$ threshold of 10~GeV, and having applied some other baseline substructure cuts. This figure includes as well the track-counts for quark-jets, which are indeed much more similar to top-jets than to gluon-jets.

This simple track-counting study has been performed at particle-level. But given that the performance is mainly driven by the wide-angle portion of the radiation, where the tracks are relatively well-separated, we do not expect track reconstruction to be a major issue. We also comment that counting of tracker hits (and perhaps even calorimeter energy) away from the jet core might be adequate to extract some immediate performance gains. There also exists far more information in the wide-angle radiation pattern than simple particle counting, as is already being harnessed in more aggressive multivariate quark/gluon taggers~\cite{Gallicchio:2012ez,Komiske:2016rsd}. 

\begin{figure*}[t!]
\begin{center}
\includegraphics[width=0.48\textwidth]{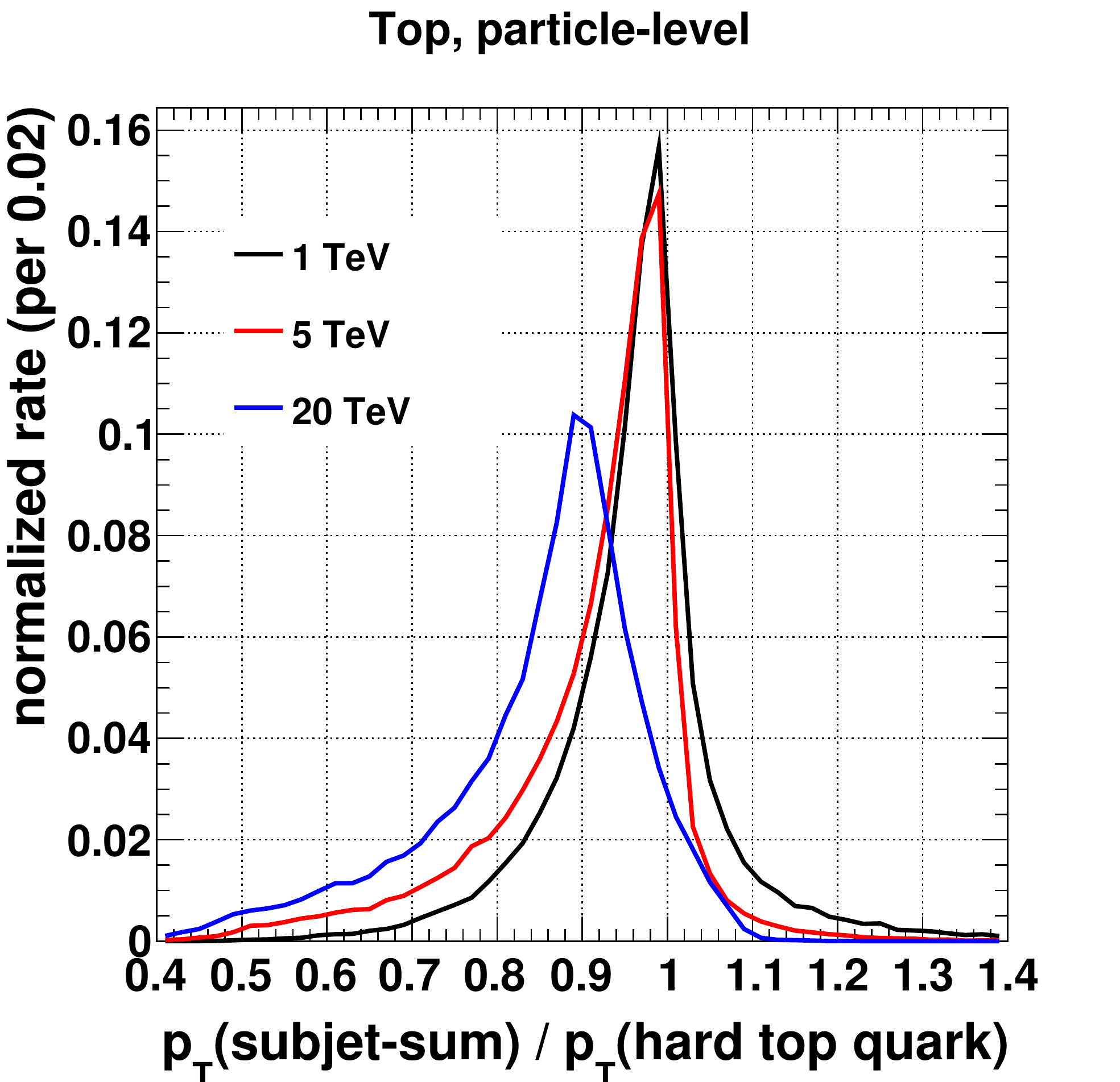} 
\caption{Distributions of top-jet $p_T$ relative to the hard parton-level top quark, for different values of the hard top $p_T$. The (de)clustering parameters are set as in Fig.~\ref{fig:topFSR}, and loose substructure cuts $N_{\rm subjets} \ge 3$ and $m_{\rm subjets} \in [140,210]$~GeV are applied.}
\label{fig:ptRel}
\end{center}
\end{figure*}

Finally, we point out that the presence of the extra radiation also somewhat complicates the measurement of the top quark's original ``parton-level'' momentum, such as would be required in reconstruction of a resonance mass or in more complicated kinematic reconstructions involving highly energetic tops. Again, such a consideration suggests that we collect as much wide-angle radiation as possible. Fig.~\ref{fig:ptRel} shows the momentum fraction carried by a loosely-tagged top quark relative to that of the corresponding hard top quark at different parton-level input energies, illustrating the cumulative effect of multiple emissions. For 1~TeV tops, the median top-jet momentum fraction is close to 0.97, whereas for 20~TeV tops, the median top-jet momentum fraction falls to 0.88.\footnote{$p_T$ fractions above one can be relatively common for the 1~TeV sample, in part because ISR effects can also become large at a 100~TeV collider.}

\subsection{Gluons splitting to $t\bar t$}
\label{sec:gttbar}

\begin{figure*}[t!]
\begin{center}
\includegraphics[width=0.48\textwidth]{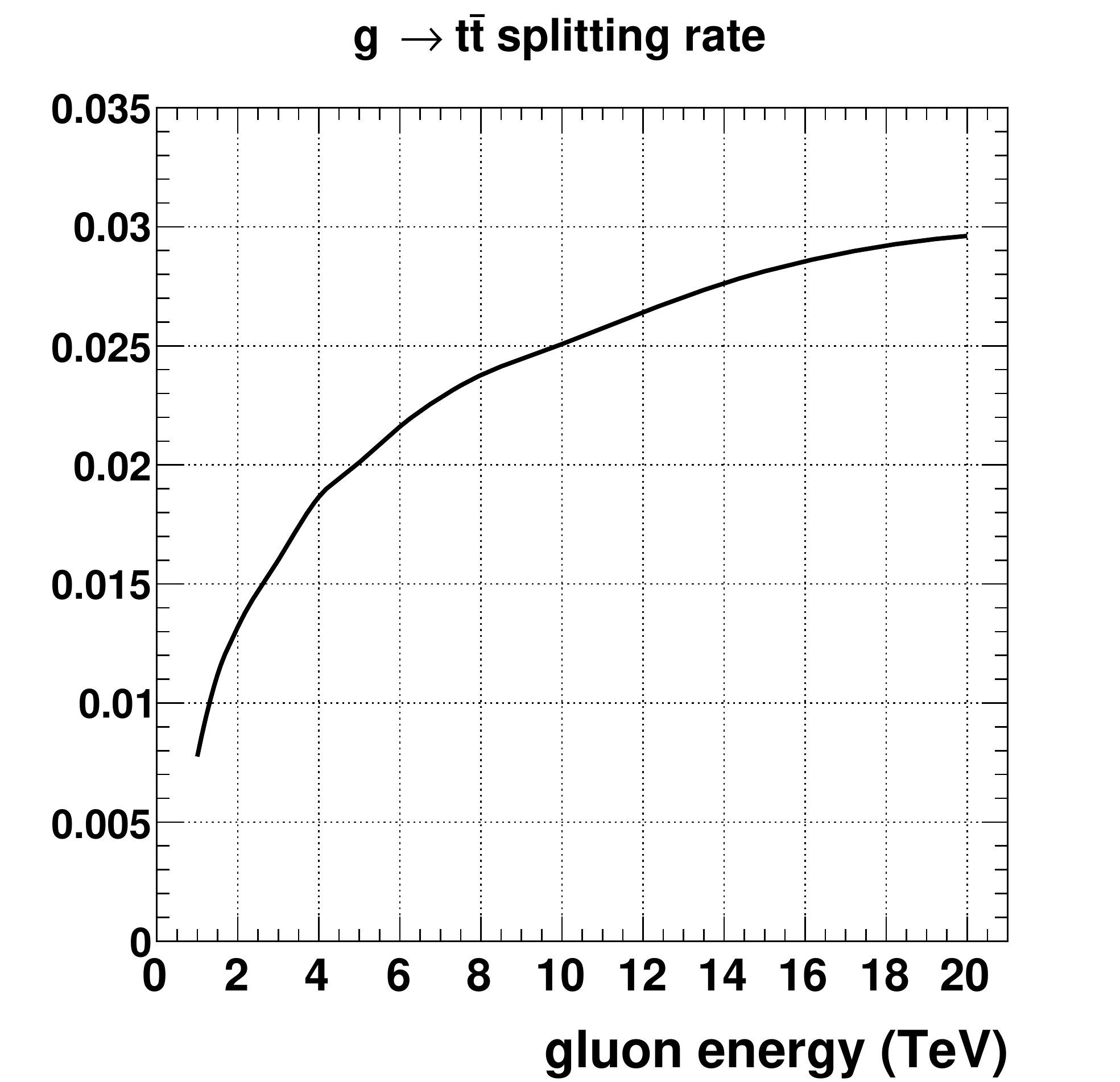} \hspace{0.2cm}
\includegraphics[width=0.48\textwidth]{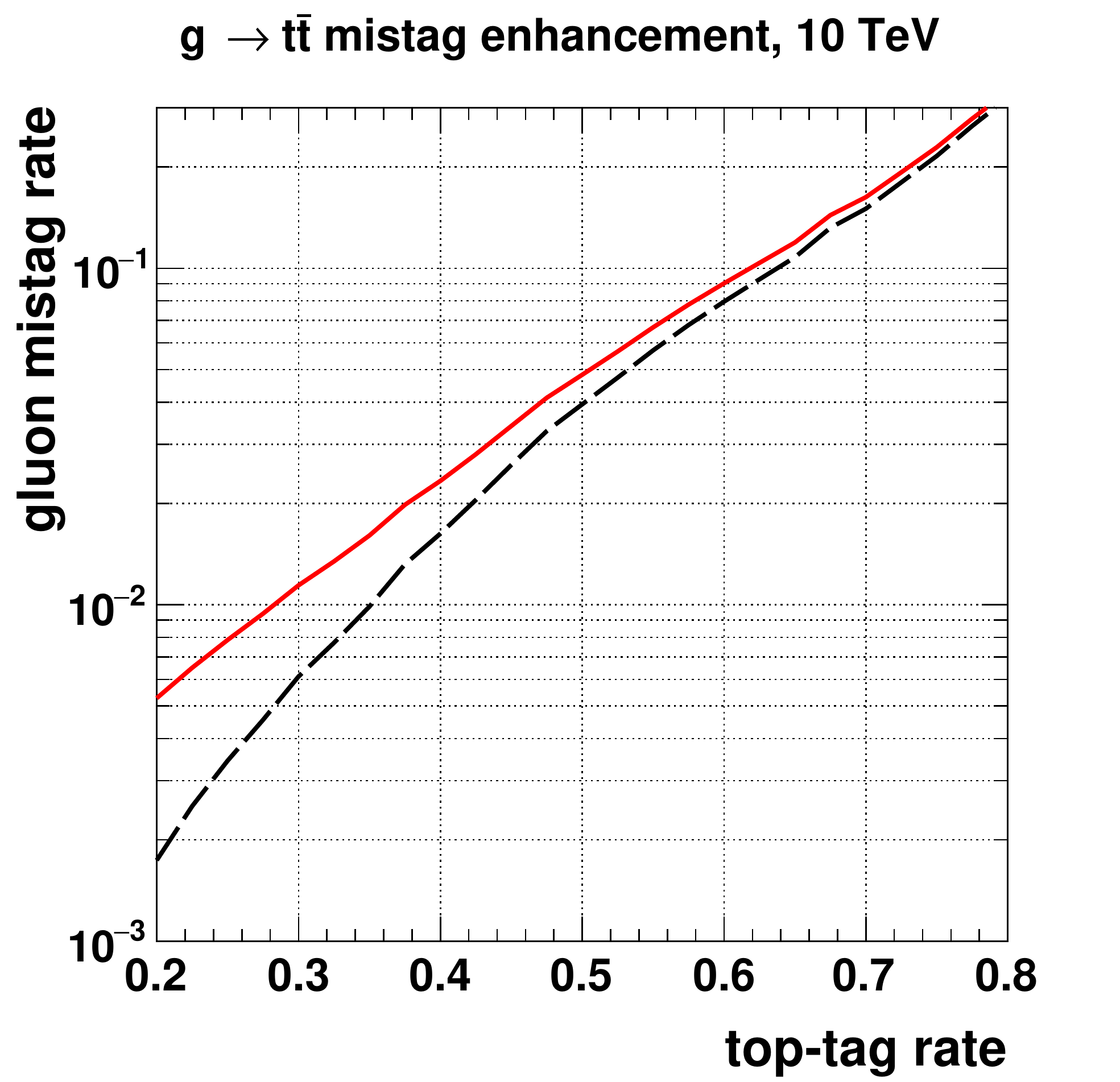}
\caption{Fixed-order splitting rates for $g\to t\bar t$ as a function of gluon energy. The strong coupling $\alpha_s$ is evaluated at $\mu_R = \sqrt{E\times m_t}$~({\bf left}). Enhancement of the particle-level gluon mistag rate, using the nominally-optimized combined tagger, at 10~TeV~({\bf right}).}
\label{fig:g-tt}
\end{center}
\end{figure*}

The standard mechanism for a gluon to end up mistagged as a hadronic top quark is for that gluon to undergo a sequence of two collinear QCD splittings at $k_T \sim m_t$. At extremely high energies, another mechanism opens up: a gluon can directly split into a $t\bar t$ pair, with the leading top in the pair decaying hadronically. A fixed-order calculation yields the integrated splitting rates shown in Fig.~\ref{fig:g-tt}.\footnote{We have computed these rates within a custom code for the full Standard Model shower at high energies~\cite{Chen:2016wkt}, with only the $g\to t\bar t$ splitting process activated. We have confirmed that the rates change only modestly when full QCD is turned on. The code does not include color connections or a model of QCD hadronization, and has only been used for these simple rate calculations. Differential rates and spin correlations have been validated against {\tt MadGraph}.} At the scale of $O$(10~TeV), the rates are 2--3\%. This is comparable to the mistag rates that we have so far estimated using the {\tt PYTHIA} shower, based purely on light QCD splittings. There is therefore a need to better understand this overlooked contribution.

The current version of {\tt PYTHIA8} does not include $g\to t\bar t$ as a splitting process. To have a baseline sample, we instead generate $pp \to t\bar t Z_{\rm inv}$ with 100~TeV beam CM energy, and top quarks decoupled from the $Z$ boson. This sample is then passed into {\tt PYTHIA8} for showering. The subsequent gluon radiation from the $t\bar t$ pair at large angles should very roughly model that from a hard gluon. We focus on $t\bar t$ pairs with $p_T$ near 10~TeV and at central rapidity, $|\eta| < 1$. The splitting rate $g\to t\bar t$ at this energy is 2.5\%.

Fixing combined tagger parameters to 50\% hadronic top-tag efficiency (optimized for discrimination against gluons), we find a 35\% efficiency for tagging these ``di-top'' jets. This is consistent with a 2/3 probability for the leading top to decay hadronically, times a roughly 50\% probability to successfully pass that top through the combined tagger. Therefore, the net rate for a gluon to pass as a top quark via this splitting channel is just under 1\%. The nominal mistag rate, without this contribution, had been estimated at just under 4\%. The correction is indeed non-negligible.

We also show in Fig.~\ref{fig:g-tt} the approximate enhancement of the gluon mistag rates for arbitrary working points at 10~TeV, optimized as above without the $g\to t\bar t$ contribution. Tighter working points for the tagger enhance the relative contribution. In the case of mistag rates near or below 1\%, the relative increase due to $g\to t\bar t$ is $O(1)$.

\begin{figure*}[t!]
\begin{center}
\includegraphics[width=0.48\textwidth]{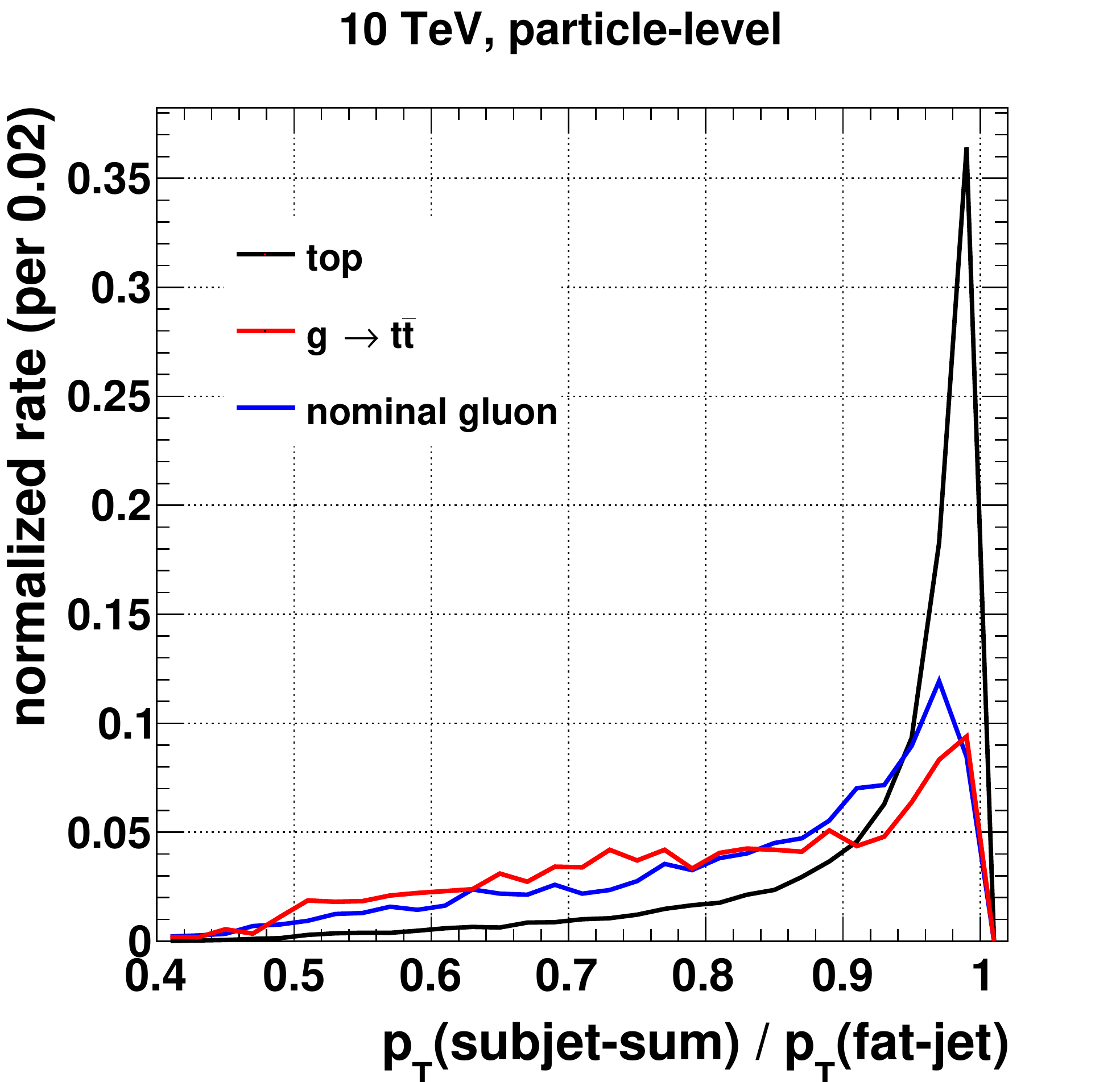} 
\caption{Normalized distributions of top-jet $p_T$ relative to its host fat-jet, at the 50\% combined-tagger working point, to illustrate the possible discrimination between prompt tops and tops from $g\to t\bar t$ splittings. Shown also is the distribution for nominal gluons, mistagged via light QCD splittings.}
\label{fig:g-tt_ptRel}
\end{center}
\end{figure*}

Of course, top quarks produced inside of gluon-jets should tend to be surrounded by more activity than prompt top quarks, and carry a smaller fraction of the total jet energy.\footnote{A similar situation already appears in $b$-tagging~\cite{Goncalves:2015prv,ATLAS:2012xna}. A gluon splitting into two collinear $b$ quarks, $g\rightarrow b\bar{b}$, within a jet can cause the gluon-jet to be mistagged as a $b$-jet. In~\cite{Goncalves:2015prv}, it was found that track counting and the fragmentation fraction of $b$-quarks are effective in isolating single $b$-jets from merged $b$-jets. ATLAS~\cite{ATLAS:2012xna} performed a multivariate analysis using jet track multiplicity, track-jet width, and the angle between two $k_T$ subjets within a jet. They found that the mistag rate for $g\to b\bar b$ at a 70\% $b$-tagging working point is $O({\rm a\ few} \times10\%)$ for $p_T = 60-480$~GeV.} For the latter, we can consider the ratio between the $p_T$ of the small top-tagged jet and its host fat-jet. This is shown in Fig.~\ref{fig:g-tt_ptRel}, including as well the corresponding distribution for normally-mistagged gluons. Tops from $g\to t\bar t$ obviously have a broader, more gluon-like distribution, which could be folded into the tagger discriminator variables. The fat-jet track-counting variables considered in the previous subsection could also be used. With the admittedly coarse model of $g\to t\bar t$ that we are employing, the track-count distribution is roughly halfway between prompt top-jets and gluons.

The presence of a companion top (or antitop) might also be inferred by generalizing to a kind of di-top tagger. For the $\approx$ 20\% of companion top decays that are leptonic, a mini-isolated lepton veto should suffice.

We further point out that $g\to t\bar t$ with a leading {\it leptonic} top could also present an interesting, overlooked background for boosted leptonic top quarks. This is especially true since the absolute energy of the leptonic top-jet may not be measurable due to the presence of the neutrino.

\subsection{Weakstrahlung off of light quarks}
\label{sec:weakstrahlung}

Particles produced in multi-TeV processes will radiate weak bosons ($W$, $Z$, and even $h$) similar to the photon and gluon radiation in QED and QCD showers. Asymptotically, this can lead to some interesting percent-scale effects on signal top-jets~\cite{Chen:2016wkt}. A more pressing issue is the effects on background jets. A light quark that radiates a $W$ or $Z$ boson, which subsequently decays hadronically, could look very similar to a hadronic top-jet. (For a discussion of weakstrahlung background to leptonic top-jets, see~\cite{Rehermann:2010vq}.) Even though the total rate is only a few percent, light quark mistag rates here (Figs.~\ref{fig:5TeV},\ref{fig:FCC_50percent},\ref{fig:FCC1_ROCs}) and elsewhere are routinely predicted to extend down to the sub-percent level. This raises the question: How much are quark mistag rates modified by weak radiation?

The radiation of a massive vector boson off of a massless fermion looks rather similar to QED or QCD for $k_T \gsim m_{W,Z}$. The integrated rate is dominated by transverse bosons, and is naively divergent in both emission angle and energy fraction, leading to the usual double-logarithmic growth with partonic process energy. However, that is not actually the region that we are interested in for top-tagging. For example, the shrinking-radius clustering with $R \propto \beta_R/p_T$ eliminates the angle logarithm. Within JHU/CMS tagging, $\delta_p$ also regulates the soft logarithm. Ultimately, we are only interested in a region with $k_T$'s of order the internal momentum scale of top decay, which happens to roughly coincide with $m_{W,Z}$. This region sits at the edge of the weak emission dead cone, where the massive shower is shutting down (and where longitudinal bosons constitute an $O(1)$ fraction of the emission rate). The amount of weak emission probability captured by a sufficiently aggressive top-tagger is approximately energy-invariant. Therefore, to the extent that weakstrahlung will pose a problem to top-tagging, it is a well-contained problem.

To model the weakstrahlung, we rerun our 5~TeV $gq \to qZ_{\rm inv}$ simulations in {\tt PYTHIA8}, with its weak FSR turned on~\cite{Christiansen:2014kba}. (See also~\cite{Krauss:2014yaa}.) We find that about 5\% of the events contain a weak boson showered off of the final-state quark, and select these for further study.\footnote{We have independently verified the rates and distributions of the weak FSR using private shower code with full polarization information~\cite{Chen:2016wkt}.} While the quarks produced in the above process should nominally be biased towards left-handed polarization (especially the down quarks), {\tt PYTHIA8} assigns their polarizations randomly. Hence our results are appropriate for unpolarized quarks, as would arise from hard QCD background processes. For background processes where the quarks are indeed polarized, the rates would need to be adjusted. Also, there is technically a small difference in the $Z$ boson emission rates of up quarks versus down quarks (about 25\% relative in favor of down quarks). Since the $Z$ boson emission rate from unpolarized quarks is anyway subdominant to the $W^\pm$ rate, we do not bother to quantify this small bias. With these caveats in mind, the total rate for emission of weak bosons that decay hadronically and become caught up in the shrinking-radius top-jet clustering is roughly 1\%.

\begin{figure*}[t!]
\begin{center}
\includegraphics[width=0.48\textwidth]{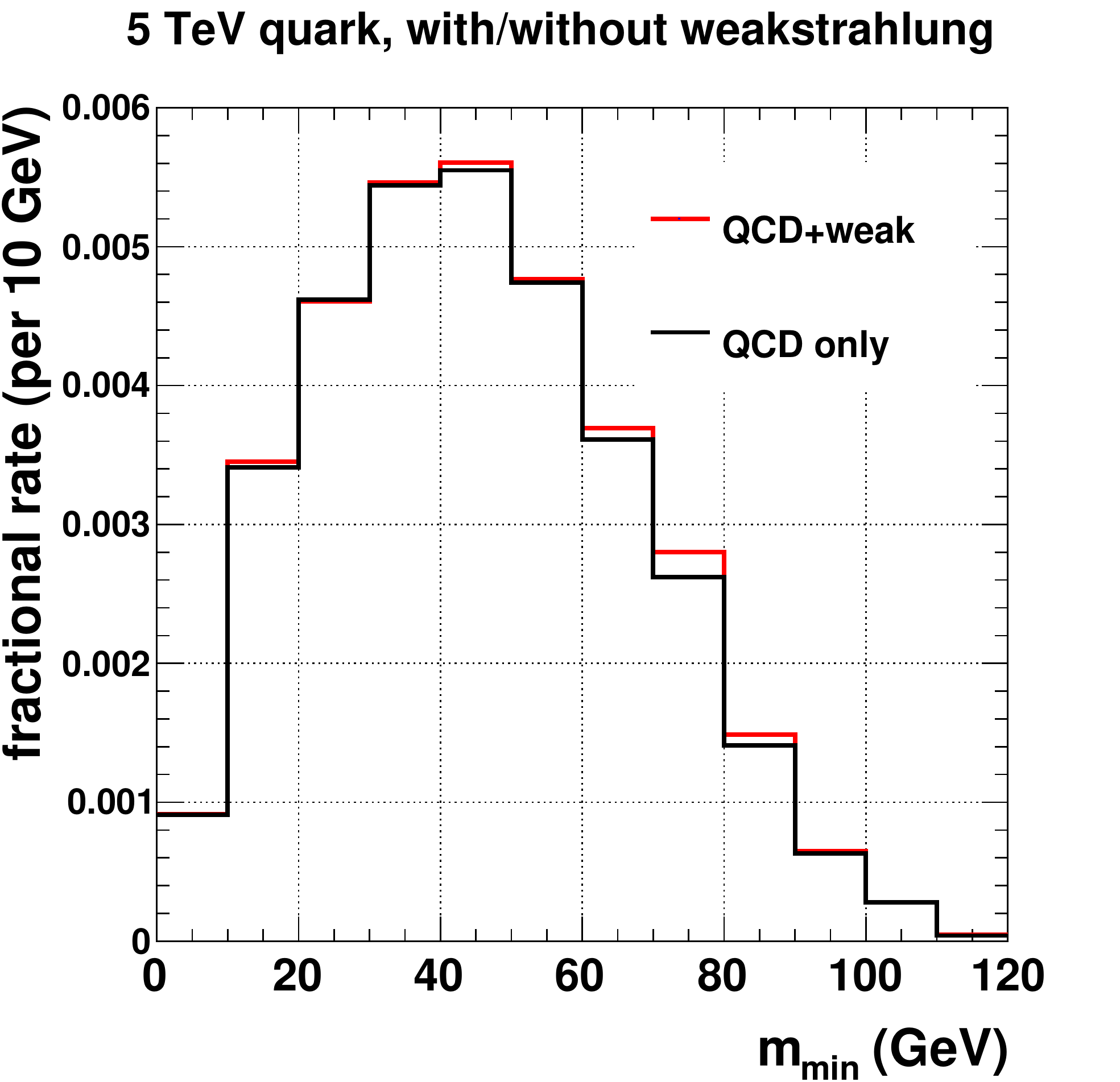} \hspace{0.2cm}
\includegraphics[width=0.48\textwidth]{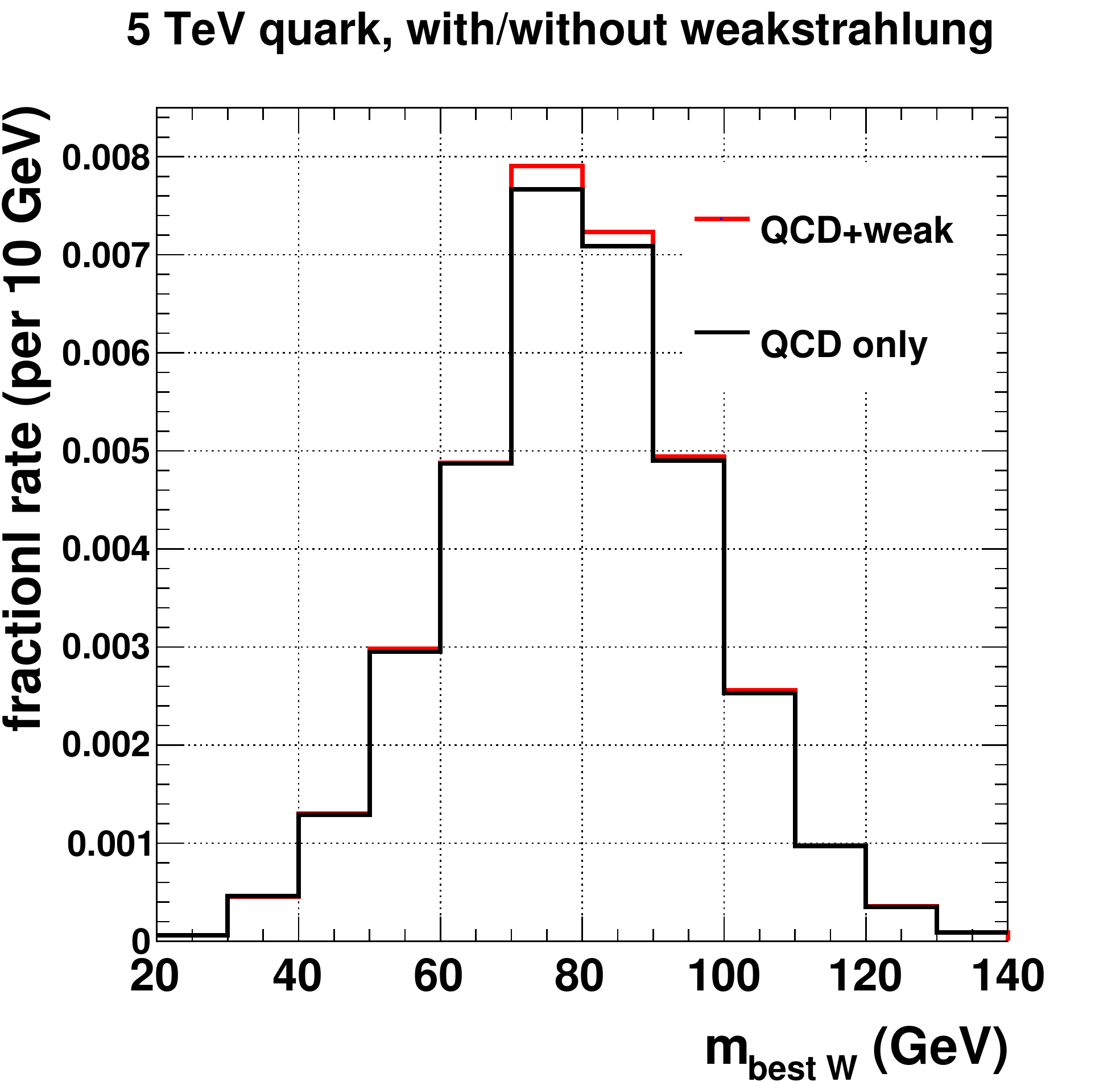}
\caption{Distribution of minimum subjet-pair mass~({\bf left}) and best-match $W$ mass~({\bf right}) for 5~TeV quark-jets with and without weakstrahlung contributions. The (de)clustering parameters, excepting $m_{\rm min}$, are set for the quark-optimized combined tagger at 50\% top-tag efficiency: $N_{\rm subjets} \ge 3$, $m_{\rm subjets} \in [140,210]$~GeV, and $\tau_{32} < 0.7$.}
\label{fig:weakstrahlung}
\end{center}
\end{figure*}

We show in Fig.~\ref{fig:weakstrahlung} the distributions of $m_{\rm min}$ and the best-$W$ mass amongst subjet pairs (as in the original JHU procedure), with (de)clustering parameters otherwise set at the 50\% working-point for the quark-optimized combined tagger.\footnote{In forming the (QCD+weak)-showered distributions, to minimize the issue of monte carlo statistical fluctuations, we have combined the original QCD-showered sample with the the subsample of (QCD+weak)-showered events that contain a radiated $W/Z$ boson. The former are reweighted by a weak Sudakov factor of $\approx$ 0.95.} Quark-jets that contain a hadronic $W$ at $k_T \sim m_t$ are almost an order-of-magnitude more likely to pass the tagger than those that do not. However, the small absolute rate for such emissions is not overcome. The presence of weakstrahlung is only visible as $\approx 5\%$ relative enhancement near $m_W$. For the 50\% working point, the quark mistag rate is approximately~2\%. Adding in weakstrahlung, this increases by a modest factor of 1.02.

As with $g\to t\bar t$ above, the relative importance of this added contribution becomes larger at tighter working points. However, in this case the size never approaches $O(1)$. For example, at a 20\% top-tag working point, with quark mistag of about 0.1\%, the weakstrahlung mistag enhancement is only 1.1.

We conclude that weakstrahlung contributions are small, and certainly justified to neglect upon a first pass.

\section{Conclusion and Outlook}
\label{sec:conclusions}

In this paper, we have explored the plausibility of top-tagging at energy frontier colliders. The LHC is poised to enter the multi-TeV regime of top-jet production, and the next generation of such machines will produce top-jets with unprecedented energies of up to $O$(10~TeV). We have categorized several correlated issues that arise at such high energies, paying special attention to substructure algorithm choices, detector reconstruction choices, detector technology, and novel QCD and electroweak showering effects.

Through the individual multivariate optimization of a JHU/CMS-type declustering top-tagger and the jet-shape variable $N$-subjettiness, we have demonstrated that discrimination against gluon-jets and quark-jets exhibits different, complementary behaviors under the two approaches. We have shown that this set of declustering and jet-shape variables can be input into a more robust combined tagger, which allows for nearly simultaneous optimization against gluons and quarks. After validating this combined tagger at idealized particle-level, we then investigated its performance on detector-level objects reconstructed according to different strategies, using toy detector simulations with semi-realistic energy deposition patterns obtained via {\tt GEANT}. Re-optimizing the combined tagger for each scenario, we quantitatively assessed how much of the discrimination power survives. For example, working at a 50\% top-tag rate at 20~TeV jet energy, mistag rates for gluons below 10\% are likely still achievable.

While tracks in recent studies~\cite{Larkoski:2015yqa} were considered as major components in establishing top-tagging at very high energy, we have pointed out here that electromagnetic calorimetry can serve a comparable and complementary role, and can also be combined to provide even more robust top-tagging. This situation would especially be facilitated by reasonable improvements in existing calorimeter technology (such as tracking calorimeters) as well as flexibility in tagging algorithm.

Our studies regarding algorithm and detector options have been fairly basic, designed only to illustrate a few of the main issues. And of course, at this point in time, we can only speculate on the specifics of possible future detectors. We expect that more sophisticated future studies of substructure approaches and detector reconstruction strategies will continue to yield useful insights and improvements, especially as more aspects of advancing detector technology and detailed detector designs are incorporated. It would be interesting as well to understand what additional improvement can be made by applying modern machine-learning techniques, which might not only pick up on subtle differences in features between top-jets and QCD-jets, but also how those features are being represented within realistic detector signals.

We have also studied novel multi-TeV physics issues related to QCD final-state radiation off of top quarks, splittings of gluons into $t\bar{t}$ pairs, and hadronic $W/Z$ weakstrahlung radiation off of light quarks. FSR from top quarks is in one sense a detrimental effect because the top mass peak becomes less well-resolved due to confusions/overlaps between decay subjets and shower subjets. But the structure of soft, wide-angle radiation from tops is different than that of gluons, as would be the case for any type of color-triplet quark. This feature can be used to construct even more powerful top-taggers by folding in ideas from quark/gluon discrimination, fractionally reducing gluon mistag rates by $\approx 20\%$ in our own simplistic fat-jet track-counting approach. But in the splittings $g\rightarrow t\bar{t}$, we also found a new, non-negligible contribution to the effective gluon mistag rate. At $O$(10~TeV) and 50\% top-tag working point, the absolute mistag contribution is about 1\%. For tighter working points, its contribution may dominate the mistag rate. More refined estimates would benefit from more systematically incorporating $g\rightarrow t\bar{t}$ into modern parton showering programs. We also pointed out that $g\rightarrow t\bar{t}$ may be a very important contribution to {\it leptonic} top-jet mistag rates. Finally, the weakstrahlung contribution, while a known major background to leptonic tops and theoretically interesting in its own right, typically remains highly subdominant to QCD splittings with top-like kinematics at all energies. It would likely only be an important consideration for precision studies.

On the physics side, there remain, as always, lingering questions about the ability to model ``gluon'' and ``quark'' jets in showering simulations, especially regarding their responses to different tagging approaches. While we have not delved into this question in any detail, a more comprehensive understanding of the possible idiosyncrasies of specific shower programs would be quite useful. Information from LHC data on top-tag performance in topologies and kinematic regions dominated by different compositions of gluons and quarks might also help resolve these questions. Any lessons learned from such studies would in principle be easy to scale up to higher energies due to the approximate scale-invariance of QCD.

These issues already illustrate the fact that top-tagging is not always simply an issue of discriminating top-jets against ``QCD jets.'' However, even the top-jets themselves come in two varieties: left-handed and right-handed chirality. Disentangling these two states can be beneficial both for new physics model discrimination as well as for further purifying out a given polarized signal hypothesis against backgrounds. However, the full interplay of discrimination between the four states ($t_L$,$t_R$,$q$,$g$) has not yet been explored. The combined robustness of top-jet polarimetry and tagging at very high energies would also be useful to study in the future.

To conclude, we have established the proof-of-concept for top-tagging at the energy frontier ranging up to $O$(10~TeV) with only modest update of the detectors. Detector limitations do not appear to present a major barrier to maintaining high-quality discrimination, and physics issues are for the most part perturbations to the main story. We hope that our results can serve as a set of conservative benchmarks for future phenomenological studies that seek to incorporate signal and background estimates that account for basic detector effects, as well as provide possible insight into future detector design.


\acknowledgments{We thank Michele Selvaggi for helping us to reproduce the tracker model of~\cite{Larkoski:2015yqa}. ZH thanks University of Oregon for logistical support through the Courtesy Postdoctoral Research Fellow appointment. MS was supported by Samsung Science and Technology Foundation under Project Number SSTF-BA1602-04. BT was supported by DoE grant No. DE-FG02-95ER40896 and by PITT PACC.}


\appendix

\section{Calorimeter Modeling and Effects on Kinematic Distributions}
\label{sec:detectorModel}

\begin{figure*}[t!]
\begin{center}
\includegraphics[width=0.48\textwidth]{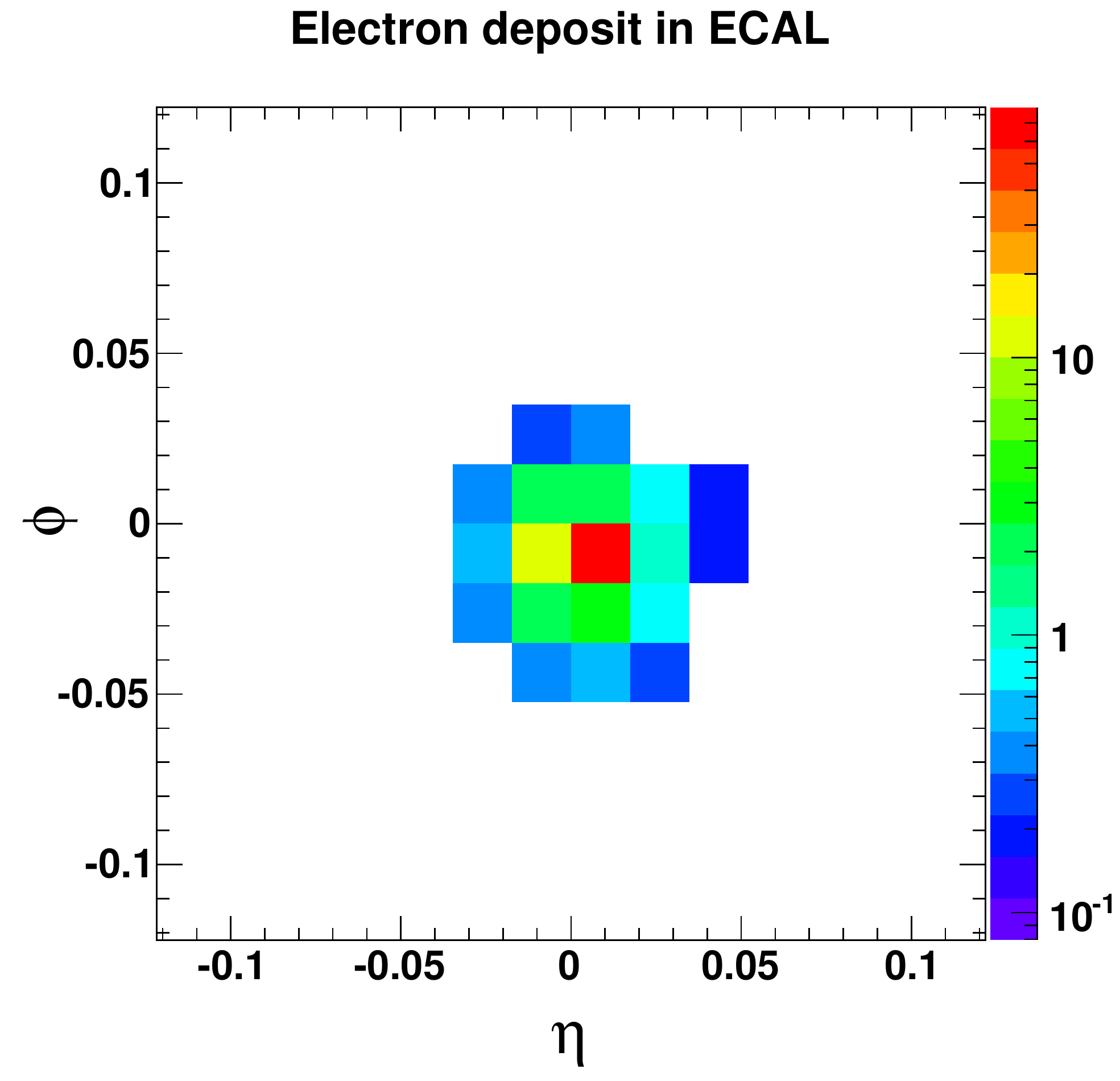} \hspace{0.2cm}
\includegraphics[width=0.48\textwidth]{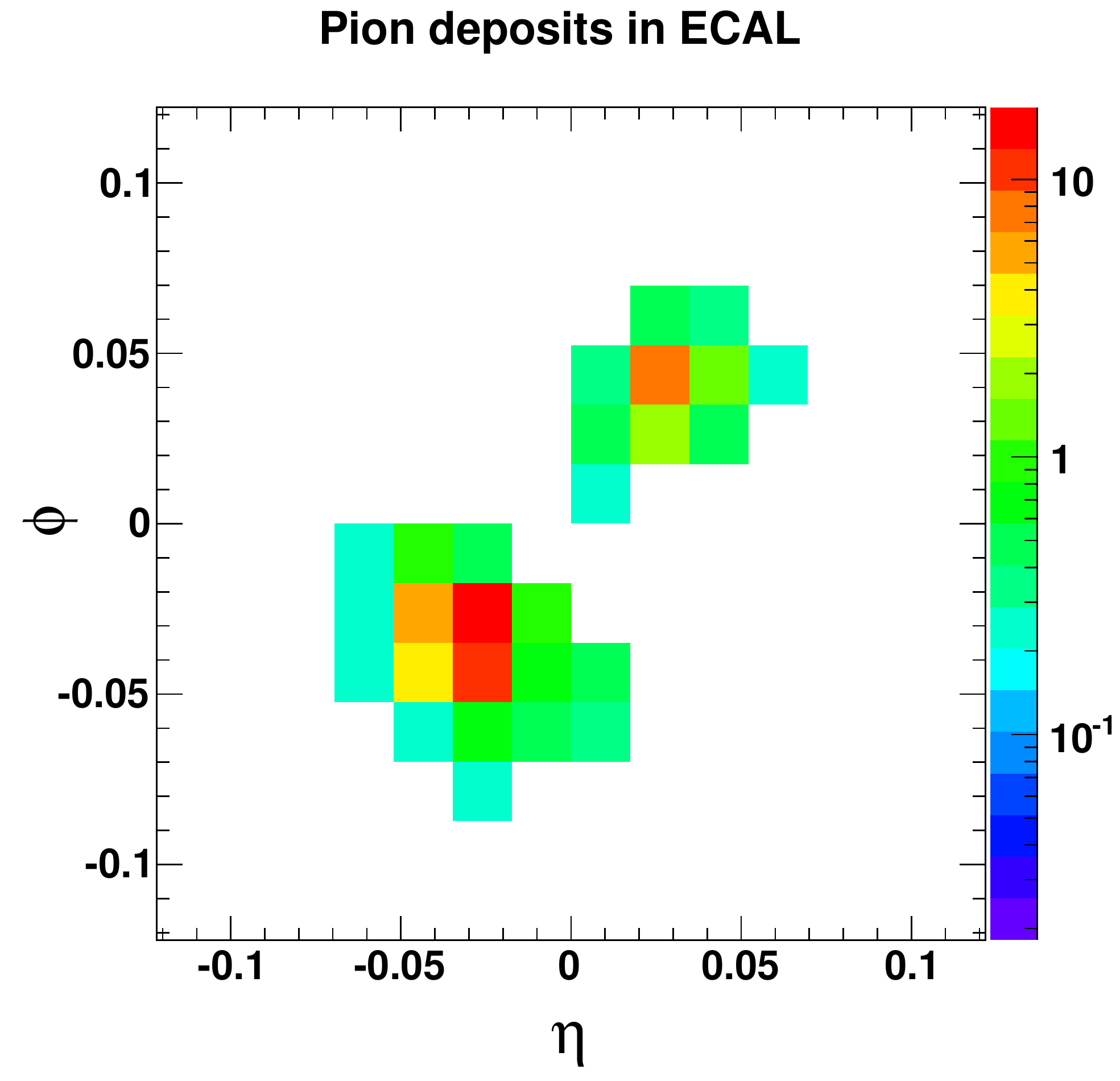}
\caption{Example ECAL energy deposition patterns of a 100~GeV electron ({\bf left}) and two 100~GeV pions ({\bf right}) in a simulated CMS-like detector. The logarithmic color scales are in GeV.}
\label{fig:ECAL}
\end{center}
\end{figure*}

We construct a set of toy ECALs in {\tt GEANT}~\cite{Agostinelli:2002hh}, each consisting of a flat wall of cells of uniform material toward which simulated particles are aimed. Collections of individual particle events, generated at different possible discrete impact points on a central cell, are stored on disk as libraries of energy deposition patterns on the calorimeter grid. In turn, taking particles from our full event simulations in {\tt PYTHIA}, each is replaced by a random impact event drawn from the library, mapped locally onto a barrel calorimeter geometry will cells of fixed $\Delta\eta$ and $\Delta\phi$.\footnote{A projective barrel geometry is actually somewhat forgiving here to our oversimplified modeling. At any value of $\eta$, the cell's physical size transverse to the particle trajectory at impact is always $\approx (\Delta\eta \cdot r) \times (\Delta\phi \cdot r)$, where $r$ is the barrel inner radius.} All particles excepting muons and neutrinos are treated in this manner. This includes hadrons, which in reality have an $O(1)$ chance of encountering a nucleus in the ECAL and depositing an $O(1)$ fraction of their energy before reaching the HCAL. Under the expectation that the shower patterns evolve only logarithmically with energy, and even then mainly only in their longitudinal profile, we use fixed particle energies of 100~GeV. We also use electron-induced showers as proxies for both electrons and photons, and $\pi^+$-induced showers as proxies for all hadrons (including neutrals). Example impacts are shown in Fig.~\ref{fig:ECAL}. Nonlinearities and sampling efficiency effects are not modeled in full detail, nor are any of the subtle aspects of the calorimeter geometry at high-$\eta$ or of impacts at non-projective angles. However, the {\tt GEANT} simulations do account for the undetected fraction of the energy from the hadron-induced events, e.g.\ lost to nuclear binding energy or soft neutrons. To approximately recover this lost energy, we universally rescale the ECAL energy by a ``calibration constant'' of $1.12$. On top of this, we also apply a naive cell-by-cell gaussian energy smearing, using the parameters recommended by~\cite{Larkoski:2015yqa} for a CMS-like detector: $\sigma(E)/E = (0.07~{\rm GeV}^{1/2})/\sqrt{E} \oplus 0.007$. We expect that this treatment conservatively double-counts some of the smearing effects. Regardless, the impact of this energy smearing (in both the ECAL and HCAL) tends to be quite subdominant to that of the fluctuations in jet energy sampled by the ECAL and the geometric smearing.

Energy flowing out of the back of the ECAL is used as input into the HCAL. We model the HCAL in a much more simplistic manner, since it catches almost all remaining energy, and the detailed angular deposition patterns at the scale of individual cells are largely integrated-out by our mini-jet clustering (described in Section~\ref{sec:detector_strategies}). We replace any incoming particle (or collectively all particles flowing out the back of an ECAL cell) with a continuous angular energy distribution according to the profile anzatz of Grindhammer, et al~\cite{Grindhammer:1989zg}: $f(r) \propto 2r/(r^2+R^2)^2$, setting $R = 1/3$ of a full cell width. Empirically, this choice reproduces the transverse shower profile observed in pion test-beam data by CMS~\cite{Baiatian:2007xva} (roughly 75\% containment in a centrally struck cell, 95\% containment in a $3\times 3$ array about it). In practice, we construct pattern libraries analogous to the ECAL, but with only one ``average'' event per discrete impact location. The HCAL cell energies are also smeared, again as in~\cite{Larkoski:2015yqa} for a CMS-like detector: $\sigma(E)/E = (1.5~{\rm GeV}^{1/2})/\sqrt{E} \oplus 0.05$. 

\begin{figure*}[t!]
\begin{center}
\includegraphics[width=0.48\textwidth]{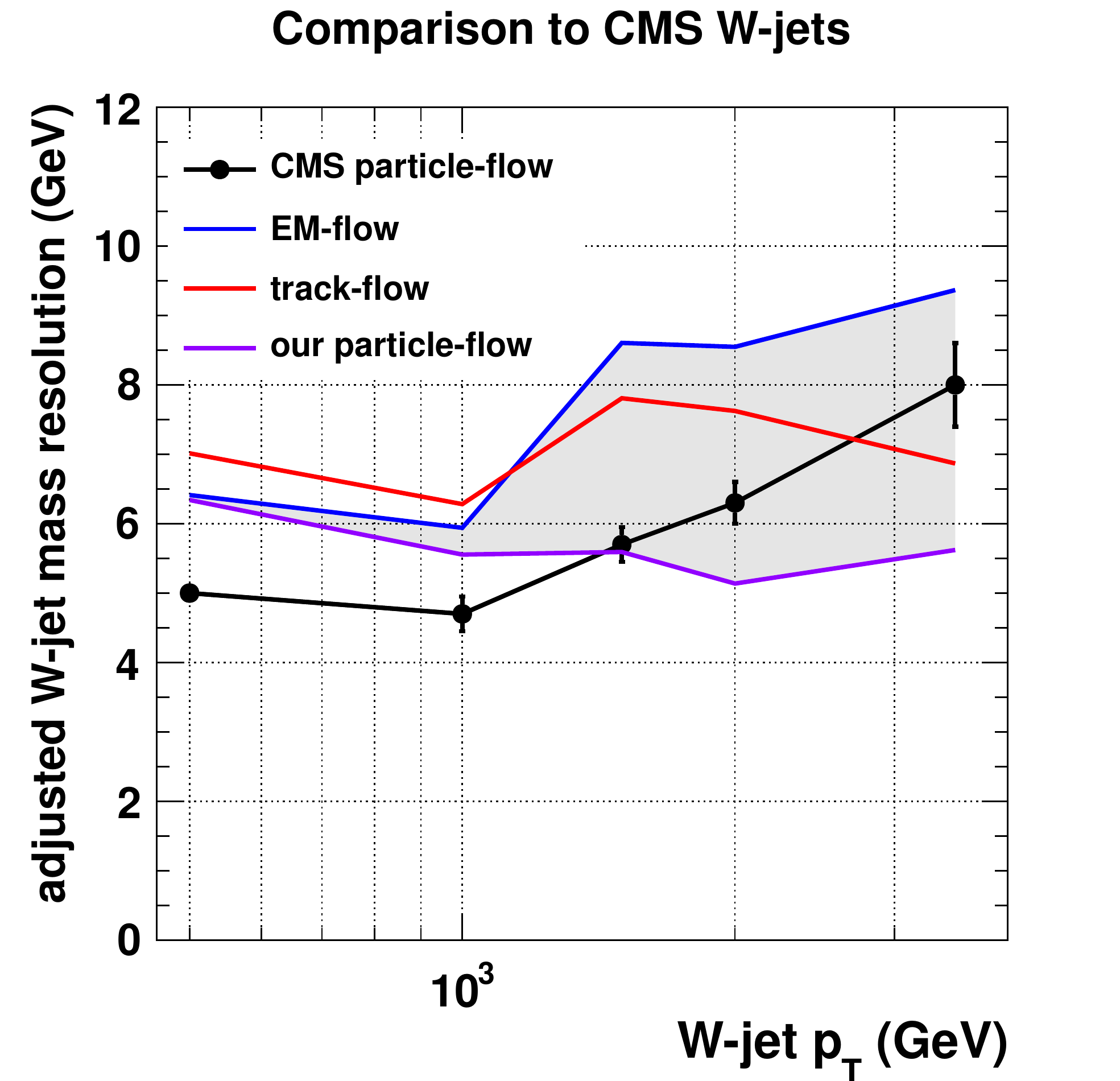}
\caption{Comparison between our detector reconstructions and CMS particle-flow~\cite{CMSboostedW}, applied to the gaussian-core mass resolution of high-$p_T$ $W$-jets. For our reconstructions, we adjust the resolution by a factor of $m_W/\vev{m}$, as our central $W$-jet mass scale can drift up to 90~GeV for EM-flow, whereas CMS stays closer to $m_W$ throughout. The light gray band represents our range of estimated optimal performances between the extreme assumptions of no tracking and perfect tracking.}
\label{fig:CMScomparison}
\end{center}
\end{figure*}

Given the existence of the CMS highly-boosted $W$ study~\cite{CMSboostedW}, we take the opportunity to compare against our approximate approach to detector modeling. We generate continuum $WZ$ events at a 13~TeV LHC in {\tt PYTHIA}, in narrow partonic $p_T$ slices.\footnote{The CMS study was run on simulation samples of Randall-Sundrum graviton decays to $WW$, which would yield mostly-transverse $W$ bosons. The $W$'s in our continuum diboson samples should similarly be mostly-transverse, an effect that {\tt PYTHIA} models through its four-fermion matrix element corrections. In both cases the high-$p_T$ $W$'s are also expected to be mostly central.} The $W$ decays hadronically, the $Z$ invisibly. We model the CMS ECAL as a uniform grid of lead tungstate, with cell width $2.2$~cm and depth 23~cm, mapped to $\eta$-$\phi$ width $0.0174$. The HCAL cell width is $0.087$ in $\eta$-$\phi$. C/A jets are formed with $R=0.8$, and we take the hardest as our $W$-jet candidate. The mini-jet radius is defined to be $1.2$ times larger than one HCAL cell width. CMS actually uses jet pruning~\cite{Ellis:2009su} with $z_{\rm cut} = 0.1$ and $D_{\rm cut} = 0.5$ before defining its $W$-jet mass, whereas we instead run our JHU declustering a single stage with $\delta_p = 0.1$. We expect the two methods to perform fairly similarly. The resolution on the $W$ is defined, as per CMS, by iteratively gaussian-fitting the mass distribution in a window $\pm 1\sigma$ about the mean, using the fit parameters of the previous iteration. (Initializing with mean and sigma near $m_W$ and a few GeV, respectively, the result usually converges within three or four steps.) We show the comparison of our three reconstruction of Section~\ref{sec:detector_strategies} to full CMS particle-flow, in Fig.~\ref{fig:CMScomparison}. It can be seen that EM-flow and track-flow almost always perform worse than CMS particle-flow, though the high stability of track-flow with perfect tracking eventually allows it to overtake. Our own idealization of particle-flow roughly straddles CMS, performing slightly worse at lower $p_T$'s and better at higher $p_T$'s. The former behavior is likely because CMS uses the detector information more intelligently than we do, and the latter behavior is probably due to the fact that CMS tracking begins to falter whereas again our tracking is perfect. Notably, CMS particle-flow lies in between our EM-flow and our particle-flow at high $p_T$, which is exactly what we would expect for a realistic particle-flow method with imperfect tracking.

\begin{figure*}[t!]
\begin{center}
\includegraphics[width=0.48\textwidth]{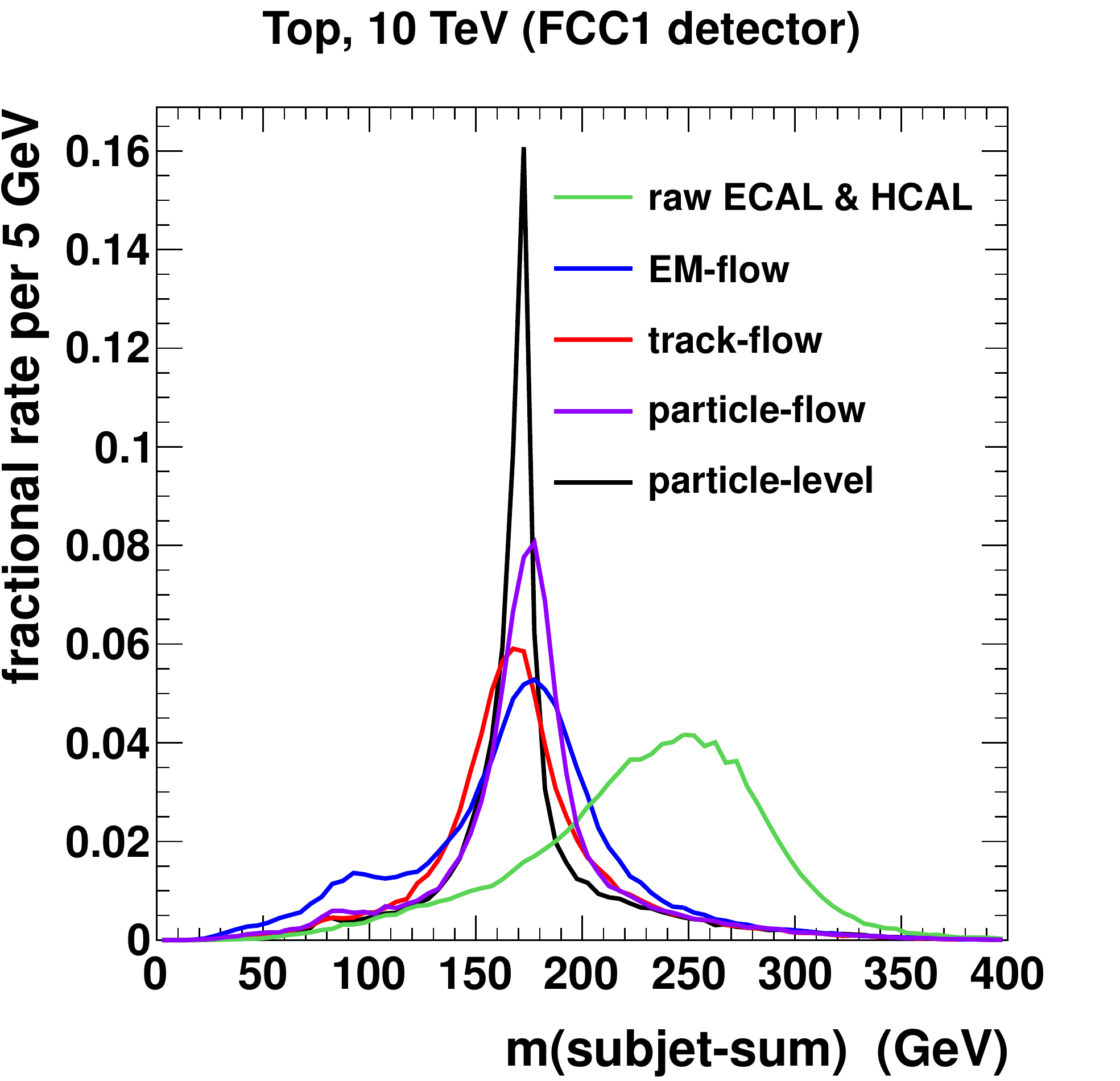} \hspace{0.2cm}
\includegraphics[width=0.48\textwidth]{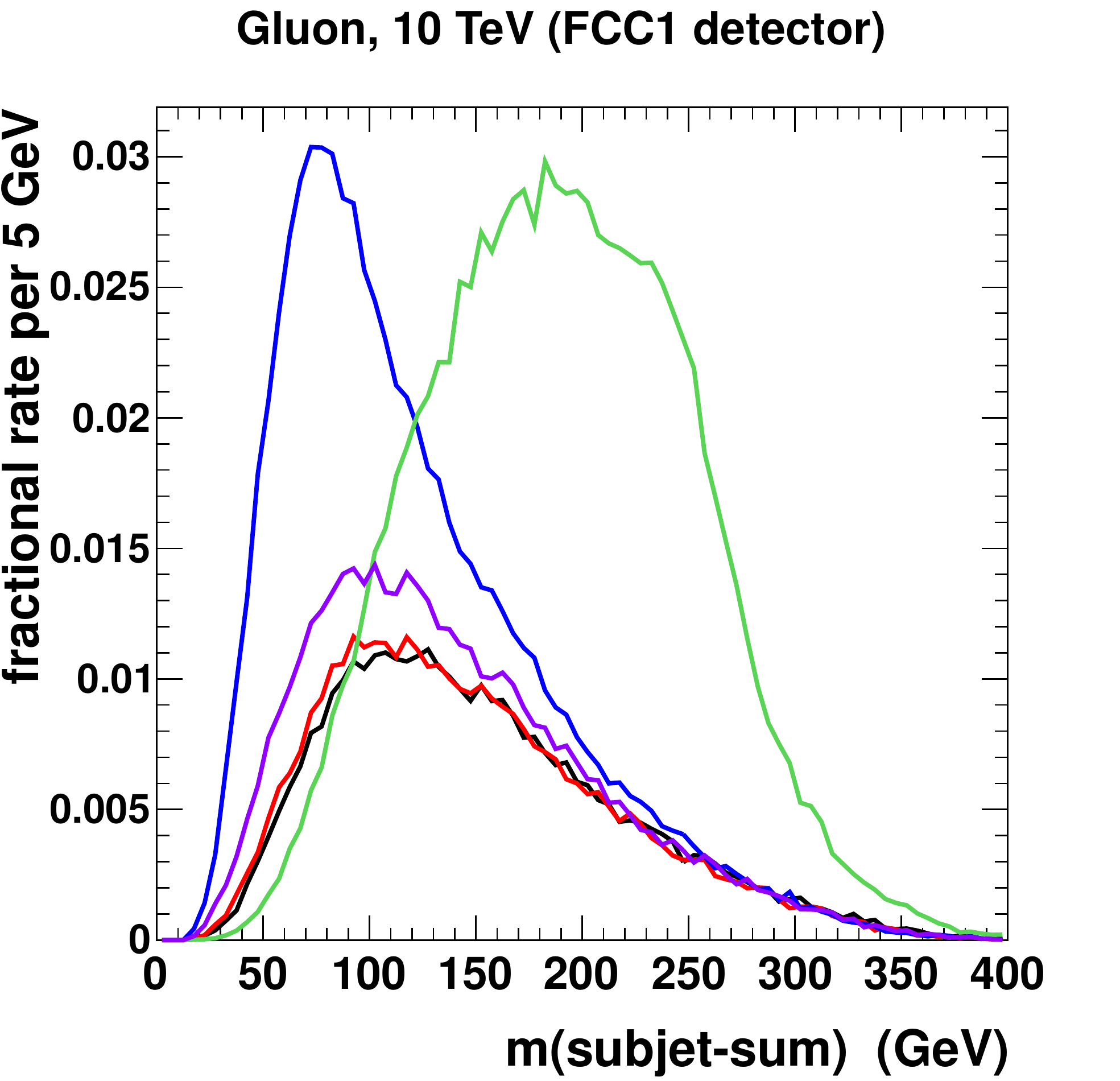}
\caption{Distributions of absolute reconstruction rate versus the subjet-sum mass for 10~TeV top-jets ({\bf left}) and gluon-jets ({\bf right}), passed through the FCC1 detector and reconstructed via different methods. A minimum requirement $N_{\rm subjets} >= 3$ has already been applied.}
\label{fig:mJ}
\end{center}
\end{figure*}

\begin{figure*}[t!]
\begin{center}
\includegraphics[width=0.48\textwidth]{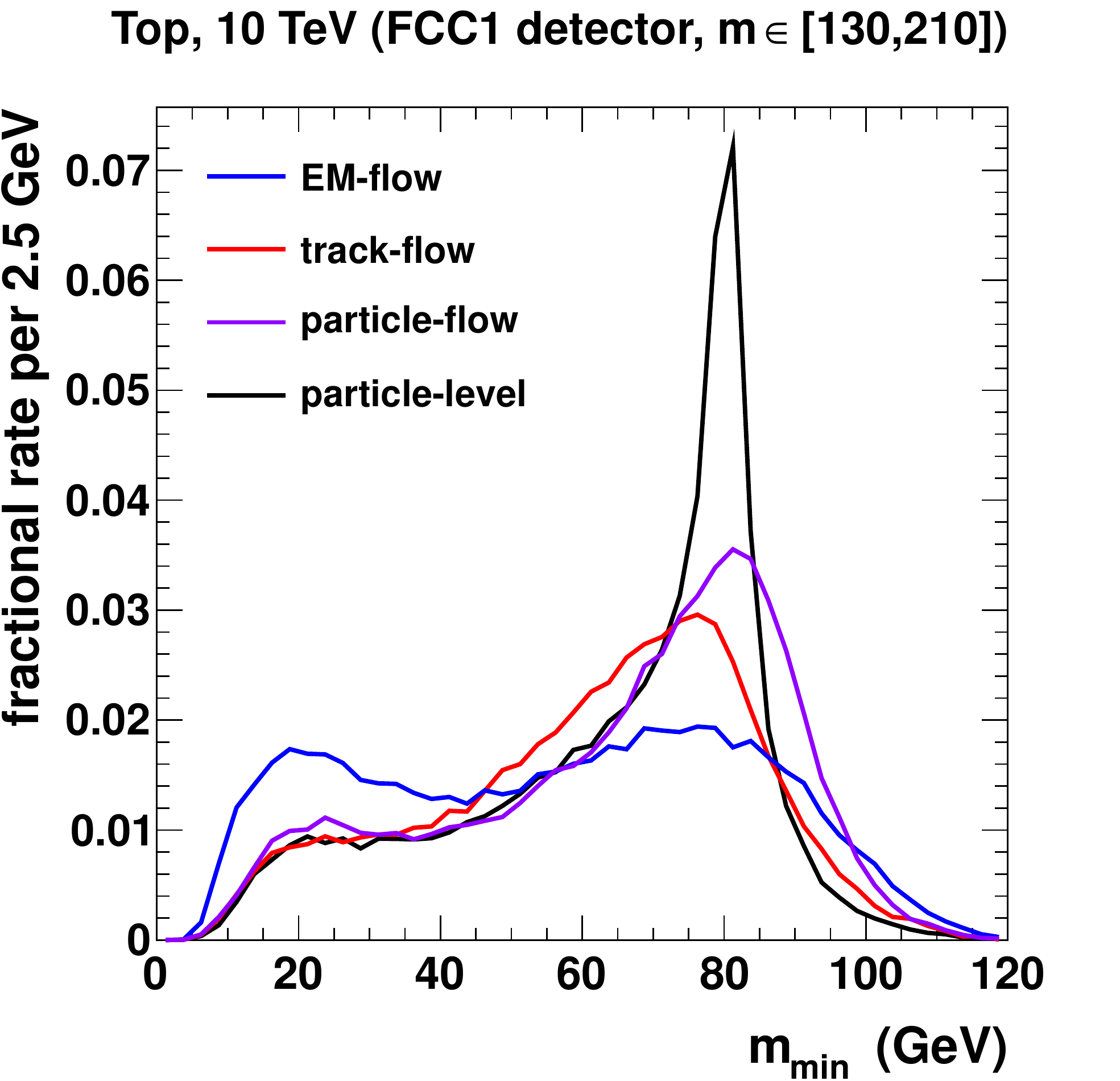} \hspace{0.2cm}
\includegraphics[width=0.48\textwidth]{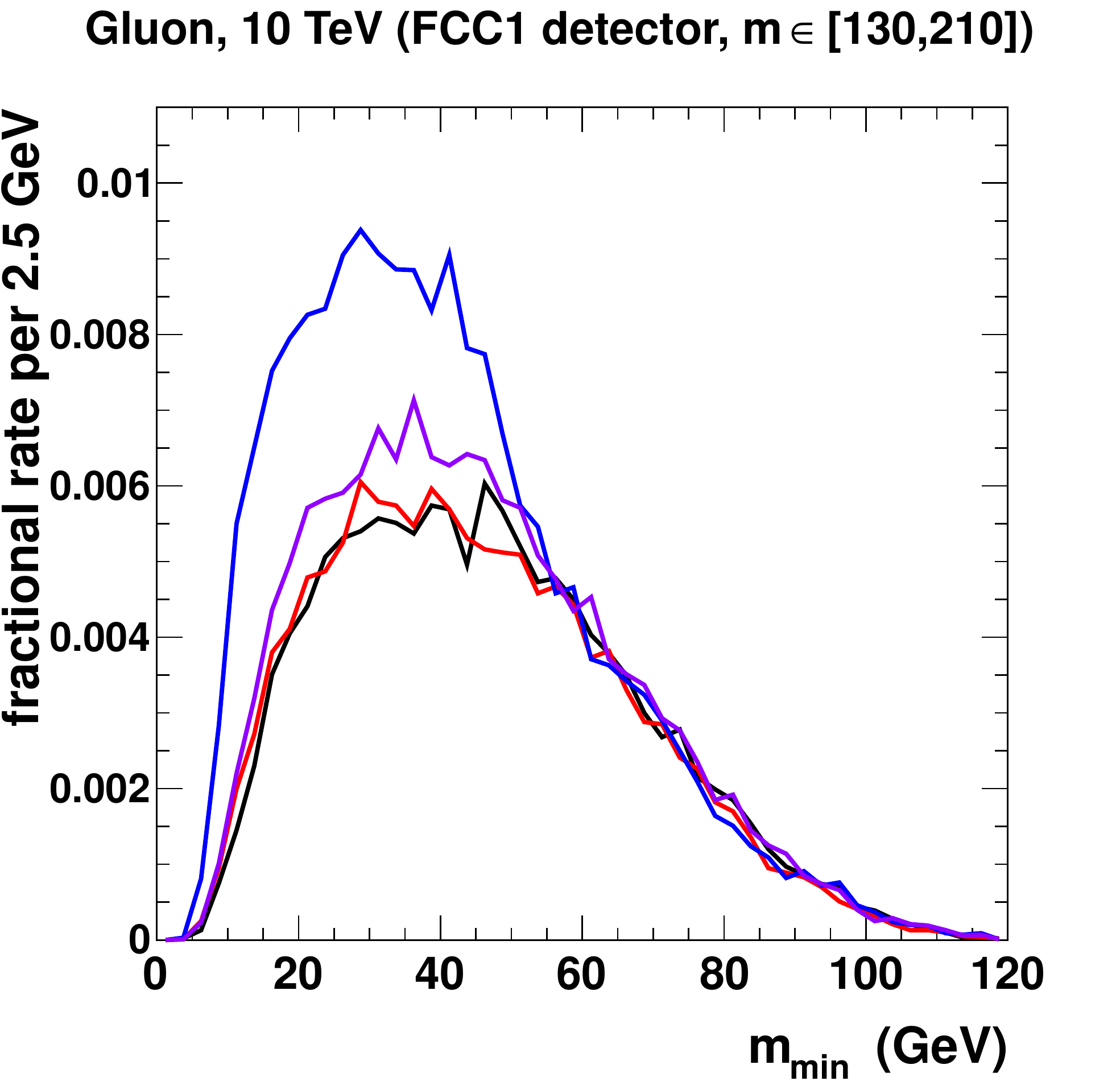}
\caption{Distributions of absolute reconstruction rate versus $m_{\rm min}$ for 10~TeV top-jets ({\bf left}) and gluon-jets ({\bf right}), passed through the FCC1 detector and reconstructed via different methods. A subjet-sum mass window requirement of $[130,210]$~GeV has been applied.}
\label{fig:mMin}
\end{center}
\end{figure*}

\begin{figure*}[t!]
\begin{center}
\includegraphics[width=0.48\textwidth]{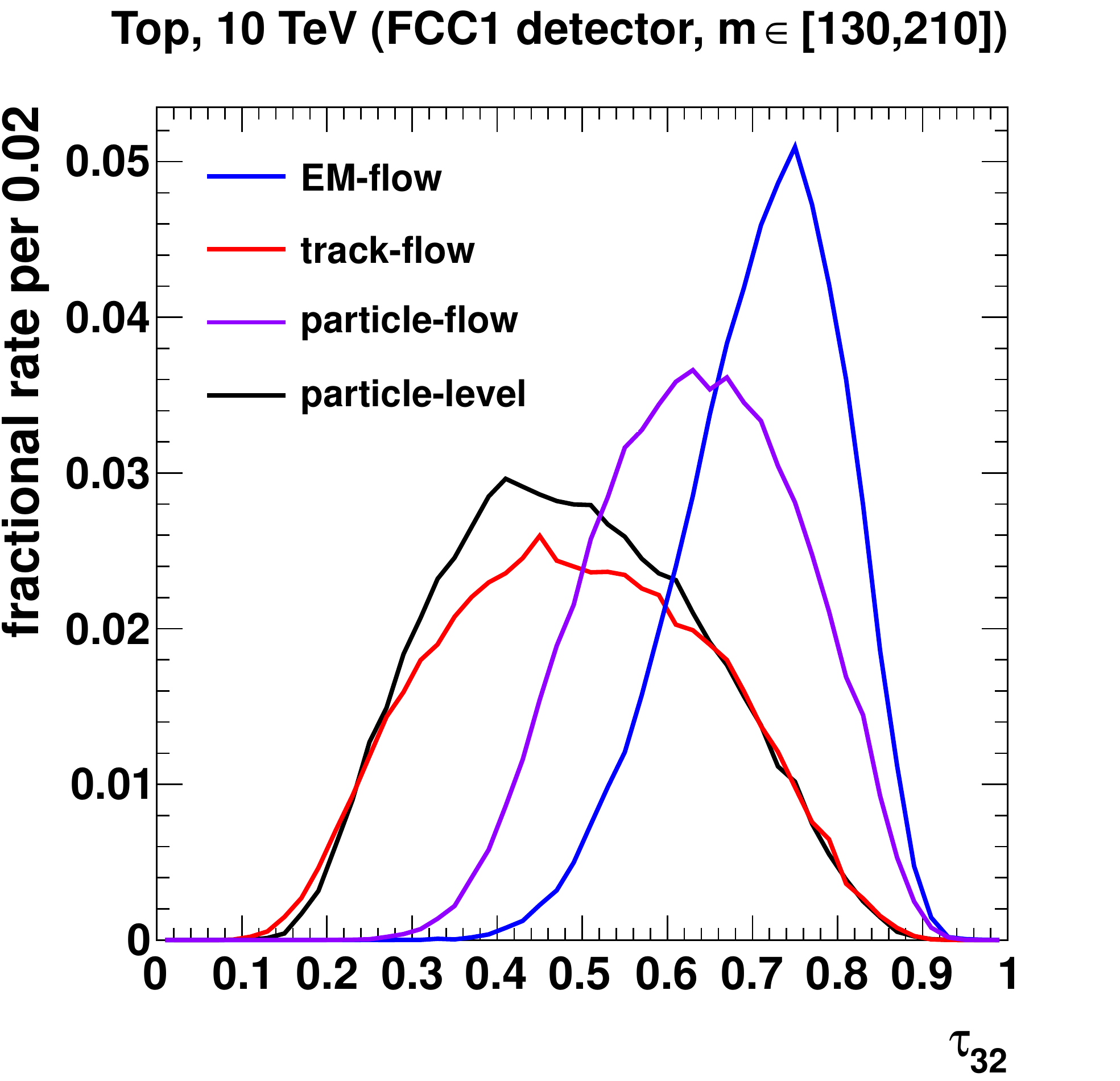} \hspace{0.2cm}
\includegraphics[width=0.48\textwidth]{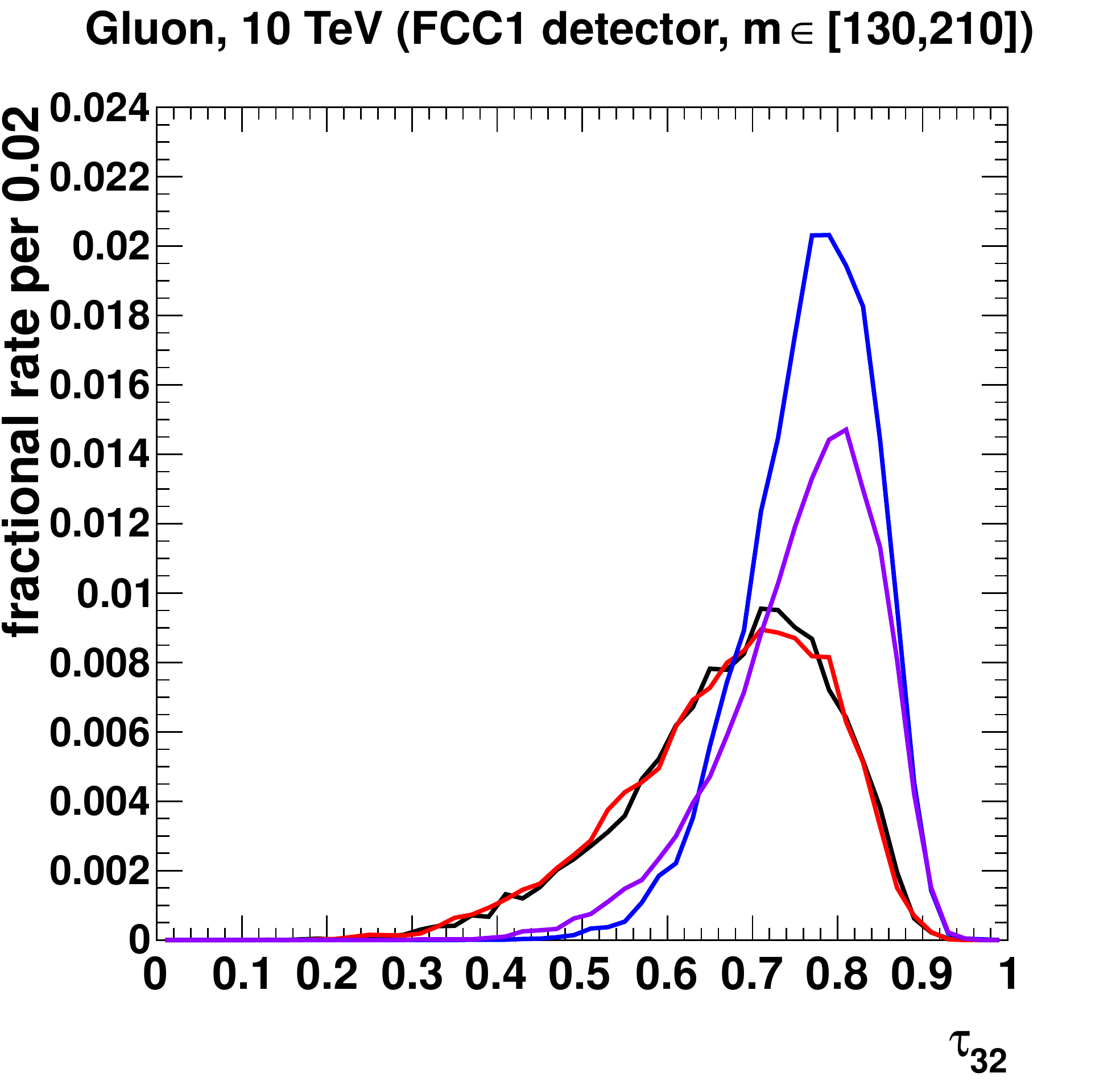}
\caption{Distributions of absolute reconstruction rate versus $\tau_{32}$ for 10~TeV top-jets ({\bf left}) and gluon-jets ({\bf right}), passed through the FCC1 detector and reconstructed via different methods. A subjet-sum mass window requirement of $[130,210]$~GeV has been applied.}
\label{fig:tau32}
\end{center}
\end{figure*}

Finally, we give some indication of how the detector model affects the reconstructed substructure observables. We use the FCC1 model introduced in Section~\ref{sec:detector}, which is essentially the CMS detector expanded in size by a factor of two. As an example $p_T$ region, we choose 10~TeV, which is where EM-flow starts to show a significant degradation with this detector choice, and our particle-flow performance becomes approximately degenerate with (perfect) track-flow. Figs.~\ref{fig:mJ} through~\ref{fig:tau32} show, respectively, the distributions of the subjet-sum mass, $m_{\rm min}$, and $\tau_{32}$ for tops and gluons. Raw calorimetry, shown only in Fig.~\ref{fig:mJ} for reference, has practically failed completely. All of the other reconstructions manage to recover a sane top mass peak, with particle-flow giving the closest approximation to particle-level. However, track-flow more closely follows the particle-level distributions for background, a result that persists for the other two observables. Cutting into the region around the top peak, Fig.~\ref{fig:mMin} shows the subsequent $m_{\rm min}$ distribution, which is more degraded for EM-flow than for the other reconstructions. Similarly, Fig.~\ref{fig:tau32} shows the $\tau_{32}$ distribution for jets near the top peak. The variable exhibits very little discrimination power for EM-flow, and discrimination power intermediate to particle-level for particle-flow. Note that, for lower $p_T$'s or more finely segmented detectors, the various reconstruction methods all approach much closer to particle-level, and to one another.


\bibliography{lit}
\bibliographystyle{apsper}

\end{document}